\documentclass[journal]{IEEEtran}
\usepackage{cite}
\ifCLASSINFOpdf
  \usepackage[pdftex]{graphicx}
  \graphicspath{{../pdf/}{../jpeg/}}
  \DeclareGraphicsExtensions{.pdf,.jpeg,.png}
\else
  \usepackage[dvips]{graphicx}
  \graphicspath{{PPLS-D_paper_IEEE_jrnl_resource/}}
  \DeclareGraphicsExtensions{.eps}
\fi
\usepackage{amsmath}
\usepackage{algorithm}
\usepackage{algorithmic}
\usepackage{array}

\usepackage{stfloats}

\usepackage{multirow}
\usepackage{subfigure}
\usepackage{amsfonts,amssymb}

\usepackage{color}

\DeclareMathOperator*{\argmax}{arg\,max}

\newcommand{\Argmax}[1]{\raisebox{0.5ex}{\scalebox{0.8}{$\displaystyle \argmax_{#1}\;$}}}

\newcommand{\themax}[1]{\raisebox{0.5ex}{\scalebox{0.8}{$\displaystyle \max_{#1}\;$}}}

\usepackage{changepage}
\usepackage{hyperref}

\begin{document}
\title{PPLS/D: Parallel Pareto Local Search based on Decomposition}

\author{Jialong~Shi,
        Qingfu~Zhang,
        and~Jianyong~Sun
        \thanks{Jialong Shi is with the School of Mathematics and Statistics, Xi'an Jiaotong University, Xi'an, China (e-mail: jialong.shi@xjtu.edu.cn).}
        \thanks{Qingfu Zhang is with the Department of Computer Science, City University of Hong Kong, Hong Kong SAR (e-mail: qingfu.zhang@cityu.edu.hk).}
        \thanks{Jianyong Sun is with the School of Mathematics and Statistics, Xi'an Jiaotong University, Xi'an, China (e-mail: jy.sun@xjtu.edu.cn).}}
\maketitle

\begin{abstract}
  Pareto Local Search (PLS) is a basic building block in many metaheuristics for Multiobjective Combinatorial Optimization Problem (MCOP). In this paper, an enhanced PLS variant called Parallel Pareto Local Search based on Decomposition (PPLS/D) is proposed. PPLS/D improves the efficiency of PLS using the techniques of parallel computation and problem decomposition. It decomposes the original search space into $L$ subregions and executes $L$ parallel processes searching in these subregions simultaneously. Inside each subregion, the PPLS/D process is guided by a unique scalar objective function. PPLS/D differs from the well-known Two Phase Pareto Local Search (2PPLS) in that it uses the scalar objective function to guide every move of the PLS procedure in a fine-grained manner. In the experimental studies, PPLS/D is compared against the basic PLS and a recently proposed PLS variant on the multiobjective Unconstrained Binary Quadratic Programming problems (mUBQPs) and the multiobjective Traveling Salesman Problems (mTSPs) with at most four objectives. The experimental results show that, no matter whether the initial solutions are randomly generated or generated by heuristic methods, PPLS/D always performs significantly better than the other two PLS variants.
\end{abstract}

%\begin{IEEEkeywords}
%keywords
%\end{IEEEkeywords}

% \fi
\IEEEpeerreviewmaketitle

\section{Introduction}
A Multiobjective Combinatorial Optimization Problem (MCOP) is defined as follows:
\begin{equation}\label{eq:MCOP}
  \begin{array}{ll}
    \mbox{maximize~/~minimize} & \ \ F(x) = (f_1(x),\dots,f_m(x) )\\
    \mbox{subject to} & \ \ x\in\mathcal{S}\\
  \end{array}
\end{equation}
where the solution space $\mathcal{S}$ is a finite set and $F:\mathcal{S}\to\mathbb{R}^m$ is the objective vector function which contains $m$ objectives. In the following discussions we focus on the maximization MCOPs. There exists a trade-off between different objectives. Usually, no single solution can optimize all the objectives. The following definitions are used in multiobjective optimization:
\begin{itemize}
  \item \textbf{Definition 1} A vector $u=(u_1,\dots,u_m)$ is said to \emph{dominate} a vector $v=(v_1,\dots,v_m)$, if and only if $u_k\geq v_k,\ \forall k\in\{1,\dots,m\}\ \land\ \exists k\in\{1,\dots,m\}: u_k>v_k$, denoted as $u \succ v$.
  \item \textbf{Definition 2} If $u$ is not dominated by $v$ and $v$ is not dominated by $u$, we say that $u$ and $v$ are \emph{non-dominated} to each other, denoted as $u \nprec v$ or $v \nprec u$. A solution set $A$ is called a \emph{non-dominated set} if and only if for any $x,y\in A$, $F(x)\nprec F(y)$.
  \item \textbf{Definition 3} A feasible solution $x^* \in \mathcal{S}$ is called a \emph{Pareto optimal solution}, if and only if $\nexists y \in \mathcal{S}$ such that $F(y) \succ F(x^*)$.
  \item \textbf{Definition 4} For problem (\ref{eq:MCOP}), the set of all the Pareto optimal solutions is called the \emph{Pareto Set} (PS), denoted as $\mbox{PS} = \{x\in\mathcal{S}|\nexists y \in \mathcal{S}, F(y) \succ F(x)\}$ and the Pareto front (PF) is defined as $\mbox{PF}=\{F(x)\vert x\in \mbox{PS}\}$.
\end{itemize}
The PS represents the best trade-off solutions and the PF represents their objective values. Hence the PS and PF constitute highly valuable information to the decision maker. The goal of a multiobjective metaheuristic is to approximate the PS and PF.

Many state-of-the-art multiobjective metaheuristics use Pareto Local Search (PLS), which was first proposed in \cite{paquete2004pareto}, as a basic building block\cite{dubois2015anytime,liefooghe2012dominance,ke2014hybridization}. PLS can be applied at the very beginning of a metaheuristic to obtain a set of high quality solutions as the starting points of the following optimization procedure. PLS also can be applied in the middle or final stage of a metaheuristic to further improve a set of high quality solutions. The main drawback of PLS is that it requires a long time to reach a good approximation of the PF. To overcome this drawback, several sequential speed-up strategies have been proposed~\cite{dubois2015anytime,inja2014queued,liefooghe2012dominance}. In this paper, we propose to use parallel computation and problem decomposition to speed up PLS. The proposed parallel PLS variant is called Parallel PLS based on Decomposition (PPLS/D), which has the following features:
\begin{itemize}
  \item PPLS/D decomposes the original search space into $L$ subregions by defining $L$ weight vectors in the objective space.
  \item PPLS/D executes $L$ processes simultaneously to exploit the computational resource of multi-core computers. Each process maintains a unique archive of solutions and searches in its own subregion.
  \item During the search, each PPLS/D process is guided by a unique scalar objective function (generated by Tchebycheff approach) in a fine-grained manner. PPLS/D is different from the well-known Two-Phase Pareto Local Search (2PPLS)~\cite{lust2010two} which combines a scalar objective heuristic phase and a PLS phase in a coarse-grained manner. In PPLS/D, the scalar function influences every move of the algorithm during the entire search process.
\end{itemize}

In the experimental studies, the multiobjective Unconstrained Binary Quadratic Programming problem (mUBQP) and the multiobjective Traveling Salesman Problem (mTSP) with $m$ = 2, 3 and 4 are selected as the test suites. In addition, two application scenarios are considered: starting from randomly generated solutions and starting from high quality solutions. In both scenarios, the performance of PPLS/D with different process numbers is compared with that of the basic PLS and a speed-up PLS variant proposed by~\cite{dubois2015anytime}. The results show that PPLS/D is significantly faster than the other two PLS variants on both problems and both scenarios. The influence of the process number in PPLS/D also is investigated.

Some preliminary work of this paper has been published in \cite{shi2017using}. The work in this paper differs from the work in \cite{shi2017using} in the following aspects.
\begin{itemize}
  \item The PLS speed-up strategies proposed in \cite{shi2017using} can only handle bi-objective MCOPs. In this paper, the proposed PPLS/D can handle more than two objectives by using the region decomposition method proposed by Liu et al.~\cite{liu2014decomposition}.
  \item In \cite{shi2017using}, the scalar objective functions that guide the search are generated by the weighted sum approach, while in this paper the scalar objective functions are generated by the Tchebycheff approach. The advantages of using the Tchebycheff scalar objective function are discussed in Section~\ref{sec:decomp}.
  \item In \cite{shi2017using}, several speed-up strategies are proposed for PLS, but the details of the strategies are not given. In this paper, a formal algorithm, PPLS/D, is proposed based on the strategies proposed in \cite{shi2017using} and the aforementioned improvements. The detailed pseudo-code of PPLS/D is given in this paper.
  \item In the experimental studies of \cite{shi2017using}, the test problem is the bi-objective mUBQP (maximization problem) and the algorithms are started from randomly generated solutions. In this paper, the mUBQP and the mTSP (minimization problem) with two, three and four objectives are used as the test suites, and the algorithms are started from randomly generated solutions and high quality solutions.
  \item In \cite{shi2017using}, there is no investigation about how the algorithm performance changes with different process numbers, while in this paper we analysis the influence of the process number.
  %\item In this paper, a variant of PPLS/D, called PPLS/D-AE is proposed.
\end{itemize}

The rest of this paper is organized as follows. Section~\ref{sec:related_work} introduces the recent works on improving the basic PLS algorithm. In Section~\ref{sec:PPLSD} the details of the proposed PPLS/D are presented. In Section~\ref{sec:epm} the experimental studies have been conducted to show that PPLS/D can significantly speed up the basic PLS. Section~\ref{sec:conclusion} concludes this paper.

\section{Related Works}\label{sec:related_work}
PLS is a problem-independent search method to approximate the PF. It is an extension of the single objective local search on MCOPs. %Given a pre-defined neighborhood structure in the solution space, PLS iteratively improves an archive of non-dominated solutions by exploring the neighborhood of the solutions in the archive. If a candidate solution is not dominated by any solutions in the archive, it will be accepted by the archive. The solutions in the archive that are dominated by the newly accepted solution will be deleted.
Algorithm~\ref{alg:PLS} shows the procedure of the basic PLS~\cite{paquete2004pareto}. In Algorithm~\ref{alg:PLS}, Update($x^\prime$, $A$) means adding $x^\prime$ to the archive $A$ and deleting the solutions in $A$ that are dominated by $x^\prime$.
%%\vspace{-0.1in}
\begin{algorithm}
    \begin{algorithmic}[1]
        \STATE \textbf{input:} An initial set of non-dominated solutions $A_0$
        \STATE $\forall x \in A_0$, set $Explored(x) \gets \mbox{FALSE}$
        \STATE $A \gets A_0$
        \WHILE {$A_0 \neq \emptyset$}
            \STATE $x_0\gets $ a randomly selected solution from $A_0$ %%%%%%%%%%%%\hfill \textit{// selection step}
            \FOR {each $x^\prime$ in the neighborhood of $x_0$}%%%%%%%%%%%%% \hfill \textit{// neighborhood exploration}
                \IF {$x^\prime$ is not dominated by any solution in $A$}
                    \STATE $A \gets \mbox{Update}(x^\prime,A)$
                    \STATE $Explored(x^\prime) \gets \mbox{FALSE}$ %%%%%%%%%%%%%%%%%\hfill \textit{// acceptance criterion}
                \ENDIF
            \ENDFOR
            \STATE $Explored(x_0) \gets \mbox{TRUE}$
            \STATE $A_0 \gets \{x \in A ~\vert~ Explored(x) \mbox{ == FALSE}\}$
        \ENDWHILE
        \STATE \textbf{return} {$A$}
    \end{algorithmic}
\caption{Pareto Local Search}
\label{alg:PLS}
\end{algorithm}

The main drawback of the basic version of PLS (Algorithm~\ref{alg:PLS}) is that it needs a large amount of time to find a good approximation of the PF. Recently, several sequential PLS variants have been proposed to speed up the basic PLS. Inja et al.~\cite{inja2014queued} proposed the Queued PLS (QPLS). In QPLS a queue of high-quality unsuccessful candidate solutions are maintained and their neighborhoods are explored. In the work of Liefooghe et al.~\cite{liefooghe2012dominance}, the PLS procedure is partitioned into three algorithmic components. For each algorithmic component, several alternatives are proposed. By combining the alternatives of different algorithmic components, a number of PLS variants are proposed and tested in \cite{liefooghe2012dominance}. Dubois-Lacoste et al.~\cite{dubois2015anytime} improved the basic PLS in a way similar to \cite{liefooghe2012dominance}, but the differences are that Dubois-Lacoste et al. decomposed the PLS into four algorithmic components and the PLS variants they proposed can switch strategies during a single run. In addition, some archive size limiting strategies are proposed in \cite{dubois2015anytime} for the bi-objective optimization case. Shi et al.~\cite{shi2017using} investigated several parallel strategies of PLS and tested their performance on bi-objective problems.

Since Zhang and Li~\cite{zhang2007moea} proposed the Multiobjective Evolutionary Algorithm based on Decomposition (MOEA/D), the decomposition based framework has attracted some research effort in evolutionary multiobjective optimization community~\cite{trivedi2017survey}. Liu et al.~\cite{liu2014decomposition} proposed MOEA/D-M2M which decomposes the original search space into a number of subregions by calculating the acute angles to a set of pre-defined weight vectors. For each subregion, MOEA/D-M2M maintains a unique sub-population and all sub-populations evolve in a collaborative way. Recently, several variants of MOEA/D-M2M~\cite{liu2016evolutionary,liu2017adaptively} have been proposed in which the subregions are adaptively adjusted during the evolutionary process. In a parallel NSGA-II variant proposed by Branke et al.~\cite{branke2004parallelizing}, the search space also is decomposed into several subregions. However in \cite{branke2004parallelizing} the decomposition is based on cone separation and it only can be applied to the problems with not more than three objectives. Derbel et al.~\cite{derbel2016multi} combined the single-objective local search move strategies with the MOEA/D framework to optimize the bi-objective traveling salesman problems. Ke et al.~\cite{ke2014hybridization} proposed the Multiobjective Memetic Algorithm based on Decomposition (MOMAD), which is one of the state-of-the-art algorithms for MCOPs. At each iteration of MOMAD, a PLS process and multiple scalar objective search processes are executed successively.

There are some studies that try to improve the basic PLS by involving a higher-level control mechanism. Alsheddy and Tsang~\cite{alsheddy2010guided} propose the Guided Pareto Local Search (GPLS), in which a penalization mechanism is applied to prevent PLS from premature. Based on a variable neighborhood search framework, Geiger~\cite{geiger2011decision} proposed the Pareto Iterated Local Search (PILS) to improve the result quality of PLS. The Two-Phase Pareto Local Search (2PPLS) is proposed by Lust and Teghem~\cite{lust2010two}, in which the PLS process starts from the high-quality solutions generated by a heuristic method. Two enhanced 2PPLS variants can be found in \cite{lust2010speed,jaszkiewicz2017proper}. In the work of Drugan and Thierens~\cite{drugan2012stochastic}, different neighborhood exploration strategies and restart strategies of PLS are discussed.

\section{Parallel Pareto Local Search based on Decomposition}\label{sec:PPLSD}
In this section we introduce the mechanism of PPLS/D on a maximization MCOP case. For the minimization MCOPs, one can first convert it into a maximization problem and then apply the PPLS/D. The key component of PPLS/D is its decomposition strategy.

\subsection{Decomposition Strategy}\label{sec:decomp}
The decomposition strategy of PPLS/D is inspired by MOEA/D~\cite{zhang2007moea} and MOEA/D-M2M~\cite{liu2014decomposition}. In MOEA/D, the original MOP is decomposed into a number of scalar subproblems. In MOEA/D-M2M, the original search space is decomposed into a number of subregions. PPLS/D integrates these two decomposition strategies.

By defining multiple weight vectors in the objective space, PPLS/D first decomposes the original search space into several subregions. Then each subregion is searched by an assigned parallel process. Due to the partition of subregions, each parallel process only needs to approximate a part of the PF. In addition, inside each subregion, the weight vector defines a scalar objective function to guide every move of the search process. Under the guidance of the scaler objective function, each parallel process can quickly approach its corresponding PF part.

\begin{figure}
  %%\vspace{-0.1in}
  \centering
  \includegraphics[width=\linewidth]{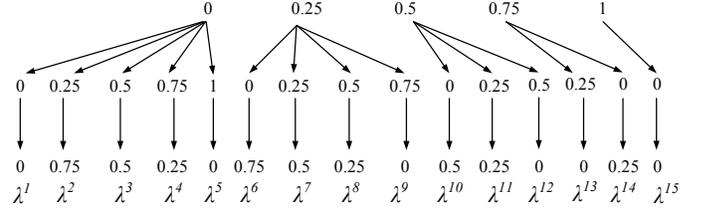}\\
  %\vspace{-0.1in}
  \caption{$L=15$ weight vectors are generated when $m=3$ and $H=4$}\label{fig:wv_generator}%%\vspace{-0.7cm}
  %\vspace{-0.2in}
\end{figure}

PPLS/D first defines $L$ weight vectors $\lambda^1,\dots,\lambda^L$. Each weight vector $\lambda^\ell = (\lambda^\ell_1,\dots,\lambda^\ell_m)$ satisfies $\sum^m_{k=1}\lambda^\ell_k=1$ and $\lambda^\ell_k \geqslant 0$ for all $k \in \{1,\dots,m\}$. These weight vectors determine the partition of the subregions and the scalar objective function inside each subregion. Hence the number of weight vectors $L$ also is the number of subregions, the number of scalar objective functions and the number of parallel processes in PPLS/D. Here $L$ is related to the predefined parameter $H\in \mathbb{N}^+$. Given an $H$ value, $L = \binom{H+m-1}{m-1}$ weight vectors are generated, in which each individual weight takes a value from $\{0/H,1/H,\dots,H/H\}$. Fig.~\ref{fig:wv_generator} shows an example in which $L$=15 weight vectors are generated when $m$=3 and $H$=4.

\begin{figure}
  %\vspace{-0.1in}
  \centering
  \includegraphics[width=0.6\linewidth]{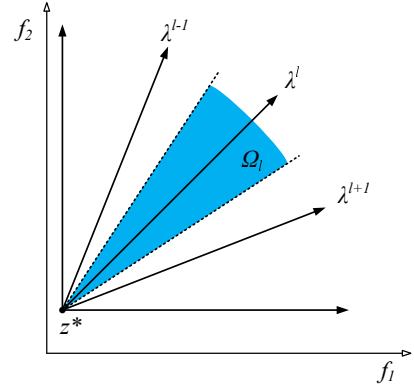}\\
  %\vspace{-0.1in}
  \caption{Relationship between the weight vector $\lambda^\ell$ and the subregion $\Omega_\ell$ in the two objective case}\label{fig:b_wv_z}
  %\vspace{-0.2in}
\end{figure}

For each parallel process $\ell$, a subregion $\Omega_\ell$ is defined in the objective space:
\begin{equation}\label{eq:omega}
\displaystyle
\begin{split}
&\Omega_\ell =\\
&\{u \text{$\in$} \mathbb{R}^m ~\vert~ \langle u\mbox{$-$}z^*,\lambda^\ell \rangle \text{$\leq$} \langle u\mbox{$-$}z^*,\lambda^j \rangle \text{, for any $j$=1,$\dots$,$L$}\},
\end{split}
\end{equation}
where the operator $\langle \cdot,\cdot \rangle$ calculates the acute angle between two vectors, $\lambda^\ell$ is the weight vector of the process $\ell$, $z^*=(z^*_1,\dots,z^*_m)$ is the reference point. Note here that $z^*$ must be the same in all processes. In this way, the objective space can be perfectly separated even when the objective number $m$ is larger than 2. Fig.~\ref{fig:b_wv_z} shows the relationship between the weight vector $\lambda^\ell$ and the subregion $\Omega_\ell$ when $m=2$.
For convenience, in the following when we state that a solution $x$ is in a subregion $\Omega_\ell$, we mean that $F(x) \in \Omega_\ell$.

Inside the assigned subregion, each process is guided by a scalar objective function. In PPLS/D, the scalar objective function of process $\ell$ is generated by the Tchebycheff approach:
\begin{equation}\label{eq:fte}
  \mbox{maximize:}~~~ f^{te}(x \vert \lambda^\ell,z^*) = \min\limits_{1\leq k\leq m}\{\frac{1}{\lambda^\ell _k}(f_k(x)-z^*_k)\},
\end{equation}
where $z^*$ is the reference point also defined in Equation~(\ref{eq:omega}). Compared to the widely-used weighted sum scalar objective function, the Tchebycheff scalar objective function can handle the case that the shape of PF is not convex. In addition, using the Tchebycheff scalar objective function can keep each PPLS/D process on the central line of the subregion at the early stage of the search (as shown in Fig.~\ref{fig:tree_kroAB100_PPLSD_H6} and Fig.~\ref{fig:tree_m2_PPLSD_H6}).

Obviously, each scalar objective function $f^{te}$ represents a direction to approach the PF. In PPLS/D, $f^{te}$ influences the search mechanism from the following two aspects:
\begin{itemize}
  \item At the beginning of each iteration, a new solution is selected from all the unexplored solutions of the archive. Then the neighborhood of the selected solution is explored. In the basic version of PLS, this solution is randomly selected, while in PPLD/D, the solution that has the largest $f^{te}$ value is selected.
  \item When exploring the neighborhood of the selected solution, the basic PLS evaluates all of the neighboring solutions and accepts the solutions that are not dominated by any solution in the archive. PPLS/D only accepts the solutions whose $f^{te}$ value is larger than the largest $f^{te}$ value in the archive. After finding an acceptable solution, PPLS/D stops evaluating the rest neighboring solutions immediately and marks the current solution as explored.
\end{itemize}

\subsection{Procedure}

\begin{algorithm}[t]
    \begin{algorithmic}[1]
        \STATE \textbf{input: } MCOP\\ \qquad \ \ \ \ Stopping criterion\\ \qquad \ \ \ \ $N(\cdot)$: neighborhood structure\\ \qquad \ \ \ \ $A_0$: initial set of non-dominated solutions\\ \qquad \ \ \ \  $L$: process number\\ \qquad \ \ \ \  $\{\lambda^1,\cdots,\lambda^L\}$: weight vectors\\ \qquad \ \ \ \  $z^*$: reference point
        \STATE \textbf{For each process $\ell\in \{1,\dots,L\}$, do independently in parallel:}
        \STATE $A_{\ell,0}\gets A_0$\hfill// Initialization
        \STATE $\forall x \in A_{\ell,0}$, $Explored(x) \gets $ FALSE
        \STATE $A_\ell \gets A_{\ell,0}$\\
        \WHILE{ $A_{\ell,0} \neq \emptyset$ \textbf{and} the stopping criterion is not met}
            \STATE $x_0 \gets \Argmax{x\in A_{\ell,0}}f^{te}(x ~\vert~ \lambda^\ell,z^*)$\label{line:select_step}
            \STATE $SuccessFlag \gets $ FALSE
            \FOR[\hfill// 1st round exploration]{each $x^\prime \in N(x_0)$}\label{line:round1}
                \IF{AcceptanceCriterion1($x^\prime, \ell$) == TRUE}
                    \STATE $\mbox{$A_\ell$}\mbox{$\gets$}$Update($x^\prime, A_\ell$)
                    \STATE $\mbox{$SuccessFlag$}\mbox{$\gets$}$TRUE
                    \STATE \textbf{break}
                \ENDIF
            \ENDFOR
            \IF {$SuccessFlag $ == FALSE}
                \FOR[\hfill// 2nd round exploration] {each $x^\prime \in N(x_0)$}\label{line:round2}
                    \IF {AcceptanceCriterion2($x^\prime, \ell$) == TRUE}
                        \STATE $\mbox{$A_\ell$}\mbox{$\gets$}$Update($x^\prime, A_\ell$)
                    \ENDIF
                \ENDFOR
            \ENDIF
            \STATE $Explored(x_0) \gets $ TRUE
            \STATE \mbox{$A_{\ell,0}$}\mbox{$\gets$}\{\mbox{$x$} \mbox{$\in$} \mbox{$A_\ell$} $\vert$ $Explored(x)$==FALSE\}
        \ENDWHILE\\
        \STATE $A_{\ell,0} \gets A_\ell$ \hfill // Re-check the archive\label{line:recheck}
        \STATE $\forall x \in A_{\ell,0}$, $Explored(x) \gets $ FALSE
        \WHILE{ $A_{\ell,0} \neq \emptyset$ \textbf{and} the stopping criterion is not met}
            \STATE $x_0 \gets \Argmax{x\in A_{\ell,0}}f^{te}(x ~\vert~ \lambda^\ell,z^*)$
            \FOR {each $x^\prime \in N(x_0)$}
                \IF {AcceptanceCriterion2($x^\prime, \ell$) == TRUE}
                    \STATE $\mbox{$A_\ell$}\mbox{$\gets$}$Update($x^\prime, A_\ell$)
                \ENDIF
            \ENDFOR
            \STATE $Explored(x_0) \gets $ TRUE
            \STATE \mbox{$A_{\ell,0}$}\mbox{$\gets$}\{\mbox{$x$} \mbox{$\in$} \mbox{$A_\ell$} $\vert$ $Explored(x)$==FALSE\}
        \ENDWHILE
        \STATE \textbf{return} non-dominated-sol$(\displaystyle \cup_{\ell=1}^{L}A_\ell)$
    \end{algorithmic}
\caption{Parallel Pareto Local Search based on Decomposition (PPLS/D))}
\label{alg:PPLSD}
\end{algorithm}

%%\vspace{-0.1in}
\begin{algorithm}%[b!]
    \begin{algorithmic}[1]
        \STATE $AcceptFlag \gets $ FALSE
        \IF {$F(x^\prime) \in \Omega_\ell$ \textbf{or} $\forall x \in A_\ell, F(x) \notin \Omega_\ell$}
            \IF {$f^{te}(x^\prime|\lambda^\ell,z^*) > \themax{x\in A_\ell}f^{te}(x|\lambda^\ell,z^*)$}
                \STATE $AcceptFlag \gets $ TRUE
            \ENDIF
        \ENDIF
        \STATE \textbf{return} $AcceptFlag$

    \end{algorithmic}
\caption{AcceptanceCriterion1($x^\prime, \ell$)}
\label{alg:ac1}
\end{algorithm}
%%\vspace{-0.1in}

%%\vspace{-0.1in}
\begin{algorithm}%[b!]
    \begin{algorithmic}[1]
        \STATE $AcceptFlag \gets $ FALSE
        \IF {$F(x^\prime) \in \Omega_\ell$ \textbf{or} $\forall x \in A_\ell, F(x) \notin \Omega_\ell$}
            \IF {$x^\prime$ is not dominated by any solution in $A_\ell$}
                \STATE $AcceptFlag \gets $ TRUE
            \ENDIF
        \ENDIF
        \STATE \textbf{return} $AcceptFlag$

    \end{algorithmic}
\caption{AcceptanceCriterion2($x^\prime, \ell$)}
\label{alg:ac2}
\end{algorithm}
%%\vspace{-0.1in}

The pseudo-code of PPLS/D is shown in Algorithm~\ref{alg:PPLSD}. From the input initial archive $A_0$, $L$ processes are executed in parallel. At each iteration of process $\ell$, the solution $x_0$, which maximizes the scalar objective function $f^{te}$ in the unexplored solution archive $A_{\ell,0}$, is selected to be explored (Line~\ref{line:select_step} in Algorithm~\ref{alg:PPLSD}). From $x_0$, two rounds of neighborhood exploration are executed. In the first round (Line~\ref{line:round1} in Algorithm~\ref{alg:PPLSD}), the \emph{acceptance criterion 1} is applied. The decision procedures of the acceptance criterion 1 are shown in Algorithm~\ref{alg:ac1}.

The acceptance criterion 1 (Algorithm~\ref{alg:ac1}) first rejects the candidate solutions that are not in the subregion $\Omega_\ell$, except when the archive $A_\ell$ does not contain any solution in $\Omega_\ell$ (line 2 in Algorithm~\ref{alg:ac1}). This exception prevents the PPLS/D process from stopping early when the PPLS/D process starts from an archive outside the subregion $\Omega_\ell$. Then the acceptance criterion 1 judges each candidate solution $x^\prime$ based on the scalar objective function $f^{te}$. If $f^{te}(x^\prime|\lambda^\ell,z^*)$ is larger than the largest $f^{te}$ value of the archive $A_\ell$, $x^\prime$ will be accepted by the archive $A_\ell$ (line 3 in Algorithm~\ref{alg:ac1}). Once a candidate solution is accepted by the acceptance criterion 1, the first round of neighborhood exploration will be stopped immediately and the second round of neighborhood exploration will be skipped. Otherwise, if no candidate solution meets the acceptance criterion 1, the second round of neighborhood exploration (Line~\ref{line:round2} in Algorithm~\ref{alg:PPLSD}) will be applied following the \emph{acceptance criterion 2}. Algorithm~\ref{alg:ac2} shows the decision procedures of the acceptance criterion 2.

The decision procedures in acceptance criterion 2 (Algorithm~\ref{alg:ac2}) are similar to that in acceptance criterion 1 (Algorithm~\ref{alg:ac1}), except that the acceptance criterion 2 is based on non-dominance relationship (line 3 in Algorithm~\ref{alg:ac2}). In the acceptance criterion 2, if a candidate solution $x^\prime$ in $\Omega_\ell$ is not dominated by any solution in $A_\ell$, it will be accepted. Note here that, the second round of neighborhood exploration does not stop after finding the first acceptable solution. In other words, in the second round of neighborhood exploration there may be multiple new solutions that are added to the archive $A_\ell$. After the two rounds of neighborhood exploration finish, the current solution $x_0$ will be marked as explored (if $x_0$ has not been deleted from $A_\ell$) and the next iteration begins. This iterated procedure stops when the given stopping criterion is met or when all the solutions in the archive are marked as explored.

In the first round of neighborhood solution, a solution may be marked as explored before all of its neighboring solutions are evaluated. Hence, at the final stage of each parallel process there is a re-check phase (Line~\ref{line:recheck} in Algorithm~\ref{alg:PPLSD}). The re-check phase first marks all the solutions in the archive $A_\ell$ as unexplored. Then it uses the acceptance criterion 2 to check all the neighboring solutions of all the solution in $A_\ell$. The neighboring solutions that satisfy the acceptance criterion 2 will be accepted by the archive $A_\ell$ and the solutions in $A_\ell$ that are dominated by the newly accepted solutions will be removed. After the re-check phase, the archive $A_\ell$ becomes locally optimal, i.e., all the neighboring solutions of all the solutions in $A_\ell$ are dominated by at least one solution in $A_\ell$. After all the $L$ processes are finished, PPLS/D combines the $L$ archives and removes the solutions in the aggregated archive that are dominated by the other members. Then the resulting aggregated archive is the output of PPLS/D.

\subsection{Remarks}
As mentioned before, the decomposition strategy of PPLS/D is based on two aspects. Firstly, the search space is decomposed into $L$ subregions $\{\Omega_1,\dots,\Omega_L\}$, so that each process only need to focus on a small search region. Secondly, inside each subregion $\Omega_\ell$, the search is guided by the scalar objective function $f^{te}$ in a fine-grained manner. At each iteration, the solution in $A_{\ell,0}$ with the largest $f^{te}$ value is selected to be explored, because its neighboring solutions may have large $f^{te}$ values too. In the main search phase, PPLS/D only accepts the candidate solution that improves the current best $f^{te}$ value. Once an acceptable solution is found and accepted, the neighborhood exploration stops immediately and the newly accepted solution will be selected to be explored in the next iteration because it has the current largest $f^{te}$ value and it is unexplored. In such a way, the $f^{te}$ value is improved fast at the early stage. If the algorithm cannot find a candidate solution that improves the current best $f^{te}$ value, the second round of neighborhood exploration begins and tries to find and accept all of the solutions that are non-dominated by the archive $A_\ell$ in the neighborhood of the current solution. In such a way, the archive keeps accepting new solutions and the search continues. After the main search phase, there may be some solutions in the archive whose neighborhood has not been fully explored, so PPLS/D conducts the re-check phase to make sure that the final archive is truly locally optimal.

\section{Experimental Studies}\label{sec:epm}
In the experimental studies we compare the performance of PPLS/D against that of the basic PLS and the PLS with Anytime Behavior Improvement (PLS-ABI) proposed in \cite{dubois2015anytime}. The test suites are the multiobjective Unconstrained Binary Quadratic Programming problem (mUBQP) and the multiobjective Traveling Salesman Problem (mTSP). For both kinds of problems, the maximum objective number is four.

\subsection{Test Problems}
The mUBQP problem can be formalized as follows
\begin{equation}\label{eq:mubqp}
%\displaystyle
\begin{split}
\mbox{maximize} & \ \ \ f_k(x) = x^T Q_k x, \ k=1,\dots,m \\%\sum_{i=1}^{n} \sum_{j=1}^{n} q^k_{ij} x_i x_j,
\mbox{subject to} & \ \ \ x \in \{0,1\}^n,\\
\end{split}
\end{equation}
where a solution $x = (x_1,\dots,x_n)$ is a vector of $n$ binary (0-1) variables and $Q_k=[q^k_{ij}]$ is a $n \times n$ matrix for the $k$th objective. Hence the $k$th objective function is calculated by $f_k(x) = \sum_{i=1}^{n} \sum_{j=1}^{n} q^k_{ij} x_i x_j$. The UBQP is NP-hard and it represents the problems appearing in a lot of areas, such as financial analysis~\cite{mcbride1980implicit}, social psychology~\cite{harary1953notion}, machine scheduling~\cite{alidaee19940}, computer aided design~\cite{krarup1978computer} and cellular radio channel allocation~\cite{chardaire1995decomposition}. In this paper, the neighborhood structure in the mUBQP is taken as the $1$-bit-flip, which is directly related to a Hamming distance of 1.

\begin{figure}
  %%\vspace{-0.1in}
  \centering
  \includegraphics[width=0.8\linewidth]{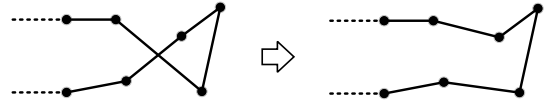}\\
  \caption{The 2-Opt neighborhood move in the TSP}\label{fig:2opt}
\end{figure}

In the mTSP, $G=(V, E)$ is a fully connected graph where $V$ is its node set and $E$ the edge set. Each edge $e \in E$ is corresponded to $m$ different costs $\{c_{e,1},c_{e,2},\dots,c_{e,m}\}$ and $c_{e,k}>0$ for all $k \in \{1,\dots,m\}$. A feasible solution $x$ of the mTSP is a Hamilton cycle passing through every node in $V$ exactly once. The mTSP problem can be formalized as follows
\begin{equation}\label{eq:mtsp}
\begin{split}
\mbox{minimize} & \ \ \ f_k(x)=\sum_{e \in x}c_{e,k}, \ k=1,\dots,m\\
\mbox{subject to} & \ \ \ \mbox{$x$ is a Hamilton cycle in $G$},\\
\end{split}
\end{equation}
The mTSP is one of the most widely used test problems in the area of multiobjective combinatorial optimization. In this paper, the neighborhood move in the mTSP is based on the 2-Opt move, in which two edges in the current solution are replaced by two other edges, as illustrated in Fig.~\ref{fig:2opt}.

In our experiment, three mUBQP instances are generated using the mUBQP generator which is available at \url{http://mocobench.sourceforge.net} and six mTSP instances are generated by combining the single objective TSP instances in TSPLIB~\cite{reinelt1991tsplib}. Table~\ref{tbl:inst} shows the information of the test instances.
\begin{table*}%[H]
\caption{Test Instances}\label{tbl:inst}
\centering
%\vspace{-0.1in}
%\resizebox{\linewidth}{!}{
\begin{tabular}{l l c c l }
\hline
Problem & Instance & m & n & Remarks \\
\hline
\multirow{3}{*}{mUBQP} & mubqp\_2\_200 & 2 & 200 & Objective correlation: 0, $Q$ matrix density: 0.8\\
& mubqp\_3\_200 & 3 & 200 & Objective correlation: 0, $Q$ matrix density: 0.8\\
& mubqp\_4\_200 & 4 & 200 & Objective correlation: 0, $Q$ matrix density: 0.8\\
\hline
\multirow{6}{*}{mTSP} & kroAB100 & 2 & 100 & Combination of TSPLIB instances kroA100 and kroB100\\
& kroAD100 & 2 & 100 & Combination of TSPLIB instances kroA100 and kroD100\\
& kroAB150 & 2 & 150 & Combination of TSPLIB instances kroA150 and kroB150\\
& kroAB200 & 2 & 200 & Combination of TSPLIB instances kroA200 and kroB200\\
& kroABC100 & 3 & 100 & Combination of TSPLIB instances kroA100, kroB100 and kroC100\\
& kroBCDE100 & 4 & 100 & Combination of TSPLIB instances kroB100, kroC100, kroD100 and kroE100\\
\hline
\end{tabular}
%}
\end{table*}

\subsection{Compared Algorithm}
Dubois-Lacoste et al.~\cite{dubois2015anytime} proposed several speed-up strategies to improve the anytime behavior of PLS. An algorithm with good anytime behavior can return high quality solutions even if it is stopped early. Besides the basic PLS, we compare PPLS/D with a sequential PLS variant which we call PLS with Anytime Behavior Improvement (PLS-ABI). PLS-ABI applies two sequential speed-up strategies proposed in \cite{dubois2015anytime}: the ``$\succ\nprec$'' strategy and the ``1*'' strategy. In the ``$\succ\nprec$'' strategy, the acceptance criterion includes two rounds. In the first round, only the candidate solutions that dominate the current solution are accepted. If no solution is accepted in the first round, then the second round begins and the candidate solutions that are not dominated by any solution in the archive are accepted. In the ``1*'', the first-improvement neighborhood exploration is applied. After all solutions in the archive have been marked as explored using the first-improvement rule, the algorithm marks all solutions in the archive as unexplored and explores them again using the best-improvement rule. Here we do not use the ``OHI'' strategy and the ``Dyngrid-HV'' strategy proposed in \cite{dubois2015anytime} because they only apply to bi-objective problems.

\subsection{Performance Metric}
The hypervolume indicator~\cite{knowles2006tutorial} is used as the metric to measure the quality of the archive obtained by different algorithms. For each test instance, let $F^{\max} = \{f_1^{\max}, f_2^{\max}, \dots, f_m^{\max}\}$ and $F^{\min} = \{f_1^{\min}, f_2^{\min}, \dots, f_m^{\min}\}$ respectively the maximum and minimum objective values ever found during all algorithm executions. For the mUBQP problem, which is a maximization problem, the reference point of the hypervolume is $$F^{\min} - 0.1\cdot (F^{\max} - F^{\min}).$$ For convenience, on the mUBQP we normalized the hypervolume value by dividing the hypervolume of the maximum objective vector $F_{max}$. For the mTSP problem, which is a minimization problem, the reference point is $$F^{\max} + 0.1\cdot (F^{\max} - F^{\min})$$ and the final hypervolume is normalized by dividing the hypervolume of $F^{\min}$.

\subsection{Algorithm Performance from Random Solutions}
We argue that PPLS/D is an efficient building block for multiobjective metaheuristics. From randomly generated solutions, PPLS/D can be used to quickly obtain a set of high quality solutions as the initial solutions of a metaheuristic. It can also be applied in the middle or final stage of a metaheuristic to further improve a high quality solution population. In this section, we test the performance of PPLS/D and the compared algorithms from randomly generated initial solutions.

In our experiment, we set $H$ = 6, 8, 10. For each $\{m, H\}$ combination, the corresponding process number $L$ is shown in Table~\ref{tbl:L}. Hence, on each test instance, five algorithms are compared, which are PLS, PLS-ABI, PPLS/D(H=6), PPLS/D(H=8) and PPLS/D(H=10). Each algorithm is executed 20 runs on each instance and the maximum runtime of each run is 100s. Each run starts from a randomly generated solution. On mUBQP instances the reference point $z^*$ is the zero vector and on mTSP instance $z^*$ is the objective vector of the initial solution. All algorithms are implemented in GNU C++. In the PPLS/D implementation, the parallel processes are executed in a sequential order and the entire search history is recorded. After all processes are finished, we integrate the recorded data to simulate a parallel scenario. The computing platform is two 6-core 2.00GHz Intel Xeon E5-2620 CPUs (24 Logical Processors) under Ubuntu system.

\begin{table}[H]
\caption{Settings of Parallel Process Number $L$ in PPLS/D}
\label{tbl:L}
\centering
%%\vspace{-0.1in}
%\resizebox{\linewidth}{!}{
\begin{tabular}{c | c c c}
\hline
 & $H=6$ & H=8 & H=10 \\
\hline
$m=2$ & $L=7$ & $L=9$ & $L=11$\\
$m=3$ & $L=28$ & $L=45$ & $L=66$\\
$m=4$ & $L=84$ & $L=165$ & $L=286$\\
\hline
\end{tabular}
%}
\end{table}

\begin{figure*}%[H]
  %\vspace{-0.1in}
  %\centering
  \subfigure[\tiny{mUBQP: mubqp\_2\_200 (m=2)}]{
    \label{fig:errbar_mubqp2} %% label for first subfigure
    \includegraphics[width=0.2\linewidth]{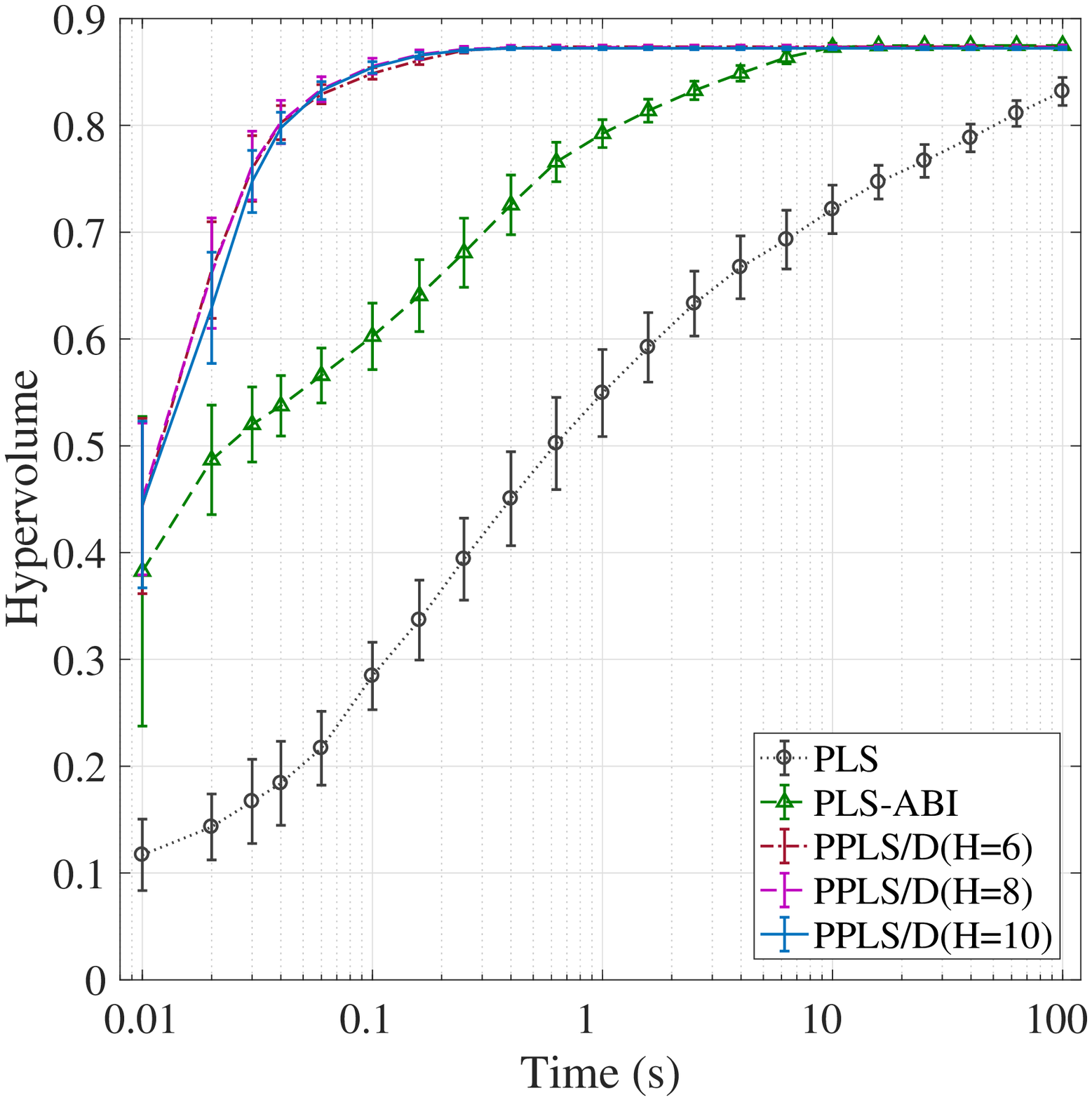}}%epm_201803052052 epm_201707261455
    \hspace{-0.10in}
  \subfigure[\tiny{mUBQP: mubqp\_3\_200 (m=3)}]{
    \label{fig:errbar_mubqp3} %% label for first subfigure
    \includegraphics[width=0.2\linewidth]{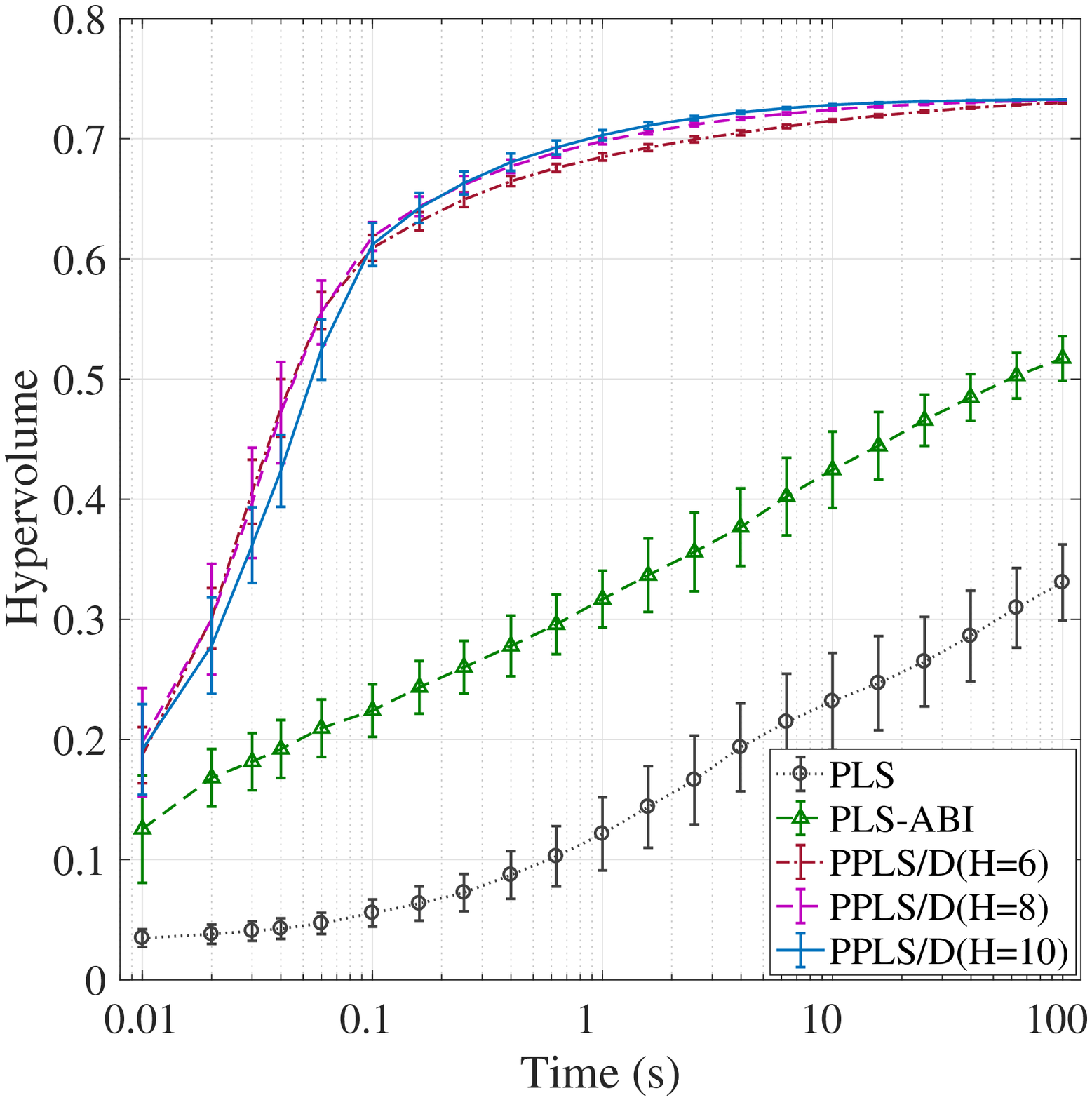}}%epm_201803052052 epm_201707261455
    \hspace{-0.10in}
  \subfigure[\tiny{mUBQP: mubqp\_4\_200 (m=4)}]{
    \label{fig:errbar_mubqp4} %% label for first subfigure
    \includegraphics[width=0.2\linewidth]{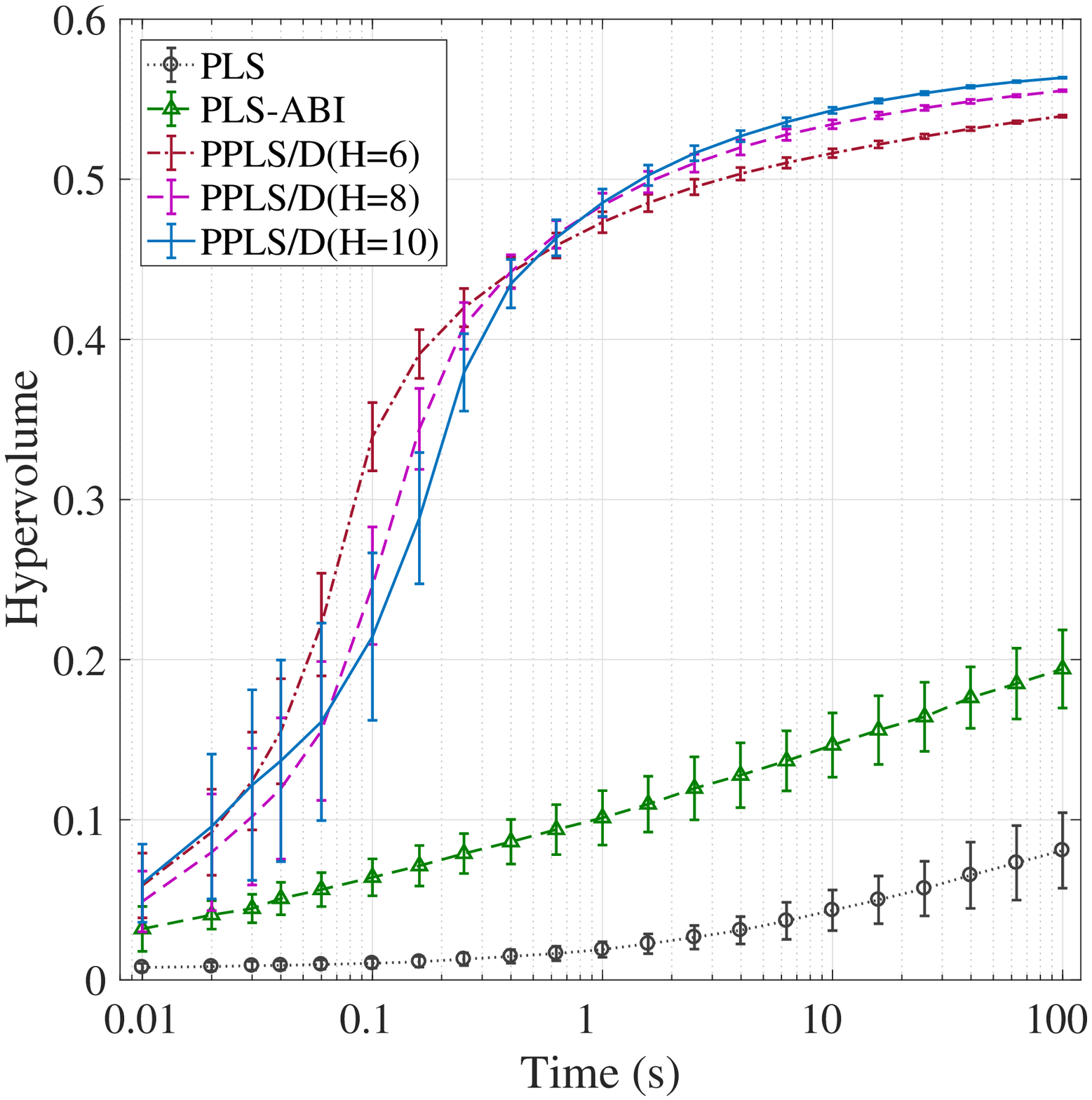}}%epm_201803052052 epm_201707261455
    \hspace{-0.10in}
  \subfigure[\tiny{mTSP: kroAB100 (m=2)}]{
    \label{fig:errbar_kroAB100} %% label for first subfigure
    \includegraphics[width=0.2\linewidth]{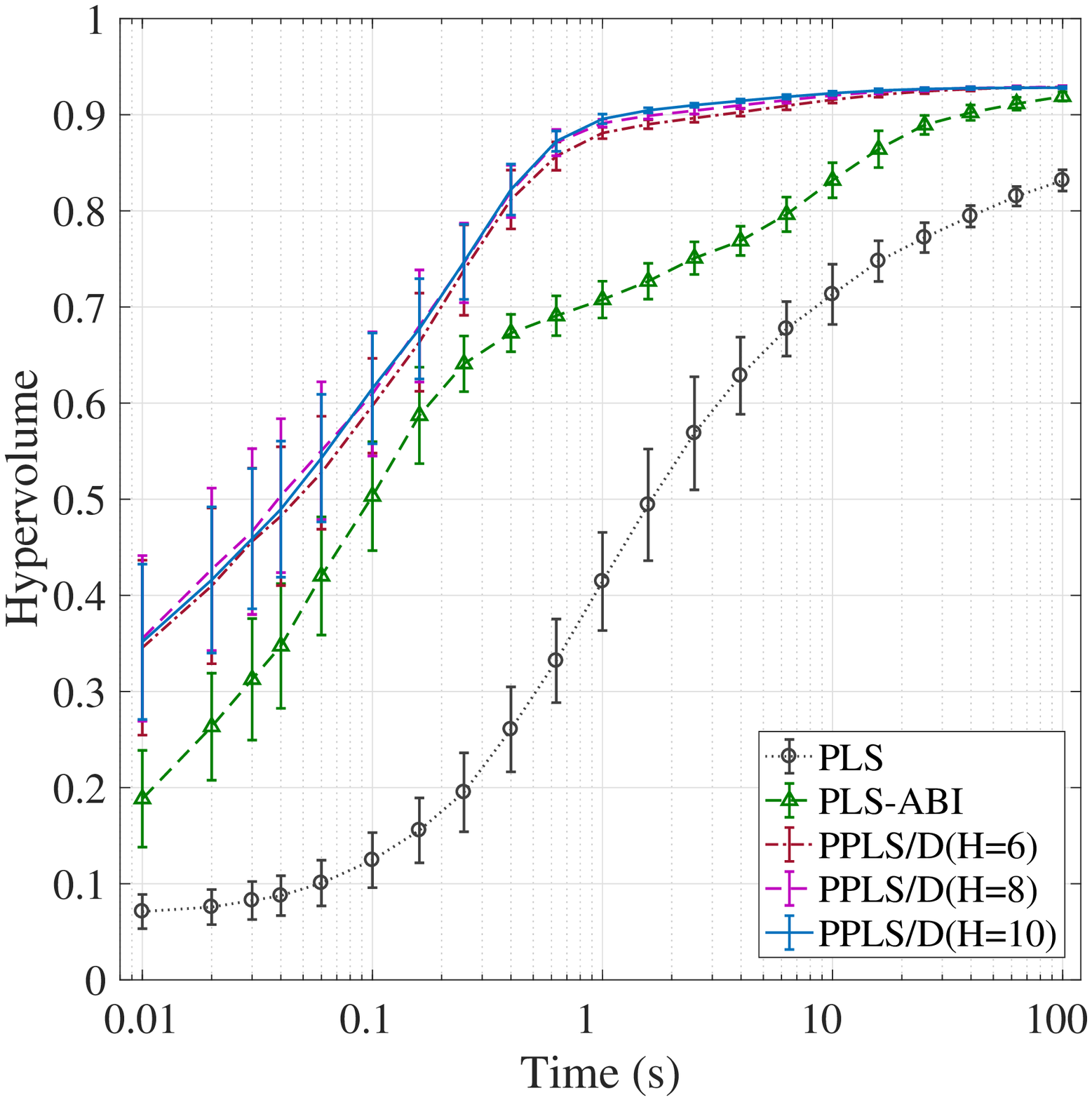}}%epm_201802141332
    \hspace{-0.10in}
  \subfigure[\tiny{mTSP: kroAD100 (m=2)}]{
    \label{fig:errbar_kroAD100} %% label for first subfigure
    \includegraphics[width=0.2\linewidth]{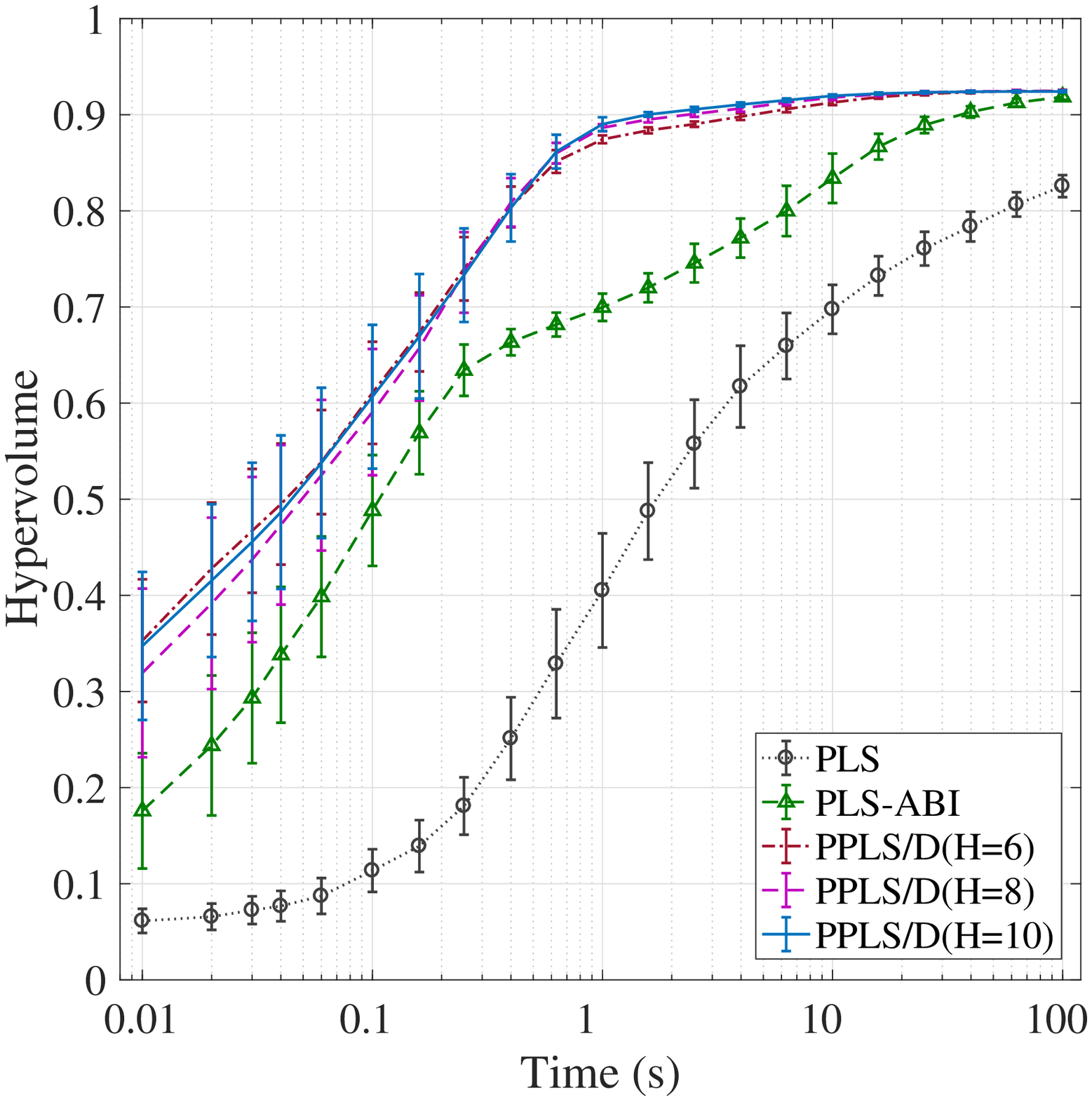}}\\%epm_201802141332
  \vspace{0in}
  \subfigure[\tiny{mTSP: kroAB150 (m=2)}]{
    \label{fig:errbar_kroAB150} %% label for first subfigure
    \includegraphics[width=0.2\linewidth]{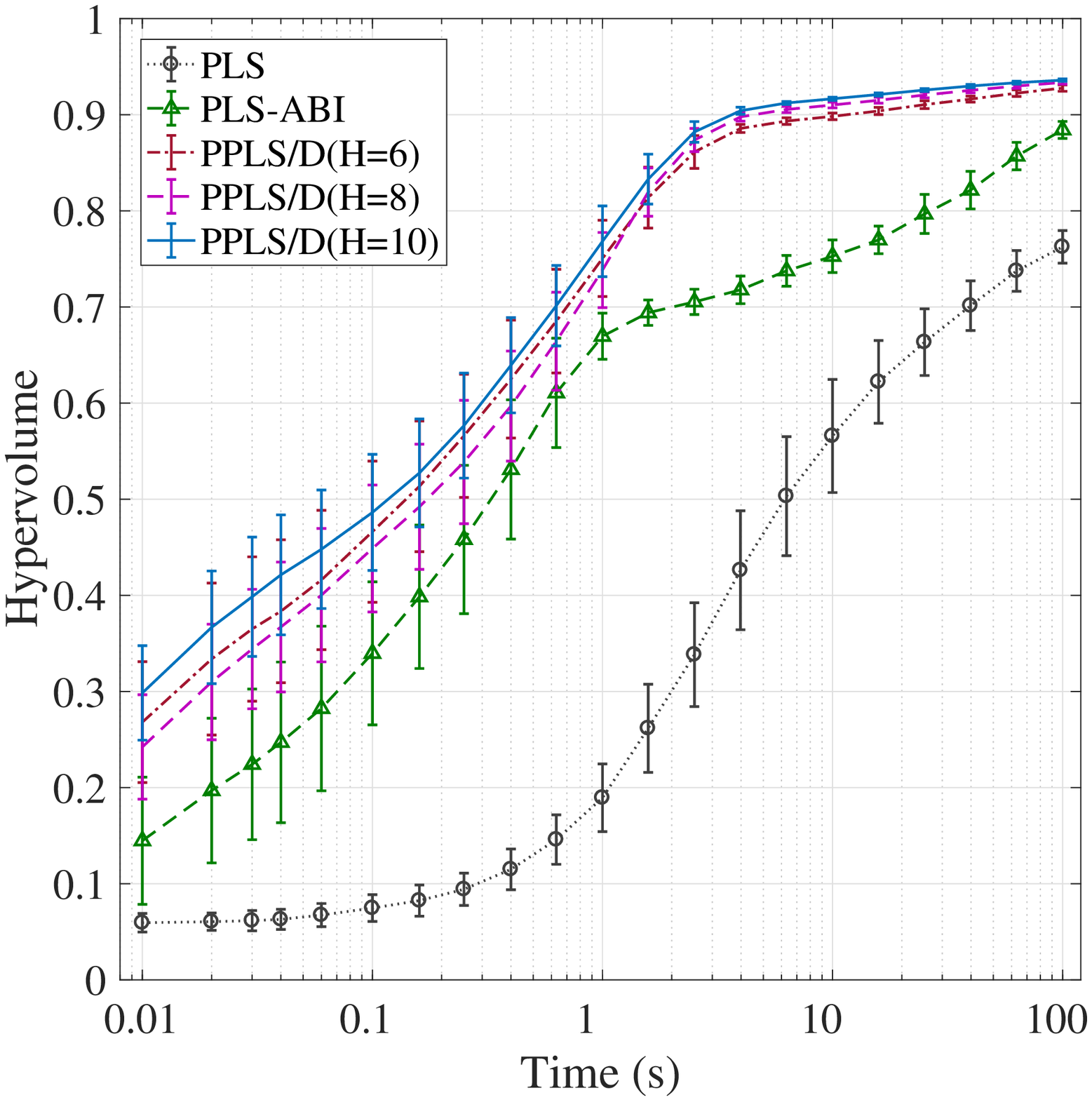}}%epm_201802141332
    \hspace{-0.10in}
  \subfigure[\tiny{mTSP: kroAB200 (m=2)}]{
    \label{fig:errbar_kroAB200} %% label for first subfigure
    \includegraphics[width=0.2\linewidth]{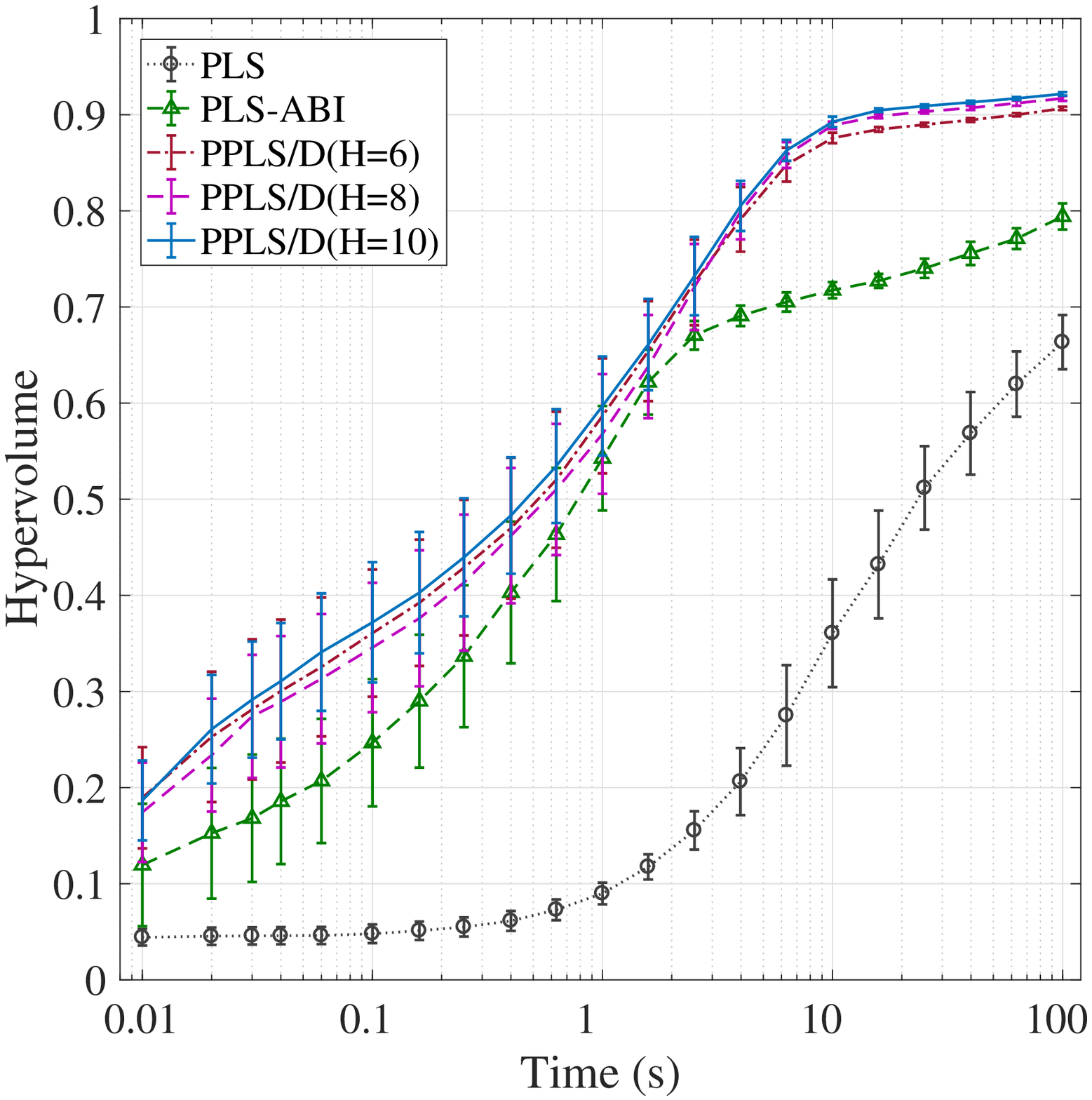}}%epm_201801291655
    \hspace{-0.10in}
  \subfigure[\tiny{mTSP: kroABC100 (m=3)}]{
    \label{fig:errbar_kroABC100} %% label for first subfigure
    \includegraphics[width=0.2\linewidth]{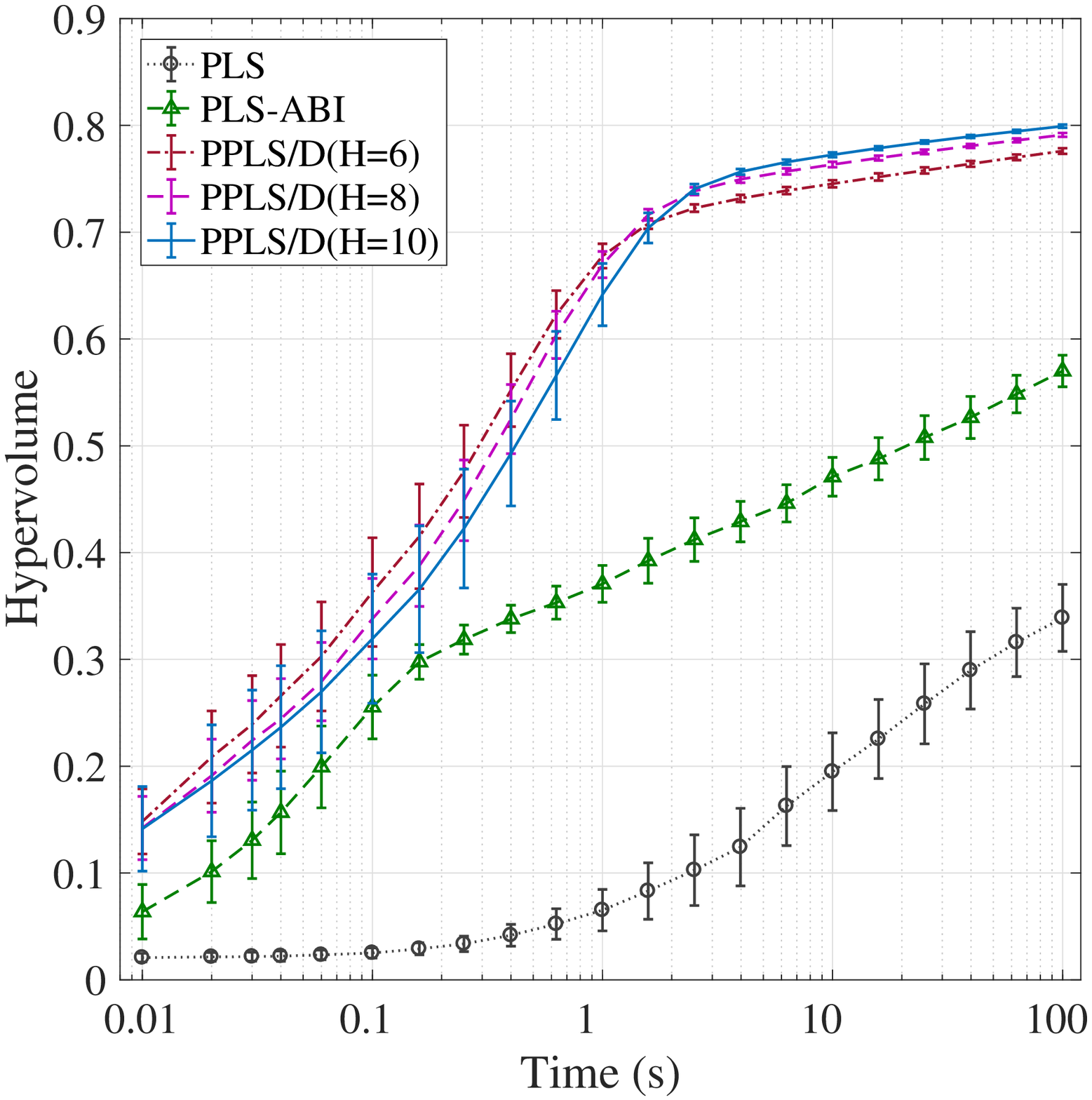}}%epm_201801291655
    \hspace{-0.10in}
  \subfigure[\tiny{mTSP: kroBCDE100 (m=4)}]{
    \label{fig:errbar_kroBCDE100} %% label for first subfigure
    \includegraphics[width=0.2\linewidth]{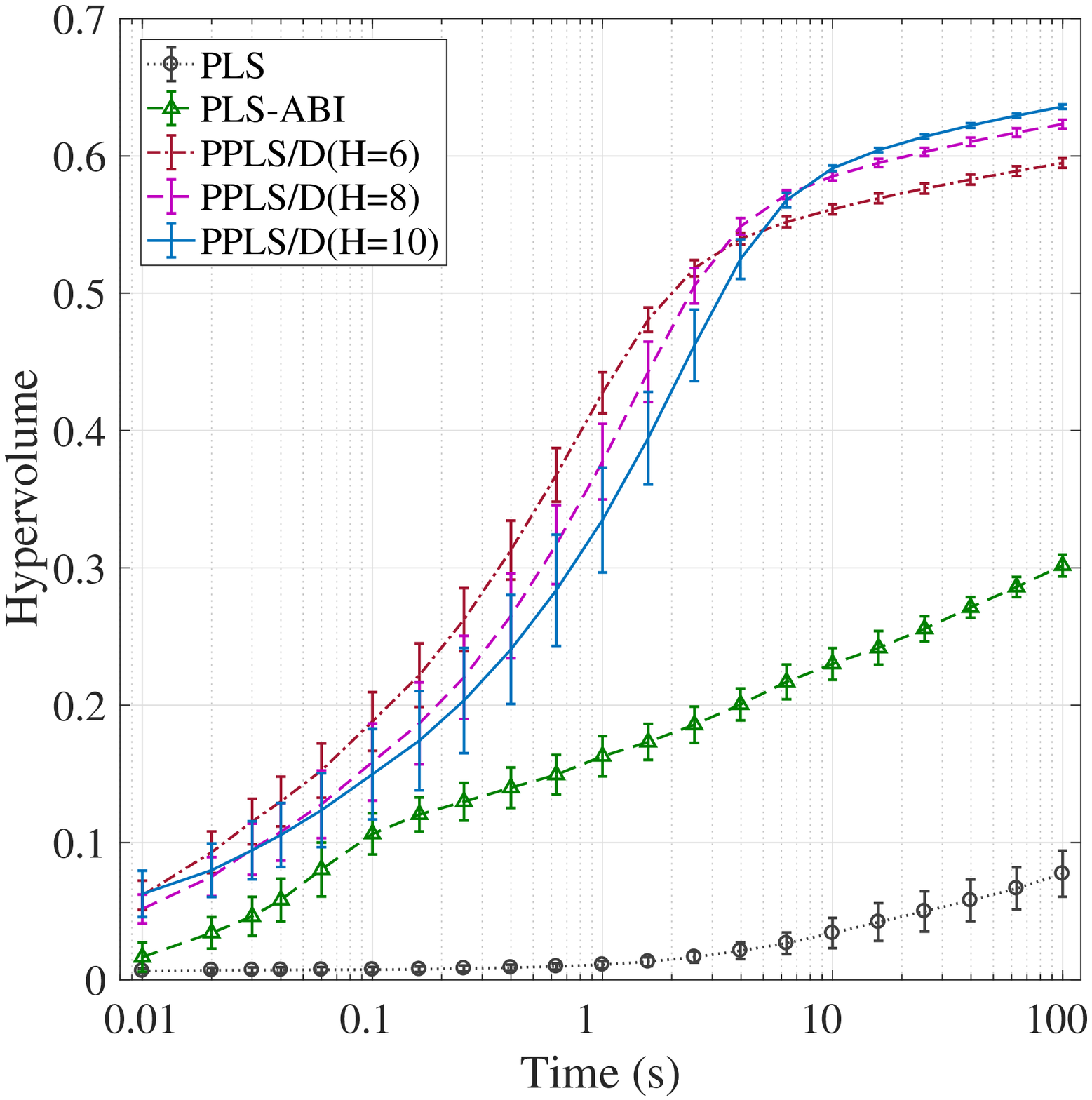}}\\%epm_201801291655
  \vspace{0in}
  \caption{The hypervolume attained by PLS, PLS-ABI, PPLS/D(H=6), PPLS/D(H=8) and PPLS/D(H=10) at different time points. Each algorithm is executed 20 runs from different randomly generated solutions.}\label{fig:errbar}
  %\vspace{-0.2in}
\end{figure*}

For each run, we record the entire search history and calculate the normalized hypervolume indicator at certain time points. Fig.~\ref{fig:errbar} shows the hypervolume values attained by the algorithms over time on the test instances. From Fig.~\ref{fig:errbar} we can see that, on all test instances the hypervolume increment speed of PPLS/D is much higher than that of PLS and PLS-ABI. For example, on the instance mubqp\_2\_200 (Fig.~\ref{fig:errbar_mubqp2}), PPLS/D(H=6) takes 0.04s to reach an average hypervolume value higher than 0.8, while PLS-ABI needs more than 1s to reach the same average hypervolume value and PLS needs more than 1.58s. On mUBQP instances, when the objective number $m$ increases, the superiority of PPLS/D against PLS and PLS-ABI increases. The same phenomenon also can be found on mTSP instances. This means that PPLS/D can handle high dimensional problems better than PLS and PLS-ABI. Hence, we conclude that the proposed PPLS/D is much more efficient than PLS and PLS-ABI when the initial solutions are randomly generated.

\subsection{Algorithm Performance from High Quality Solutions}
In this section, we test the performance of PPLS/D and the compared algorithms from high quality initial solutions. The high quality solutions are generated by executing single objective local search on the Tchebycheff scalar objective functions (Equation~(\ref{eq:fte})) with different weight vectors. Specifically, we set $H=6$ and generate $L = \binom{H+m-1}{m-1}$ weight vectors. For each test instance, a random solution is generated and $L$ different local search procedures start from it. Each local search procedure is based on the Tchebycheff scalar objective function defined by a unique weight vector. For the mUBQP instances, the 1-bit-flip first-improvement local search is applied. For the mTSP instances, the 2-Opt first-improvement local search is applied. The $L$ local search procedures return $L$ high quality solutions. After removing the solutions that are dominated by other solutions, we get the initial solution set of the algorithms.

From the high quality solution set, we test the performance of PPLS/D(H=6), PLS and PLS-ABI. Note here that, in this experiment PLS is equivalent to the well-known 2PPLS~\cite{lust2010two}, where PLS is started from locally optimal solutions of weighted sum problems. Hence this experiment also compares the proposed PPLS/D against 2PPLS. The other experiment settings are same as the settings in the previous section.

\begin{figure*}%[H]
  %\vspace{-0.1in}
  %\centering
  \subfigure[\tiny{mUBQP: mubqp\_2\_200 (m=2)}]{
    \label{fig:errbar_mubqp2_hqi} %% label for first subfigure
    \includegraphics[width=0.2\linewidth]{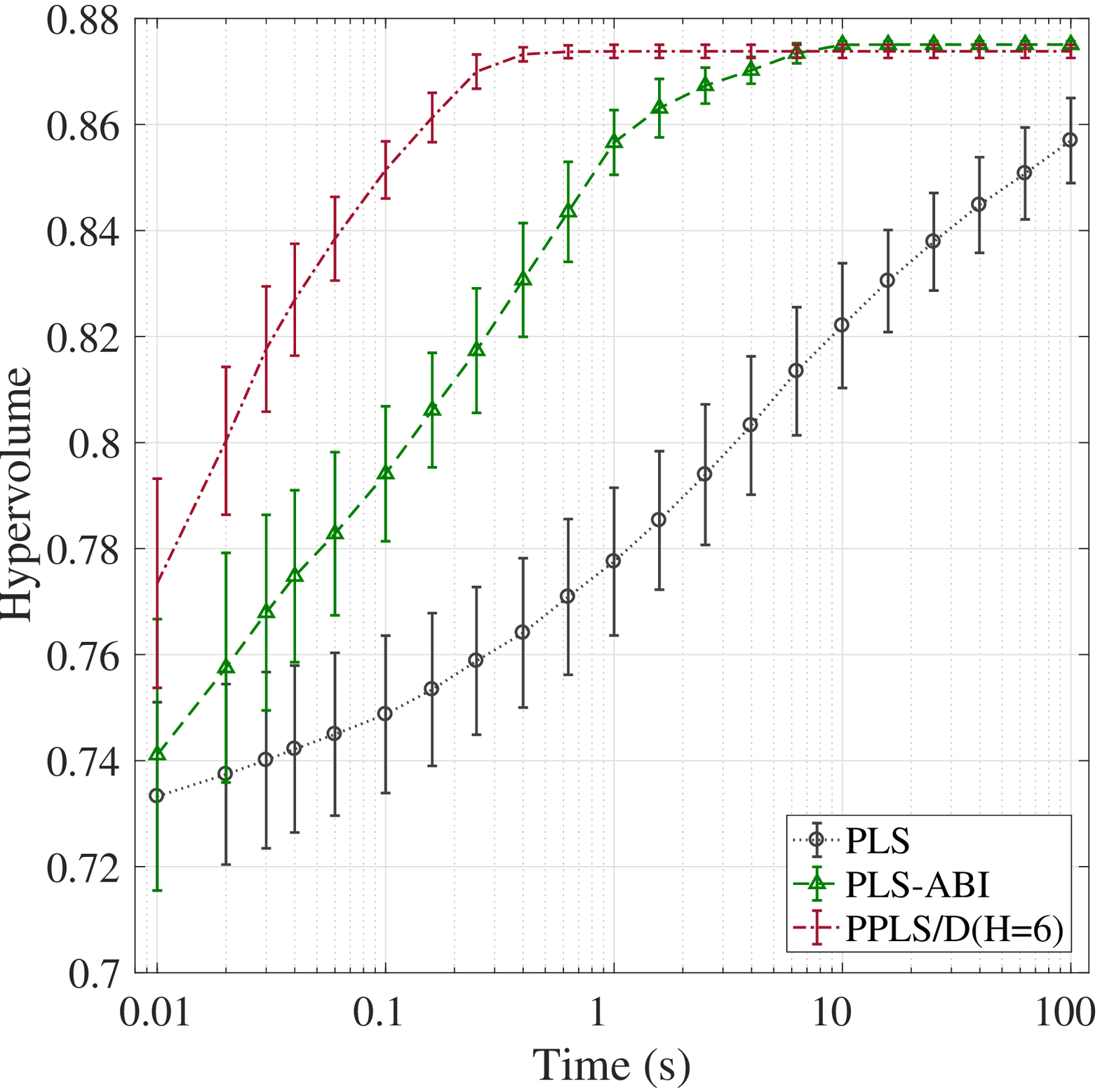}}%epm_201802051149
    \hspace{-0.10in}
  \subfigure[\tiny{mUBQP: mubqp\_3\_200 (m=3)}]{
    \label{fig:errbar_mubqp3_hqi} %% label for first subfigure
    \includegraphics[width=0.2\linewidth]{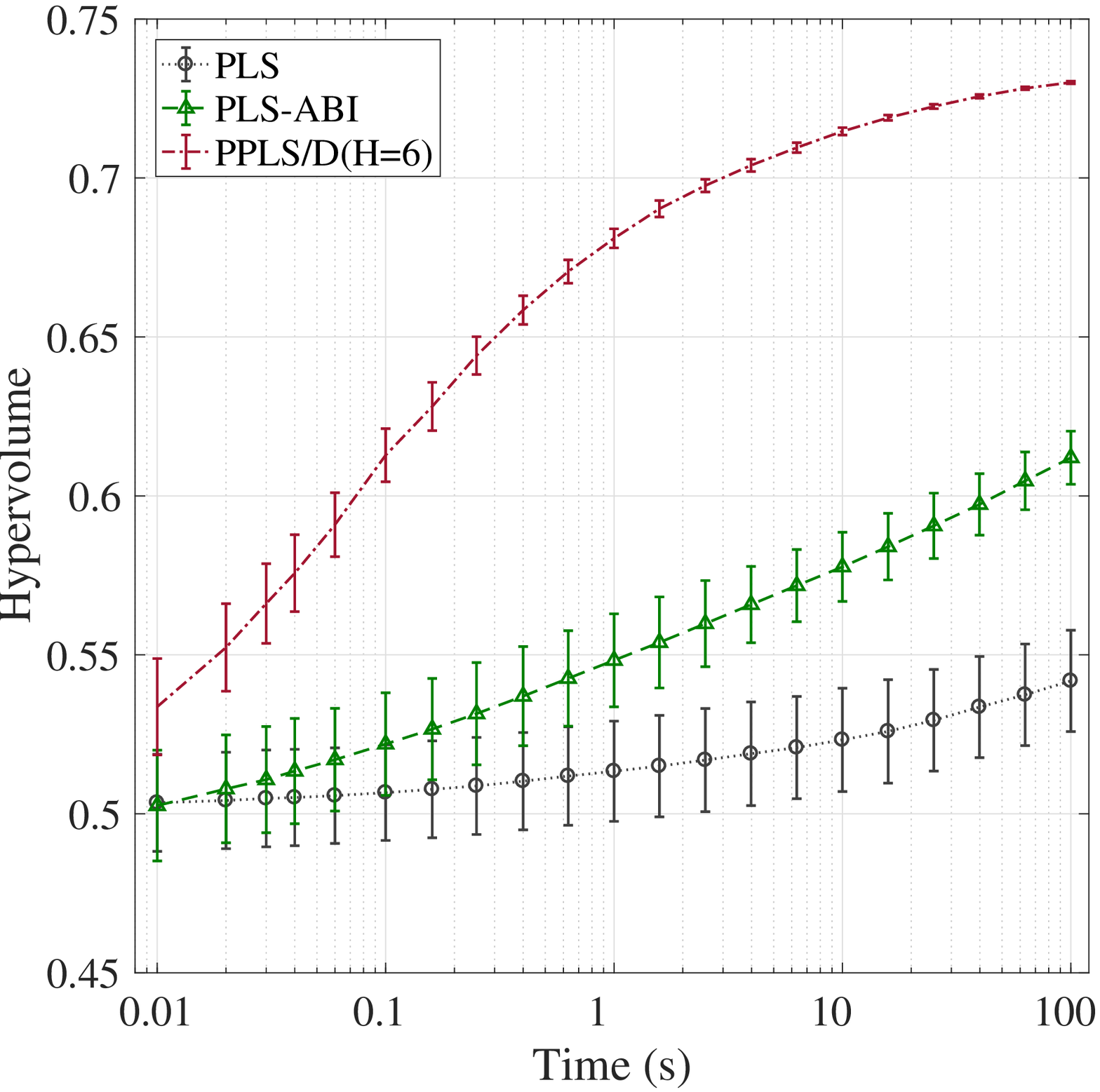}}%epm_201802051149
    \hspace{-0.10in}
  \subfigure[\tiny{mUBQP: mubqp\_4\_200 (m=4)}]{
    \label{fig:errbar_mubqp4_hqi} %% label for first subfigure
    \includegraphics[width=0.2\linewidth]{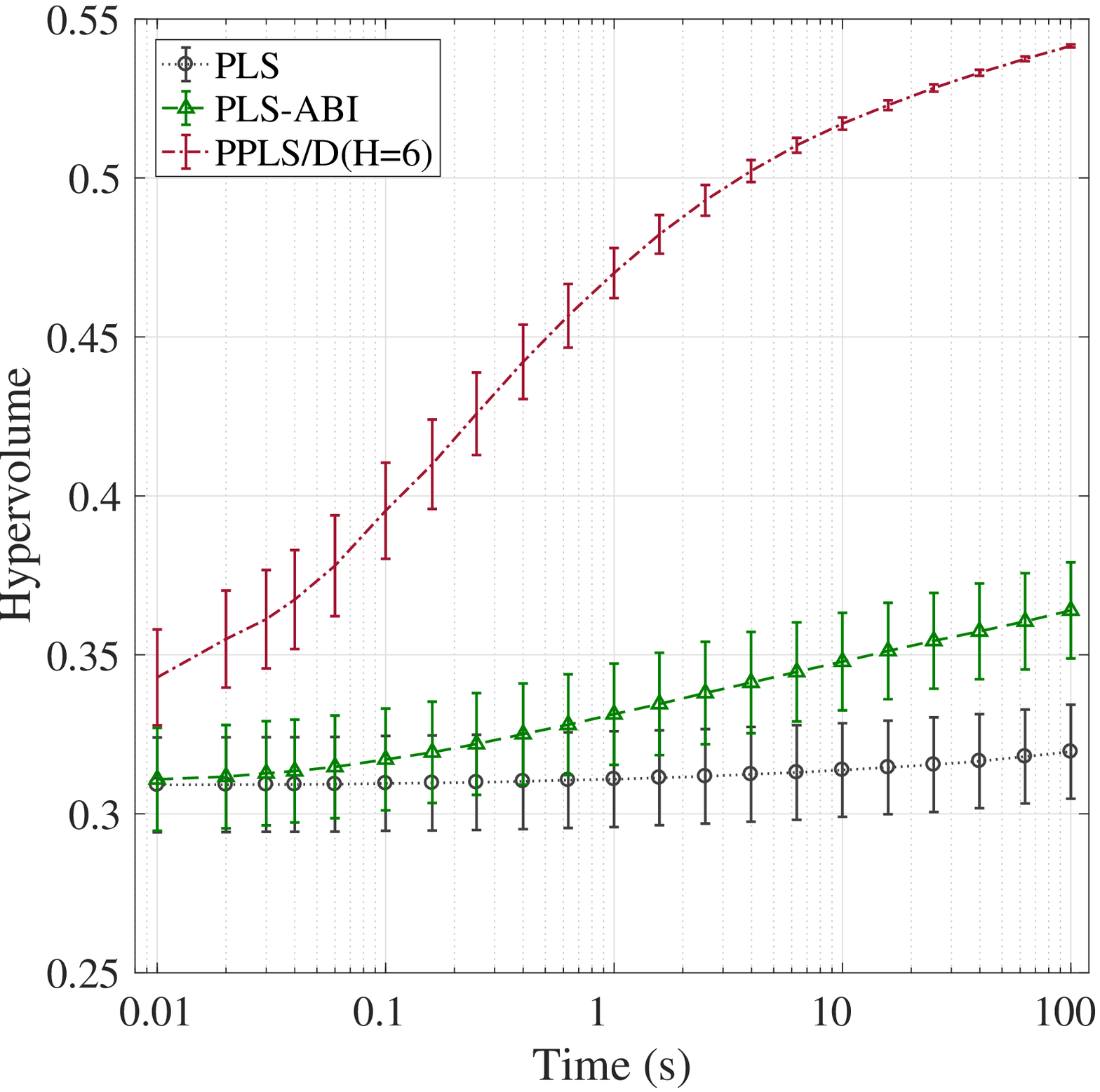}}%epm_201802051149
    \hspace{-0.10in}
  \subfigure[\tiny{mTSP: kroAB100 (m=2)}]{
    \label{fig:errbar_kroAB100_hqi} %% label for first subfigure
    \includegraphics[width=0.2\linewidth]{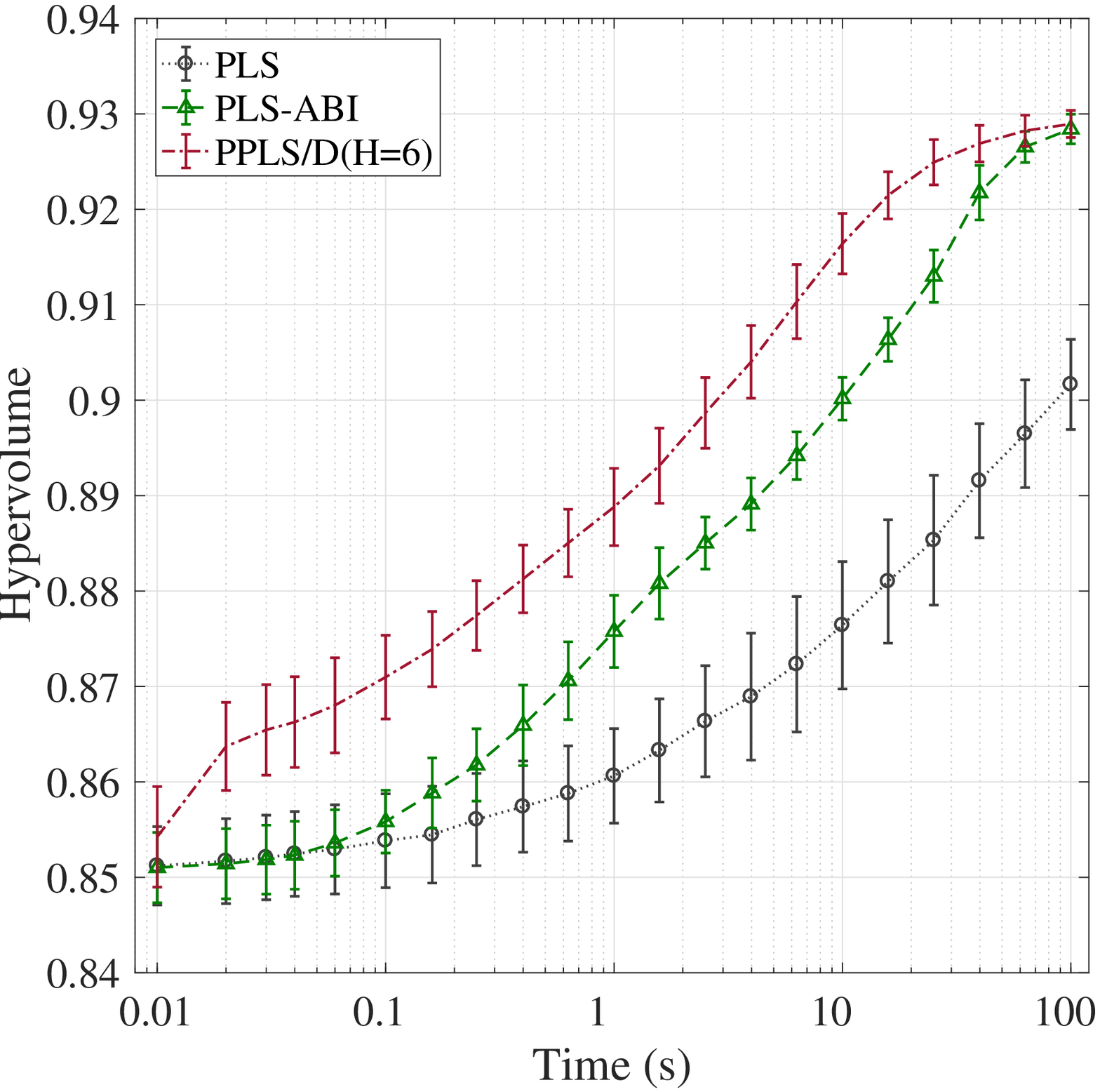}}%epm_201802261945
    \hspace{-0.10in}
  \subfigure[\tiny{mTSP: kroAD100 (m=2)}]{
    \label{fig:errbar_kroAD100_hqi} %% label for first subfigure
    \includegraphics[width=0.2\linewidth]{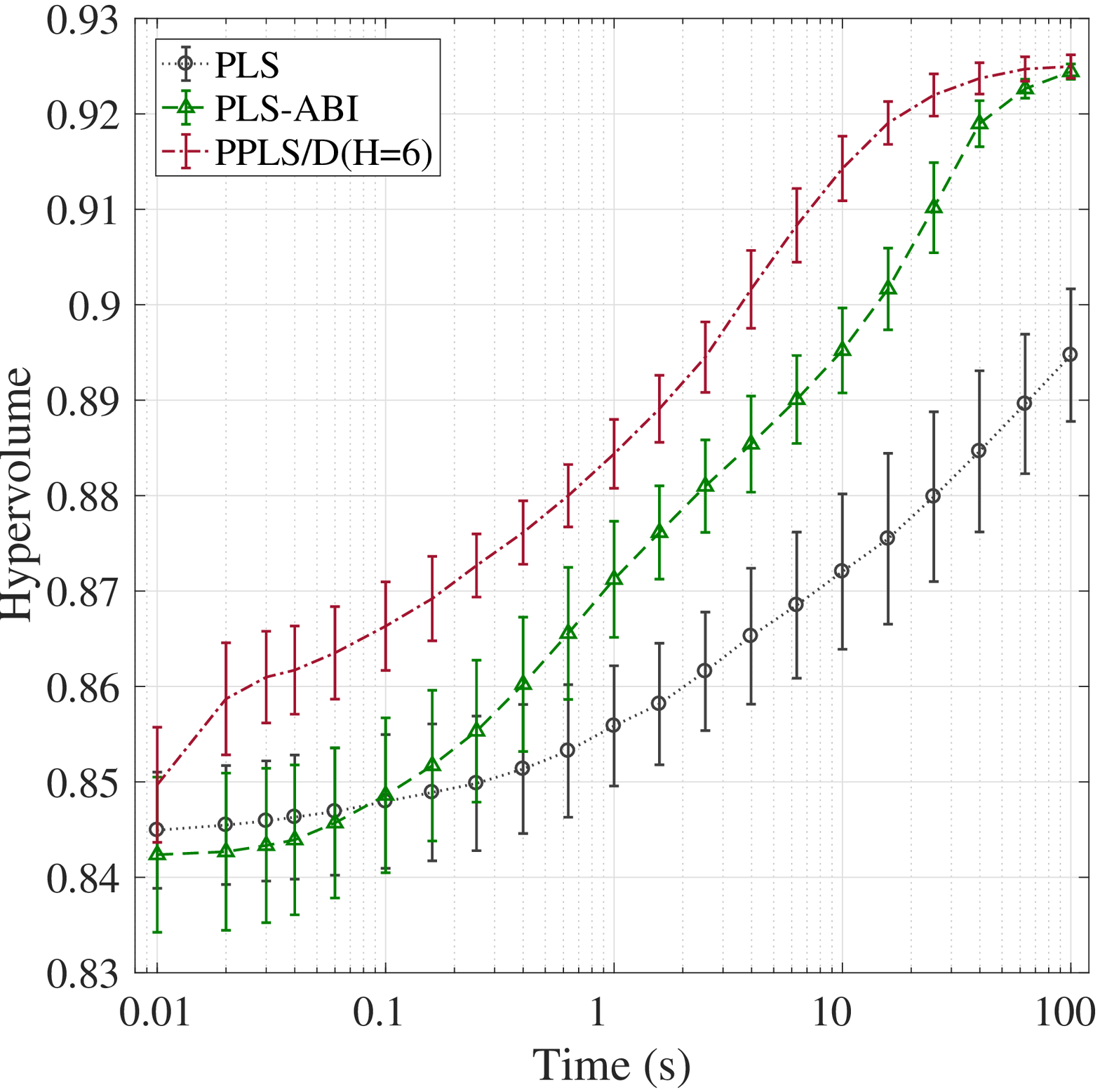}}\\%epm_201802261945
    \vspace{0in}
  \subfigure[\tiny{mTSP: kroAB150 (m=2)}]{
    \label{fig:errbar_kroAB150_hqi} %% label for first subfigure
    \includegraphics[width=0.2\linewidth]{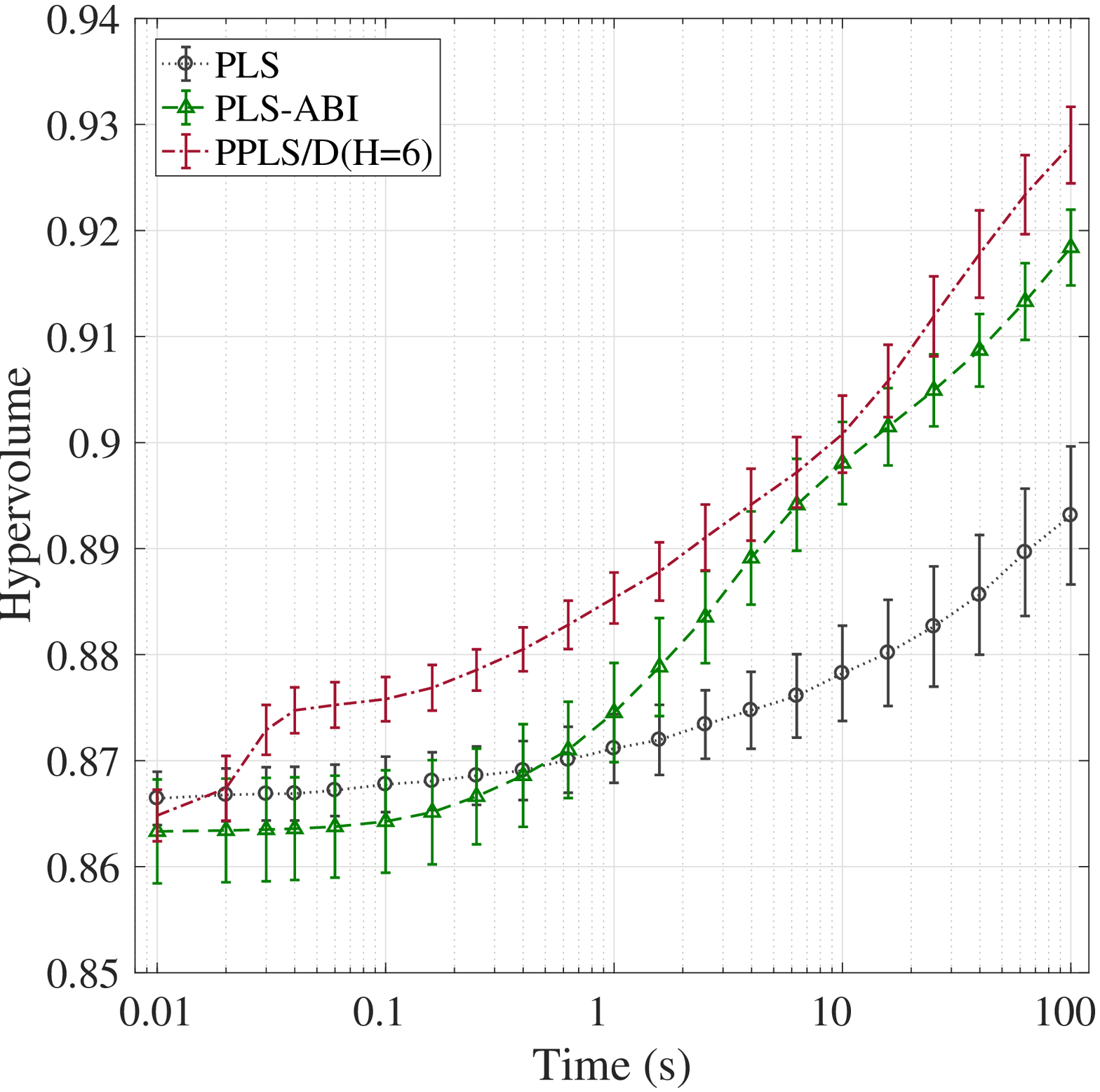}}%epm_201802261945
    \hspace{-0.10in}
  \subfigure[\tiny{mTSP: kroAB200 (m=2)}]{
    \label{fig:errbar_kroAB200_hqi} %% label for first subfigure
    \includegraphics[width=0.2\linewidth]{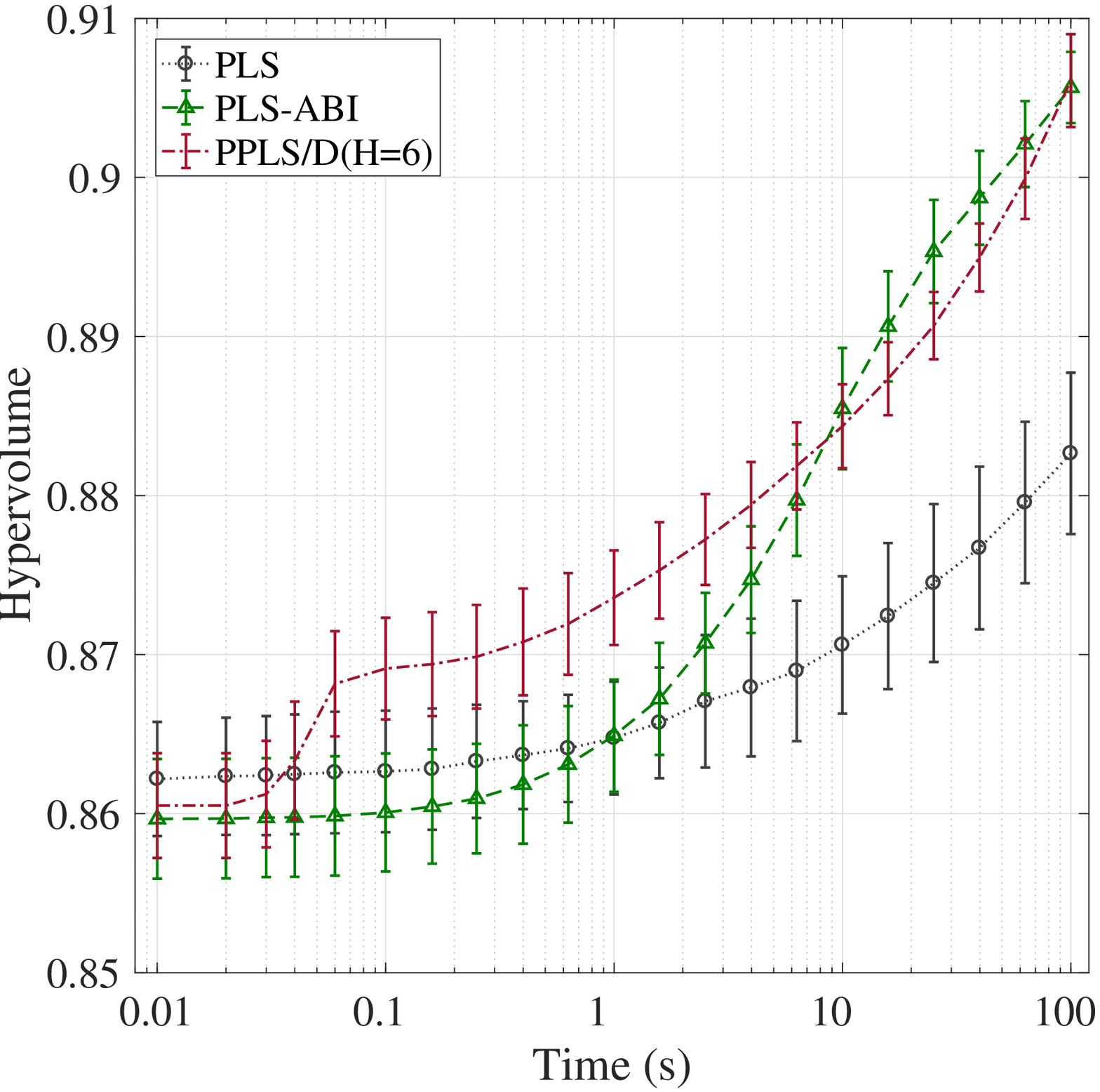}}%epm_201802121153
    \hspace{-0.10in}
  \subfigure[\tiny{mTSP: kroABC100 (m=3)}]{
    \label{fig:errbar_kroABC100_hqi} %% label for first subfigure
    \includegraphics[width=0.2\linewidth]{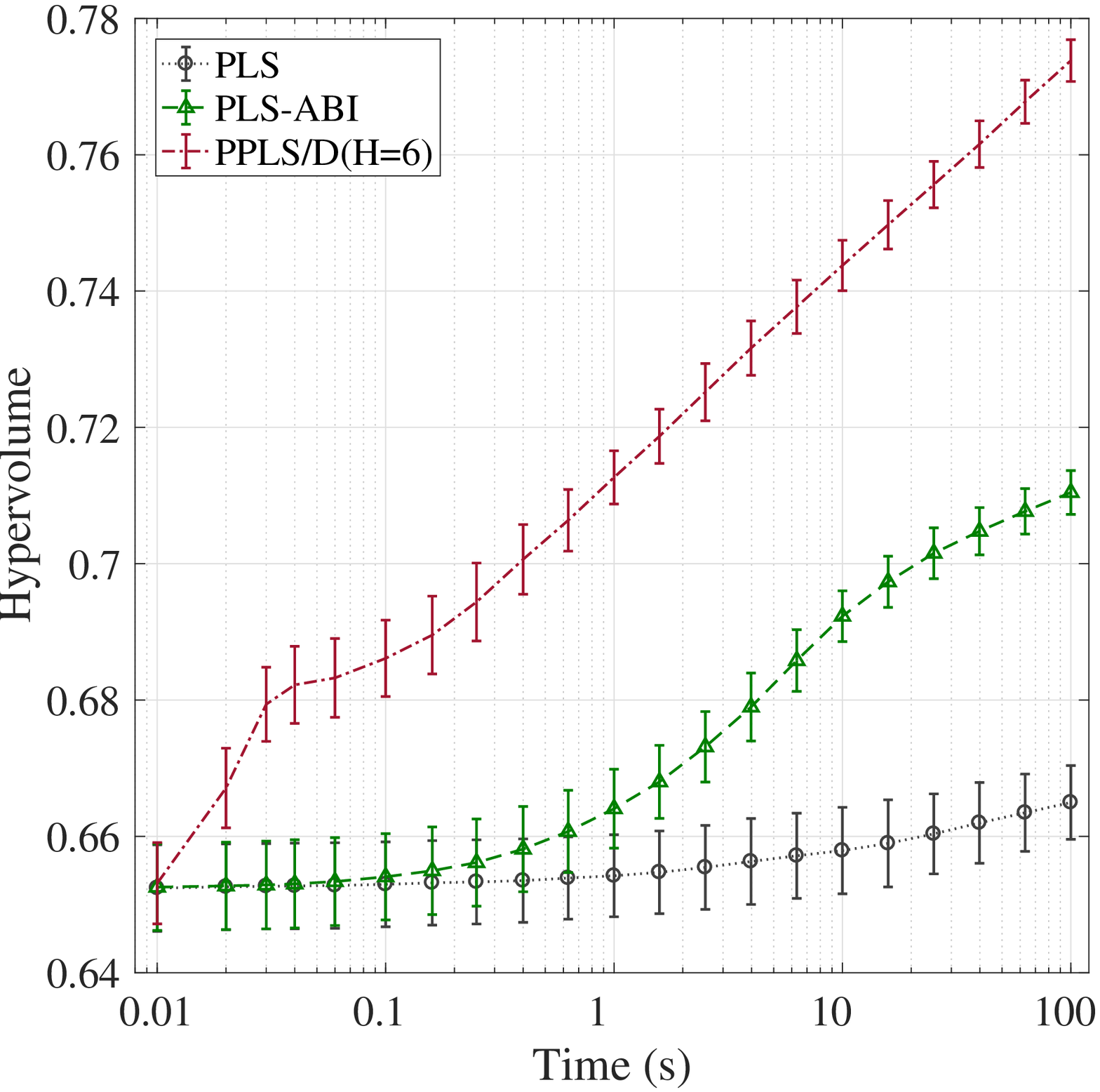}}%epm_201802121153
    \hspace{-0.10in}
  \subfigure[\tiny{mTSP: kroBCDE100 (m=4)}]{
    \label{fig:errbar_kroBCDE100_hqi} %% label for first subfigure
    \includegraphics[width=0.2\linewidth]{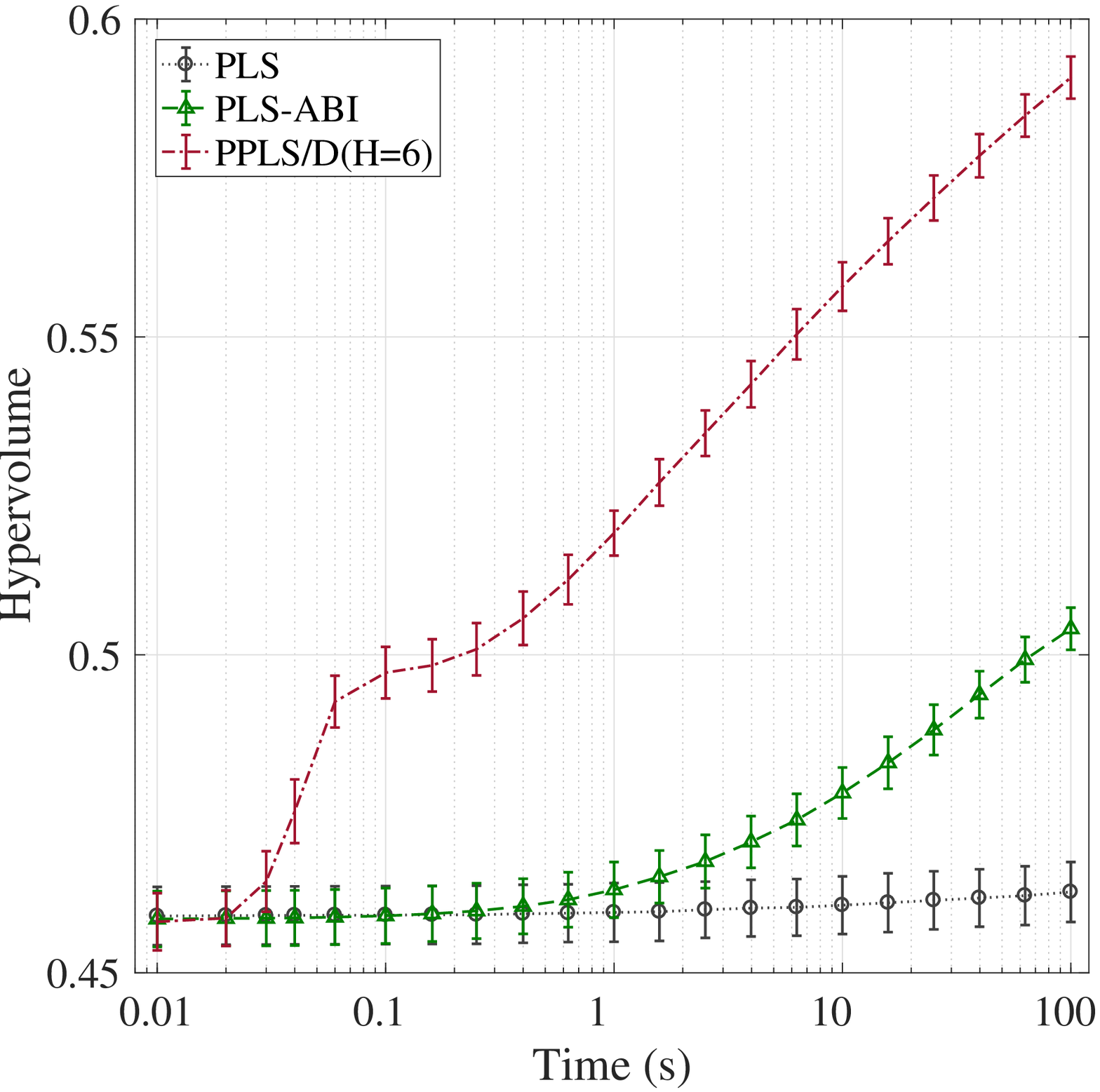}}\\%epm_201802121153
  \vspace{0in}
  \caption{The hypervolume attained by PLS, PLS-ABI and PPLS/D(H=6) at different time points. Each algorithm is executed 20 runs from high quality solutions which are generated by executing local search on Tchebycheff aggregated functions (Equation~(\ref{eq:fte})) with different weight vectors. }\label{fig:errbar_hqi}
  %\vspace{-0.2in}
\end{figure*}

Fig.~\ref{fig:errbar_hqi} shows the hypervolume values attained by the algorithms over time on the test instances. From Fig.~\ref{fig:errbar_hqi} we can see that when the initial solutions are high quality solutions, the proposed PPLS/D outperforms PLS and PLS-ABI on most instances. Considering that in this experiment PLS is equivalent to the well-known 2PPLS, the results show that PPLS/D can outperform 2PPLS on the tested mUBQP and mTSP instances. In addition, when the objective number $m$ increases, the superiority of PPLS/D against PLS and PLS-ABI increases. Hence, we conclude that PPLS/D is more efficient than PLS and PLS-ABI when the initial solutions are high quality solutions.

\subsection{Comparison against Parallel PLS-ABI}
In the previous sections, the proposed PPLS/D is compared against the sequential PLS and PLS-ABI. In this section, we compare PPLS/D against the parallel implementation of PLS-ABI (P-PLS-ABI). In P-PLS-ABI, $L$ PLS-ABI processes are executed simultaneously with different random seeds. After the $L$ parallel processes finish, P-PLS-ABI combines the $L$ sub-archives and deletes the solutions that are duplicated or dominated by other solutions to get the final archive. In the comparison experiment, both PPLS/D and P-PLS-ABI are implemented in a real parallel scenario using the widely-used parallelization tool MPI. In other words, each run of PPLS/D and P-PLS-ABI occupies $L$ cores to execute $L$ processes simultaneously. Each process is executed on one unique core. We set the stopping criterion of each process to be $10^6$ function evaluations. For the PPLS/D implementation, we set $H=6$, which means that PPLS/D contains $L=7$ parallel processes on two-objective instances, $L=28$ parallel processes on three-objective instances and $L=84$ parallel processes on four-objective instances. The P-PLS-ABI implementation contains the same number of parallel processes as the PPLS/D implementation on each instance. Both PPLS/D and P-PLS-ABI are executed 20 runs on each instance and each run is started from a randomly generated solution. The experimental platform is the Tianhe-2 supercomputer which is the 4th ranked supercomputer in the world\cite{top500.org}. The other experimental settings are same as the settings in the previous sections.

\begin{figure*}%[H]
  %\vspace{-0.1in}
  %\centering
  \subfigure[\tiny{mUBQP: mubqp\_2\_200 (m=2)}]{
    \label{fig:errbar_mubqp2_pplsabi} %% label for first subfigure
    \includegraphics[width=0.197\linewidth]{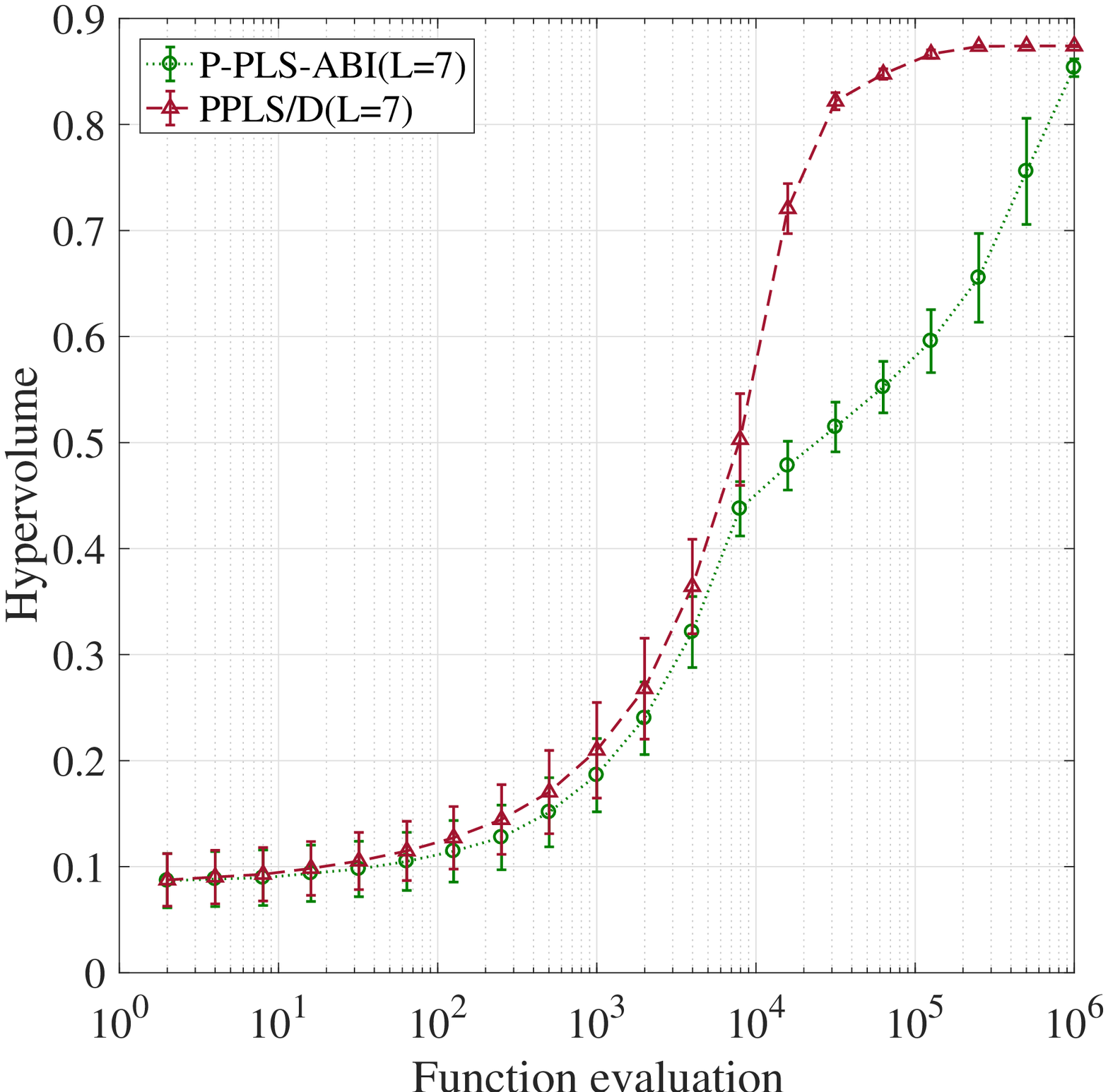}}%epm_201806121626
    \hspace{-0.09in}
  \subfigure[\tiny{mUBQP: mubqp\_3\_200 (m=3)}]{
    \label{fig:errbar_mubqp3_pplsabi} %% label for first subfigure
    \includegraphics[width=0.197\linewidth]{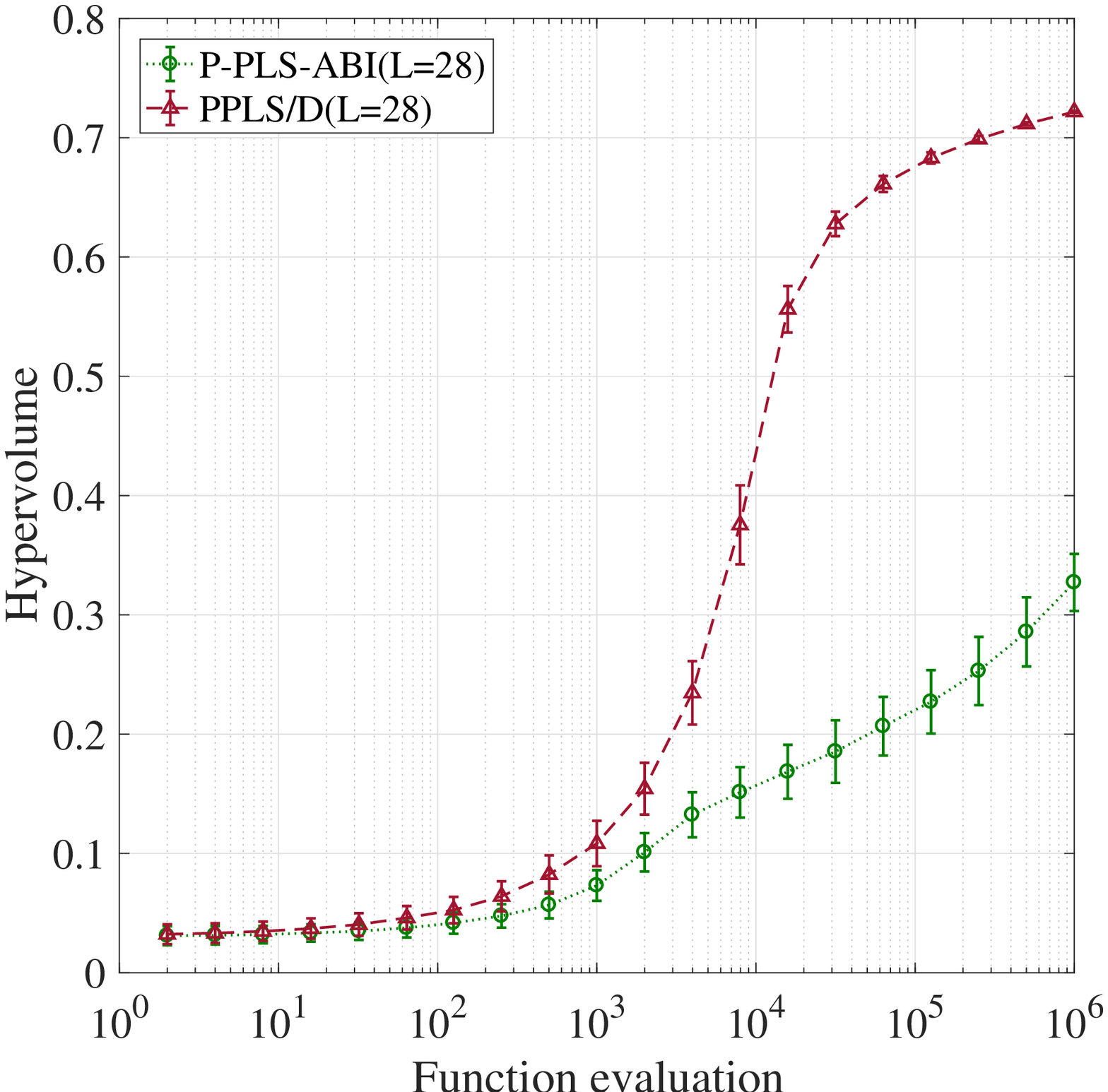}}%epm_201806121626
    \hspace{-0.09in}
  \subfigure[\tiny{mUBQP: mubqp\_4\_200 (m=4)}]{
    \label{fig:errbar_mubqp4_pplsabi} %% label for first subfigure
    \includegraphics[width=0.197\linewidth]{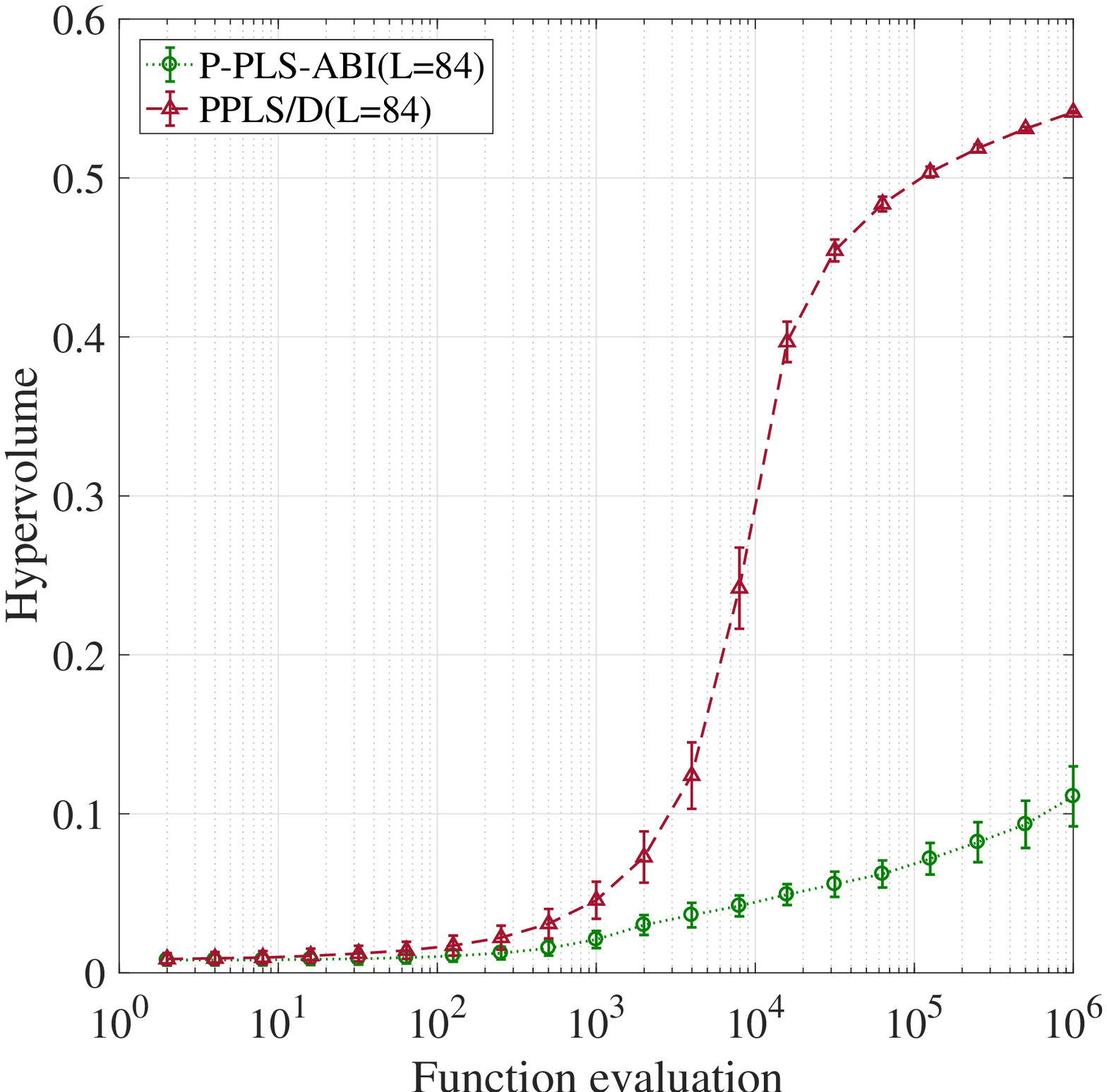}}%epm_201806121626
    \hspace{-0.09in}
  \subfigure[\tiny{mTSP: kroAB100 (m=2)}]{
    \label{fig:errbar_kroAB100_pplsabi} %% label for first subfigure
    \includegraphics[width=0.197\linewidth]{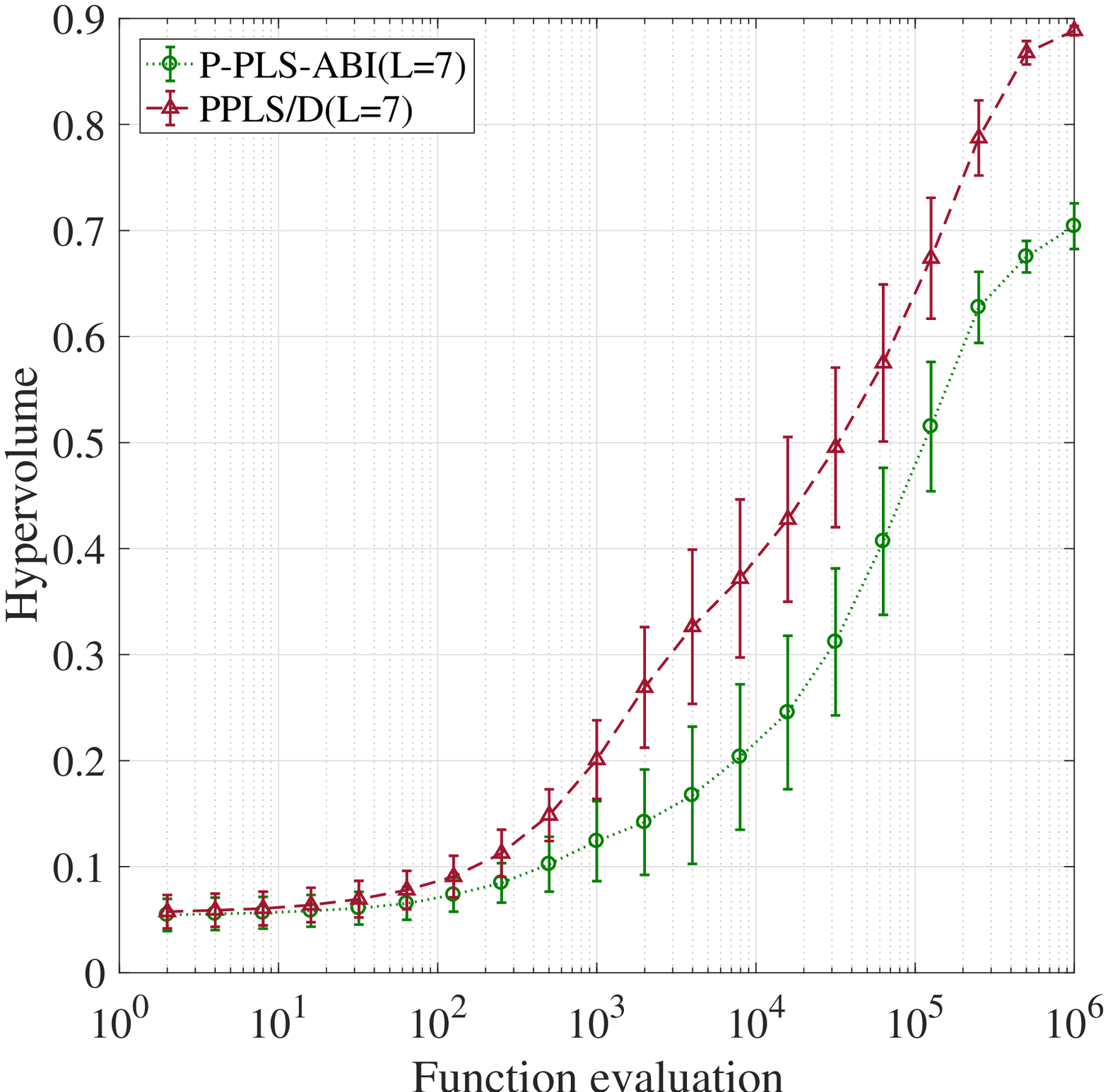}}%epm_201806141451
    \hspace{-0.09in}
  \subfigure[\tiny{mTSP: kroAD100 (m=2)}]{
    \label{fig:errbar_kroAD100_pplsabi} %% label for first subfigure
    \includegraphics[width=0.197\linewidth]{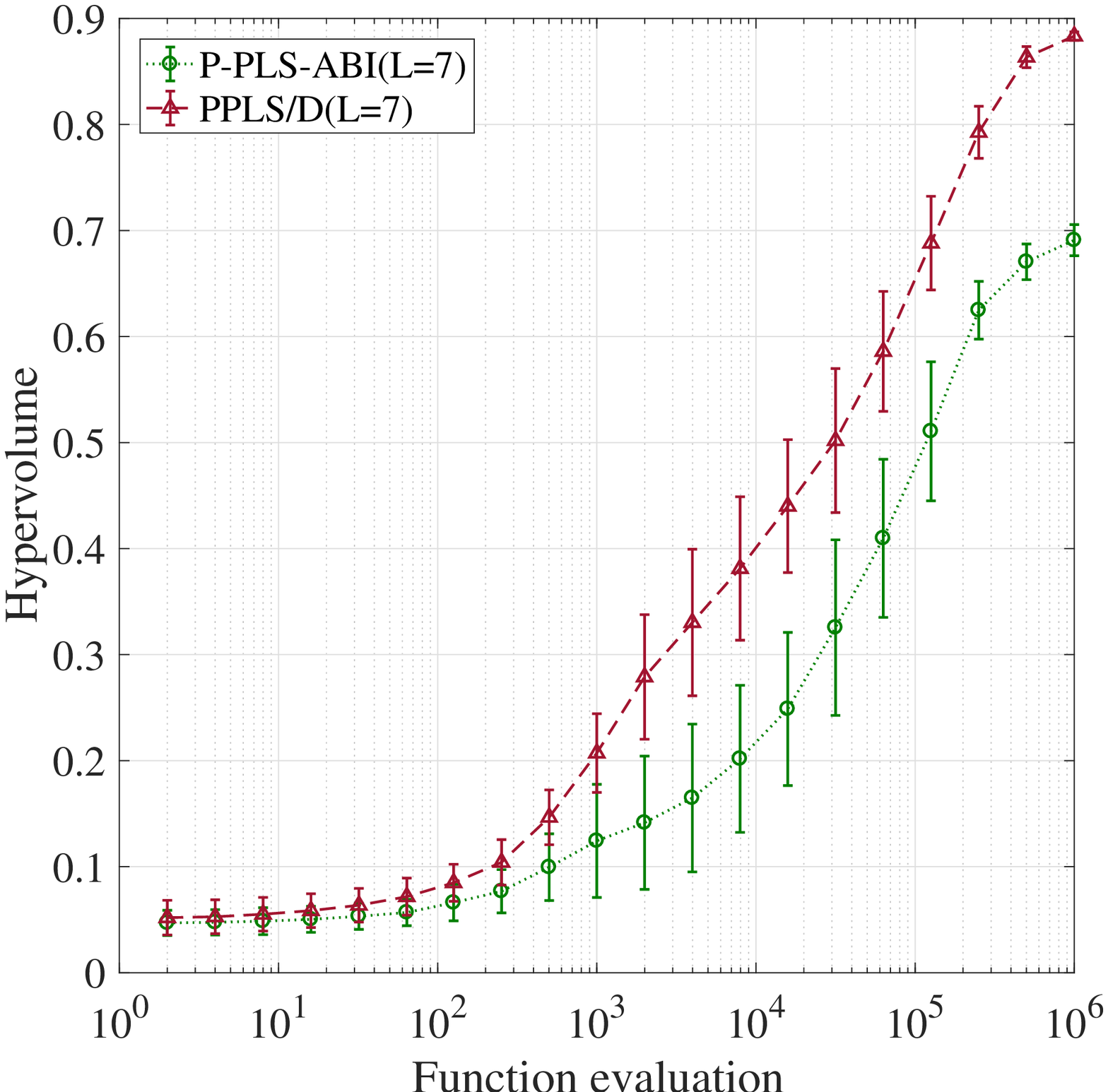}}\\%epm_201806141451
  \vspace{0in}
  \subfigure[\tiny{mTSP: kroAB150 (m=2)}]{
    \label{fig:errbar_kroAB150_pplsabi} %% label for first subfigure
    \includegraphics[width=0.197\linewidth]{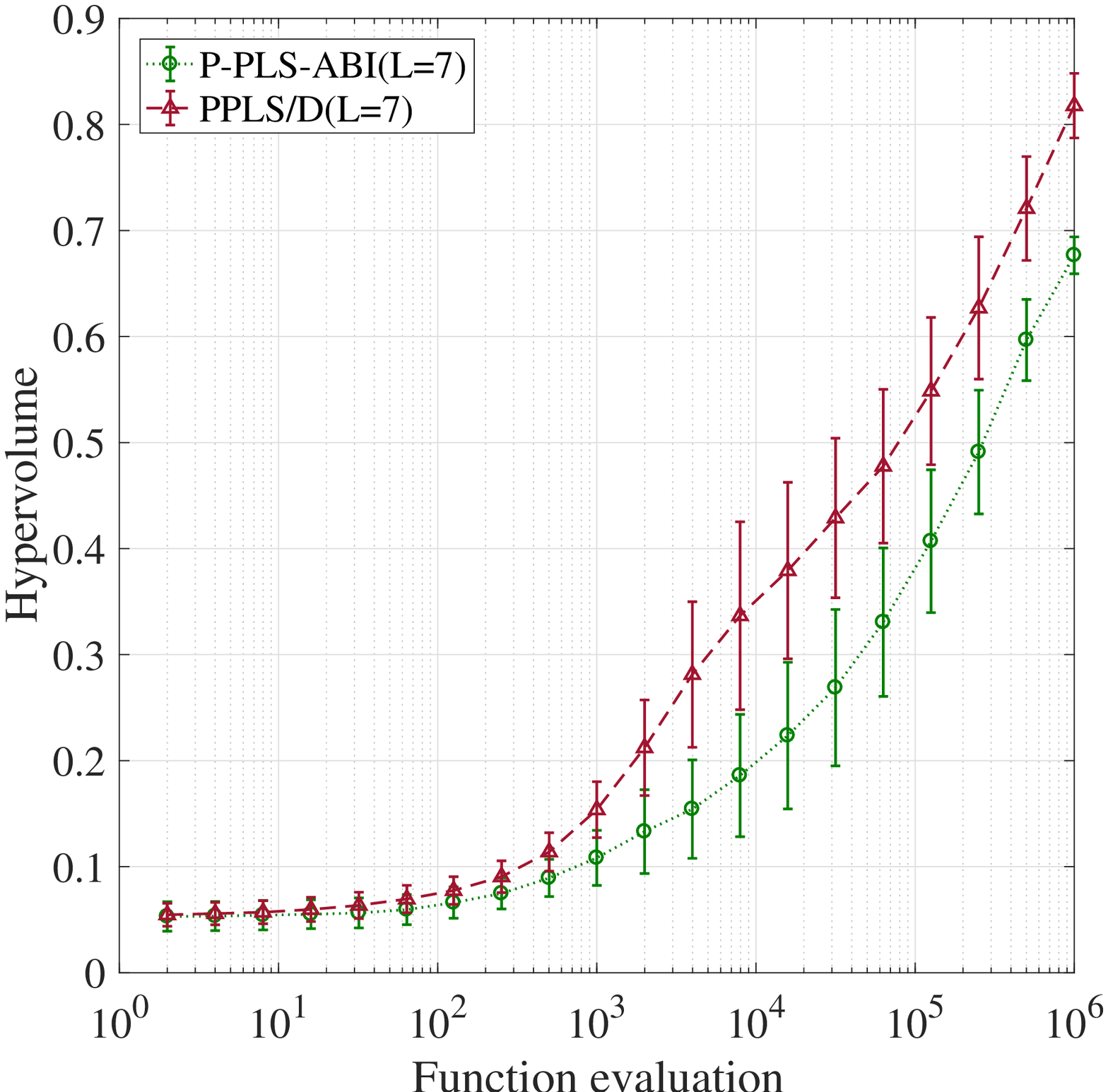}}%epm_201806141451
    \hspace{-0.09in}
  \subfigure[\tiny{mTSP: kroAB200 (m=2)}]{
    \label{fig:errbar_kroAB200_pplsabi} %% label for first subfigure
    \includegraphics[width=0.197\linewidth]{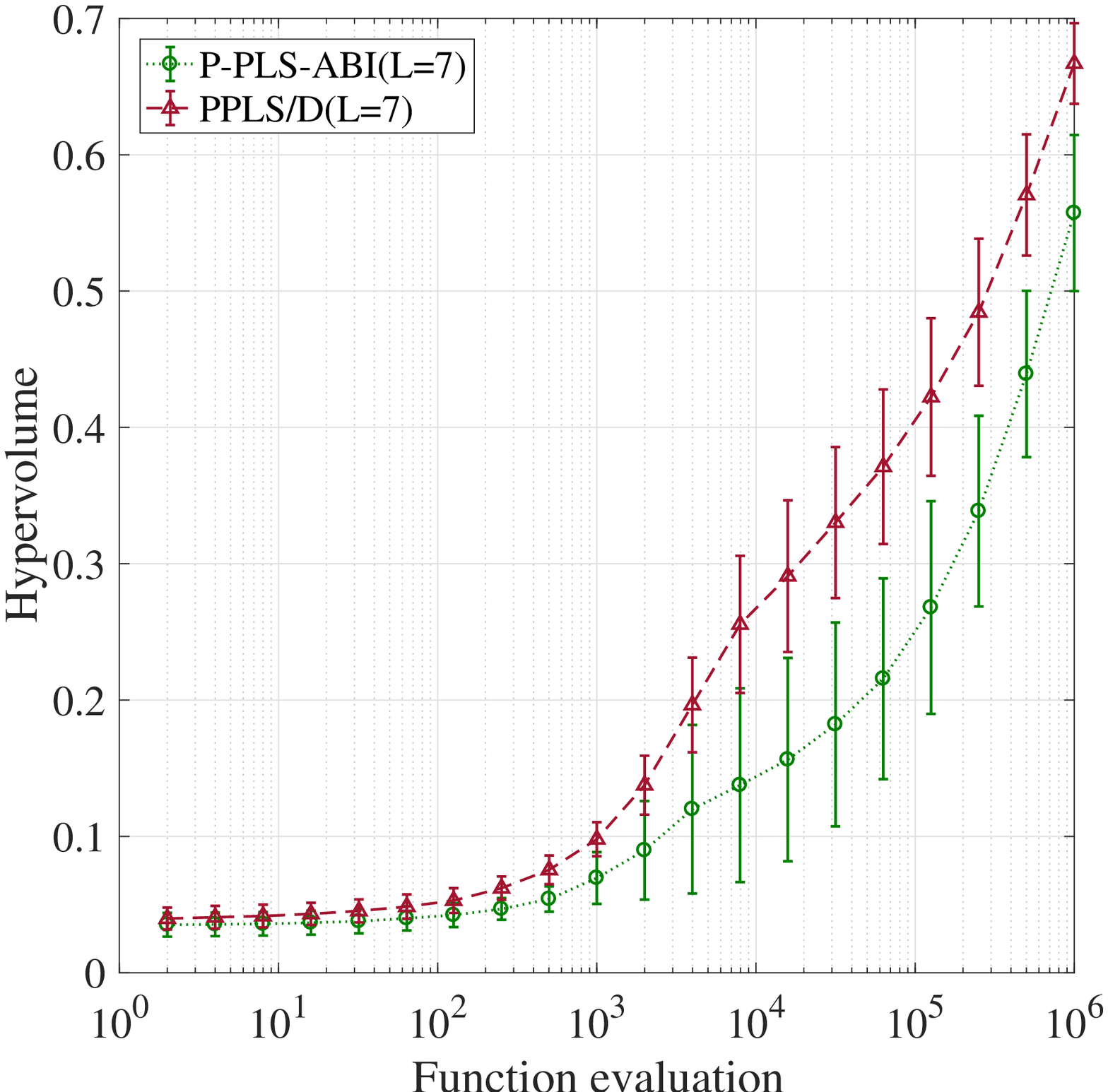}}%epm_201806141451
    \hspace{-0.09in}
  \subfigure[\tiny{mTSP: kroABC100 (m=3)}]{
    \label{fig:errbar_kroABC100_pplsabi} %% label for first subfigure
    \includegraphics[width=0.197\linewidth]{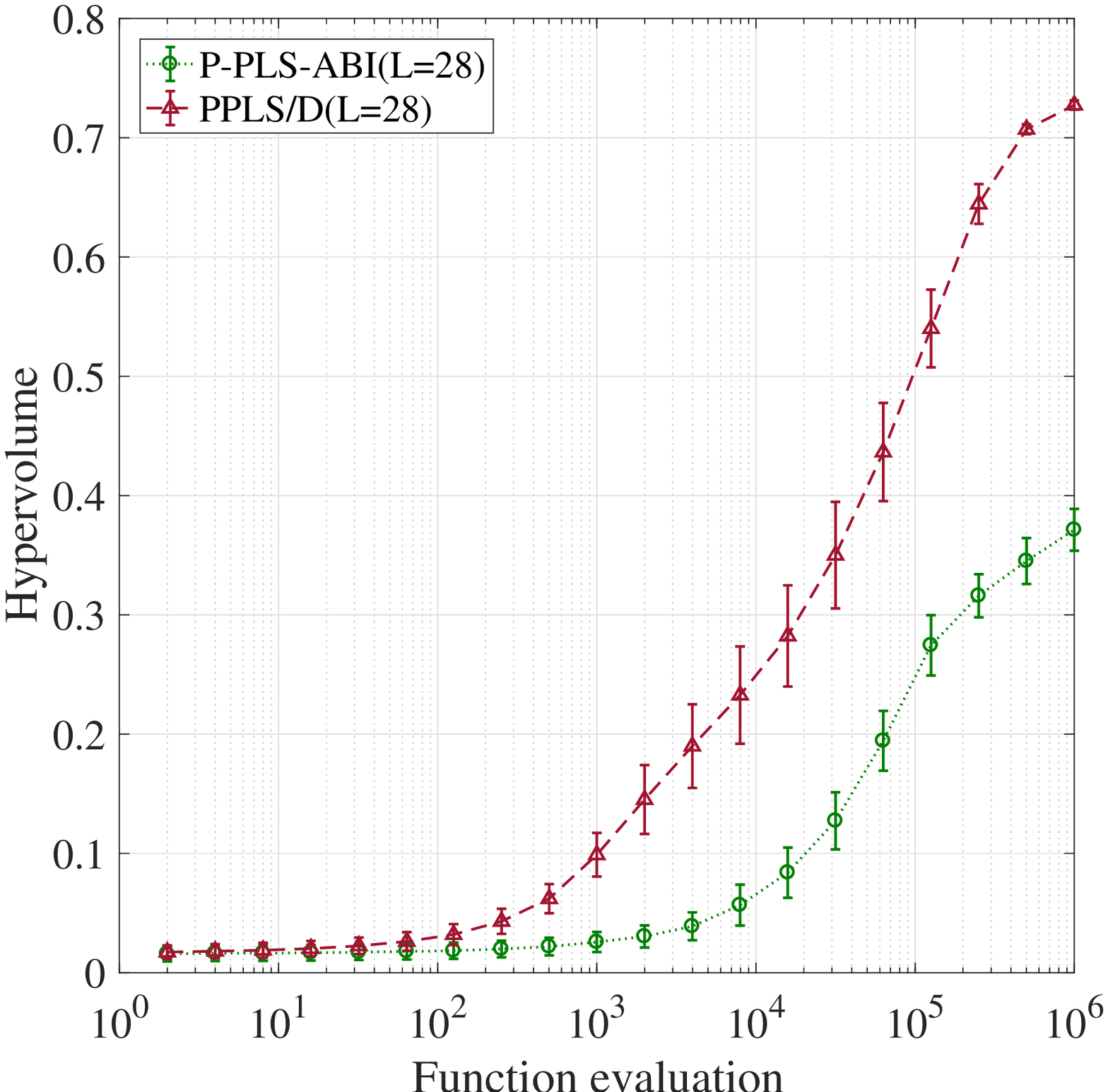}}%epm_201806141451
    \hspace{-0.09in}
  \subfigure[\tiny{mTSP: kroBCDE100 (m=4)}]{
    \label{fig:errbar_kroBCDE100_pplsabi} %% label for first subfigure
    \includegraphics[width=0.197\linewidth]{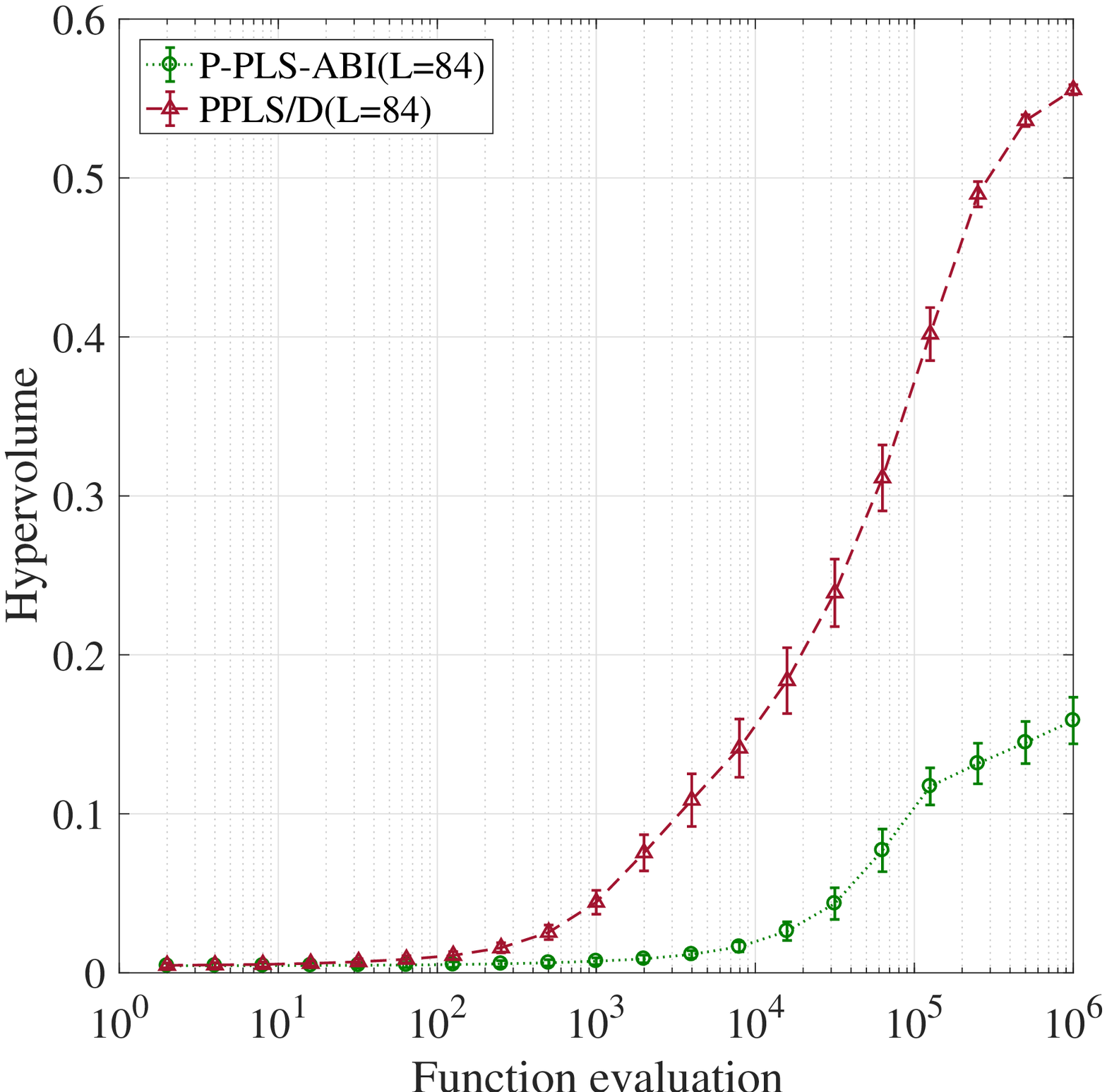}}\\%epm_201806141451
  \vspace{0in}
  \caption{The hypervolume attained by P-PLS-ABI and PPLS/D versus function evaluations. Each algorithm is executed 20 runs from different randomly generated solutions.}\label{fig:errbar_pplsabi}
  %\vspace{-0.2in}
\end{figure*}

Fig.~\ref{fig:errbar_pplsabi} shows the hypervolume values attained by PPLS/D and P-PLS-ABI versus function evaluations. From Fig.~\ref{fig:errbar_pplsabi} we can see that the proposed PPLS/D performs significantly better than P-PLS-ABI on all of the test instances. Hence, we conclude that the proposed PPLS/D can outperform both sequential and parallel implementations of PLS-ABI, which means that PPLS/D is a highly efficient building block for multiobjective combinatorial metaheuristics.

\subsection{Algorithm Behavior Investigation}
\begin{figure}%[t]
  %\vspace{-0.1in}
  \centering
    \includegraphics[width=0.35\linewidth]{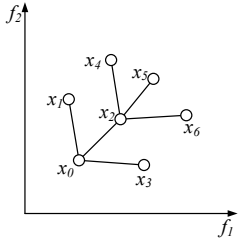}\\
  \caption{A example of the trajectory tree diagram}\label{fig:Traj_tree}
  %\vspace{-0.2in}
%\myfigureshrinker
\end{figure}

Shi et al.\cite{shi2017using} proposed a diagram called ``\emph{trajectory tree}'' to show the behavior of a PLS process. For a PLS process, the trajectory tree plots all the solutions that were ever accepted by the archive and the neighborhood relationship among them on the objective space. For example, the trajectory tree in Fig.~\ref{fig:Traj_tree} shows that from the solution $x_0$, three neighboring solutions $x_1$, $x_2$ and $x_3$ were accepted by the archive. Then the neighborhood of $x_2$ was explored and $x_4$, $x_5$ and $x_6$ were accepted by the archive. %Fig.~\ref{fig:Traj_tree_real} shows a real case of the trajectory tree on a bi-objective mUBQP instance.

\begin{figure*}%[H]
  %\vspace{-0.1in}
  \centering
  \subfigure[\tiny{PLS on mubqp\_2\_200}]{
    \label{fig:tree_m2_PLS} %% label for first subfigure
    \includegraphics[width=0.16\linewidth]{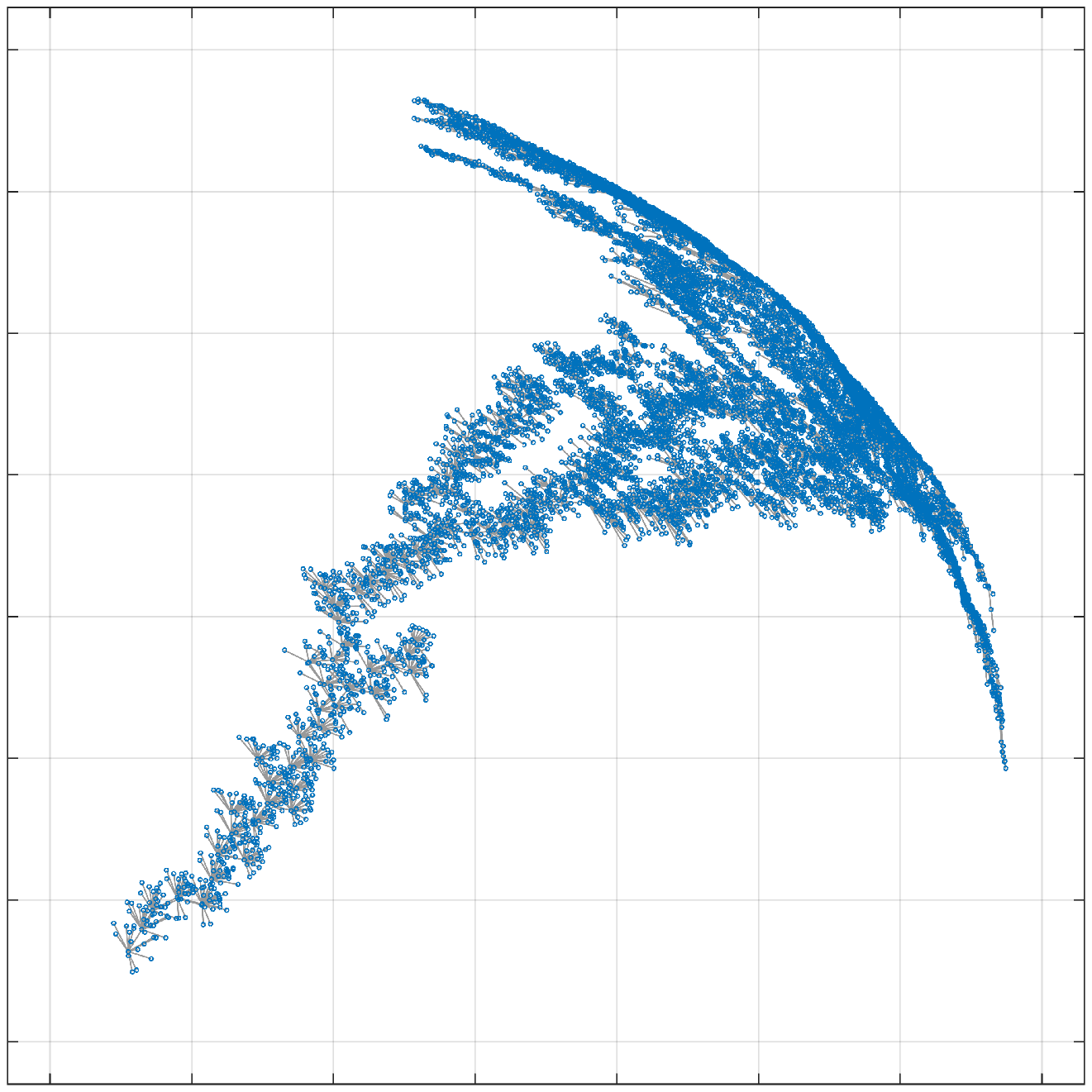}}%epm_201803052052
    %\hspace{-0.08in}
  \subfigure[\tiny{PLS-ABI on mubqp\_2\_200}]{
    \label{fig:tree_m2_PLSABI} %% label for first subfigure
    \includegraphics[width=0.16\linewidth]{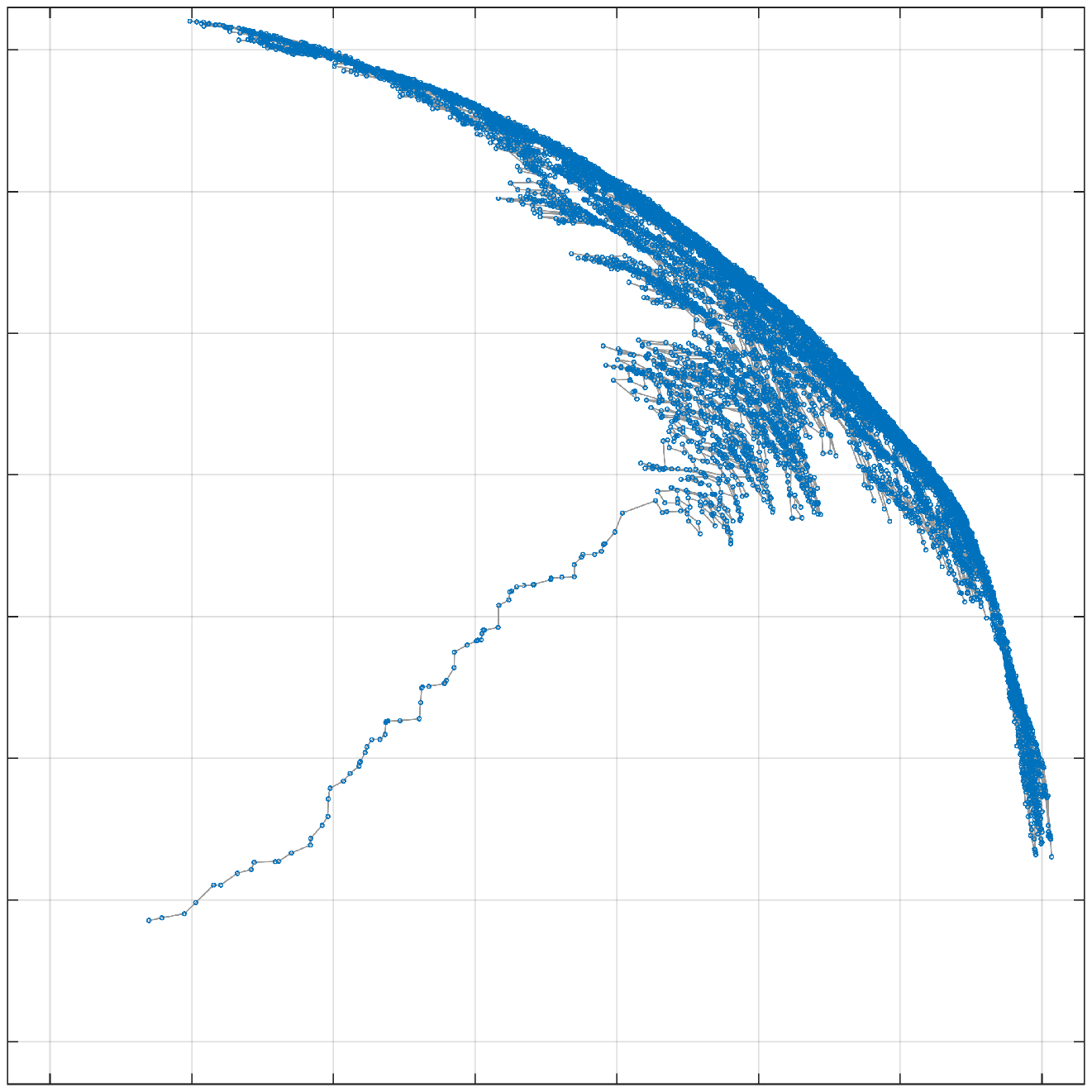}}%epm_201803052052
    %\hspace{-0.08in}
  \subfigure[\tiny{PPLS/D(H=6) on mubqp\_2\_200}]{
    \label{fig:tree_m2_PPLSD_H6} %% label for first subfigure
    \includegraphics[width=0.16\linewidth]{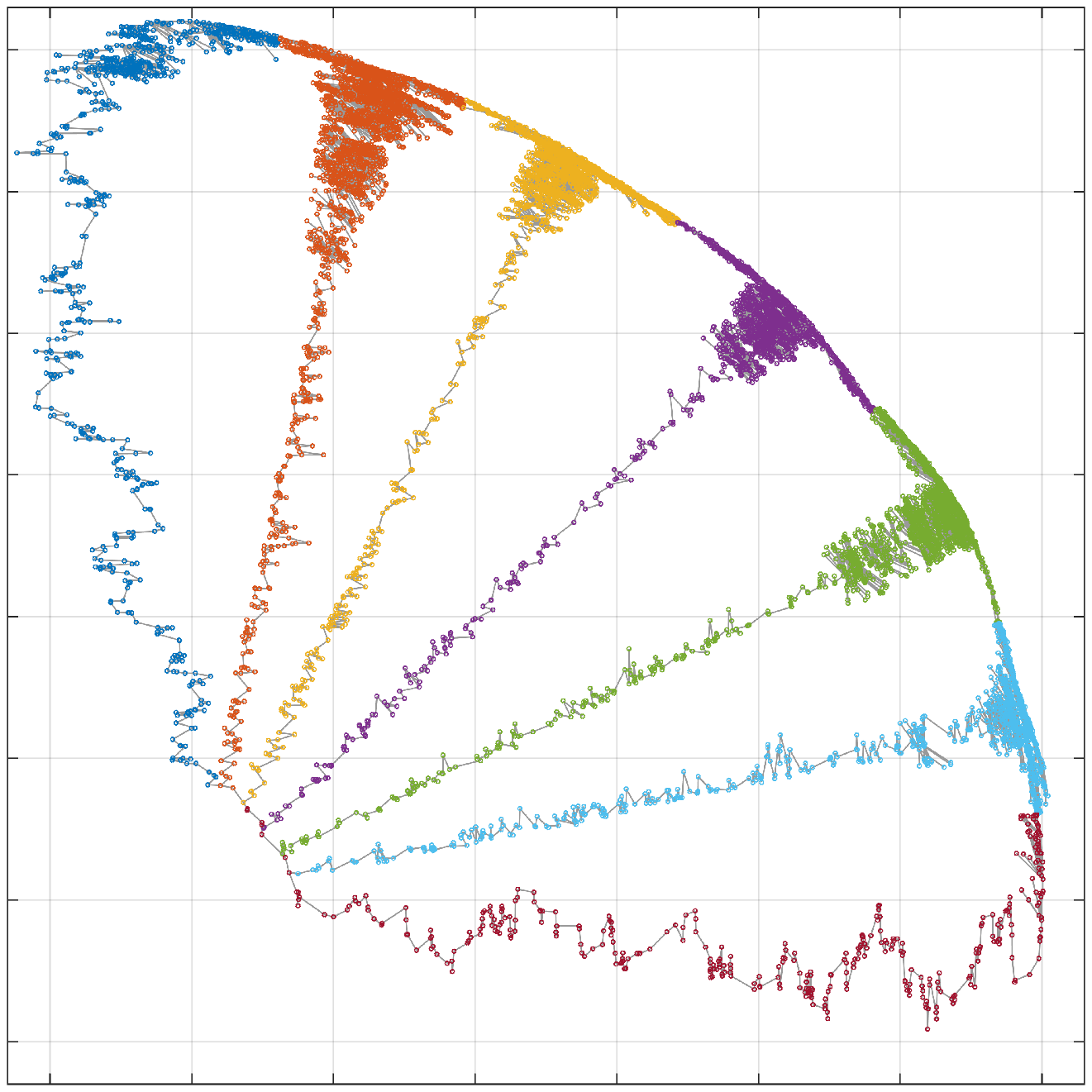}}%epm_201803052052
  \subfigure[\tiny{PLS on kroAB100}]{
    \label{fig:tree_kroAB100_PLS} %% label for first subfigure
    \includegraphics[width=0.16\linewidth]{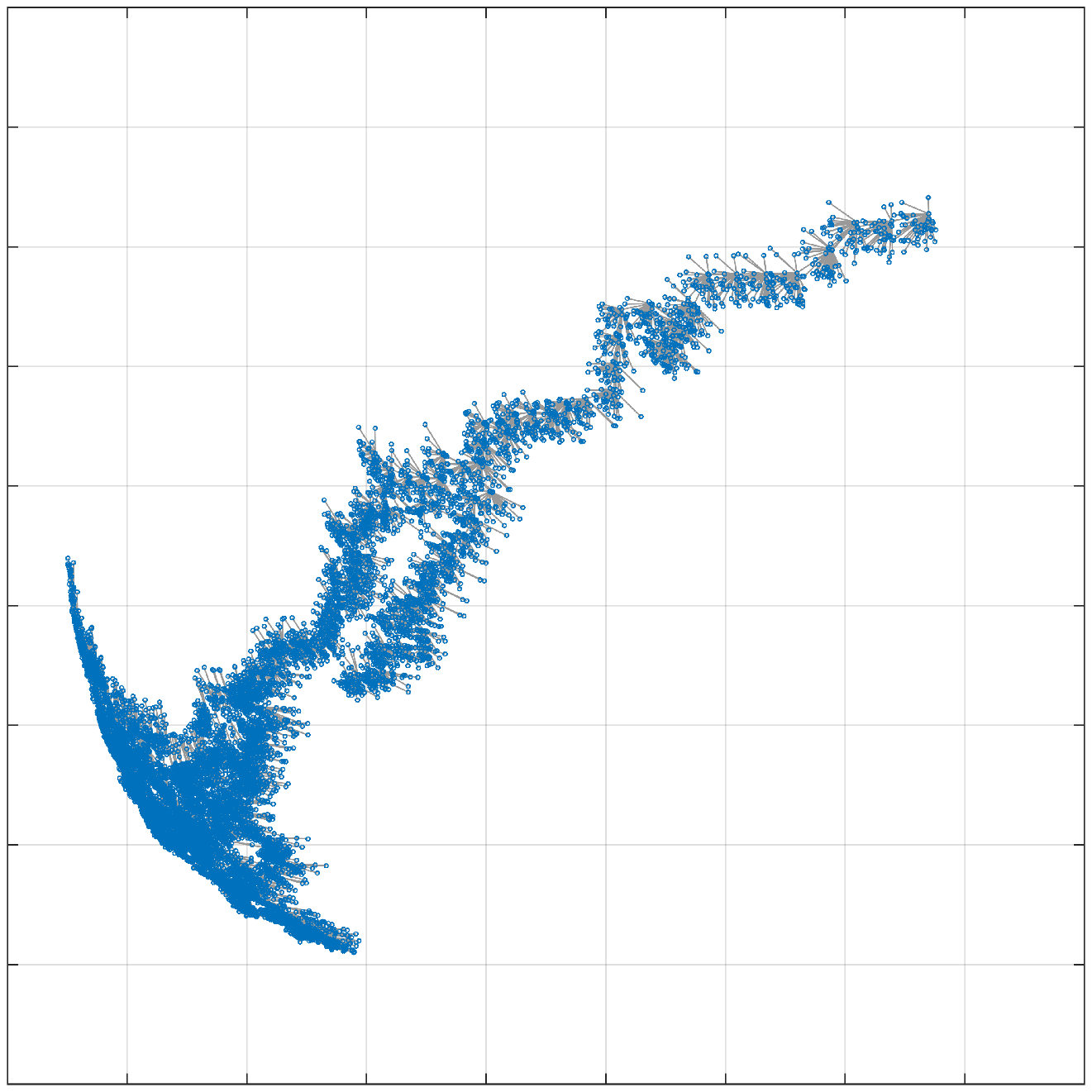}}%epm_201802141332
    %\hspace{-0.08in}
  \subfigure[\tiny{PLS-ABI on kroAB100}]{
    \label{fig:tree_kroAB100_PLSABI} %% label for first subfigure
    \includegraphics[width=0.16\linewidth]{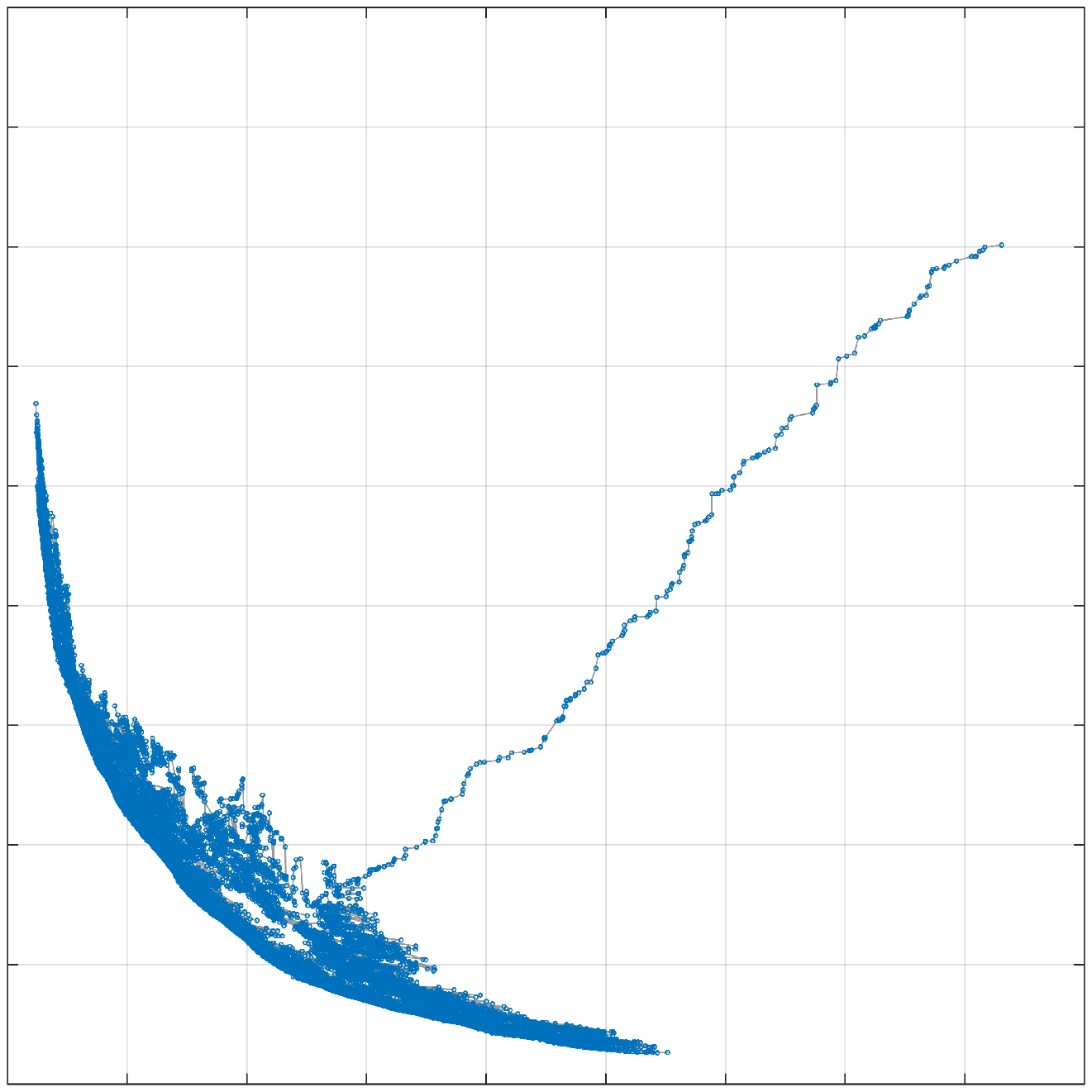}}%epm_201802141332
    %\hspace{-0.08in}
  \subfigure[\tiny{PPLS/D(H=6) on kroAB100}]{
    \label{fig:tree_kroAB100_PPLSD_H6} %% label for first subfigure
    \includegraphics[width=0.16\linewidth]{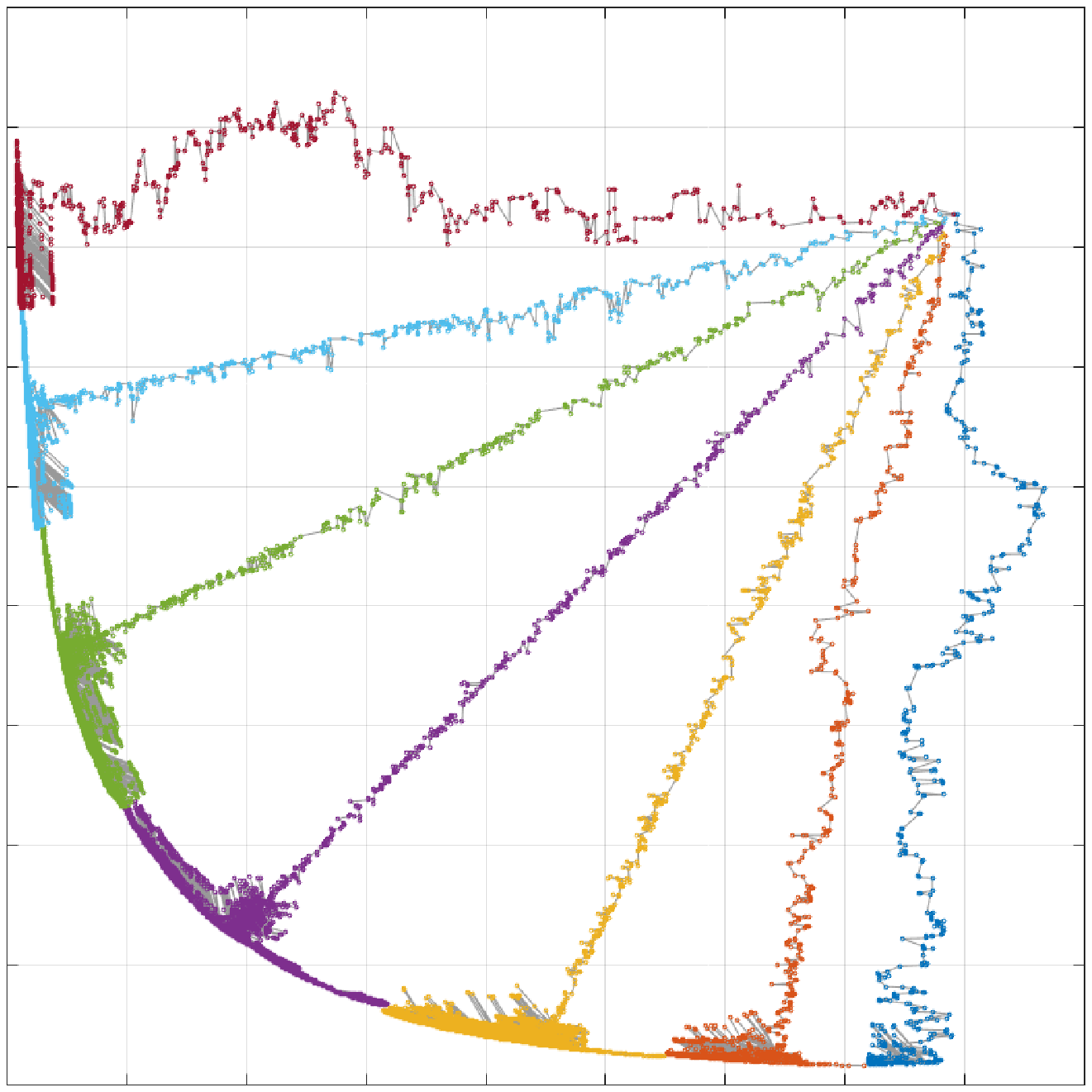}}\\%epm_201802141332
  %\vspace{-0.1in}
  \caption{The trajectory trees of a run of PLS, a run of PLS-ABI and a run of PPLS/D(H=6) on the bi-objective mUBQP instances mubqp\_2\_200 and the bi-objective mTSP instance kroAB100. The algorithms are started from randomly generated solutions. }\label{fig:trees_m2}
  %\vspace{-0.2in}
\end{figure*}

Figure~\ref{fig:tree_m2_PLS}, \ref{fig:tree_m2_PLSABI} and \ref{fig:tree_m2_PPLSD_H6} show the trajectory trees of a PLS run, a PLS-ABI run and a PPLS/D(H=6) run on the bi-objective mUBQP instance mubqp\_2\_200. Fig.~\ref{fig:tree_kroAB100_PLS}, \ref{fig:tree_kroAB100_PLSABI} and \ref{fig:tree_kroAB100_PPLSD_H6} show the trajectory trees of the three algorithms on the bi-objective mTSP instance kroAB100. We can see that the trajectory tree of PLS contains many branches at its early stage. It means that PLS accepts many solutions into its archive at the early stage. It is unnecessary to accept so many solutions at the early stage because their quality is relatively low. Meanwhile PLS-ABI leaves a single trajectory at its early stage and approaches the PF faster than PLS. However, after the early stage, the branch number of the PLS-ABI trajectory tree becomes very large. As for PPLS/D, we can see that at the early stage each parallel process leaves a single trajectory when approaching the PF. Because PPLS/D is guided by the Tchebycheff scalar functions, we can see that at the early stage the trajectory of each PPLS/D process is mainly located in the central line of the subregion. In the middle and final stage, the trajectory tree of PPLS/D contains fewer branches than PLS and PLS-ABI but it still can approximate the PF ideally. In addition, because PPLS/D contains multiple parallel processes and each process only needs to approximate a part of the PF, PPLS/D converges much faster than PLS and PLS-ABI. In the Appendix, we show more examples of the trajectory trees.

\subsection{Effect of Process Number $L$}
\begin{figure}
  %%\vspace{-0.1in}
  \centering
  \subfigure[in a wide subregion]{
    \label{fig:narrow_a} %% label for first subfigure
    \includegraphics[width=0.48\linewidth]{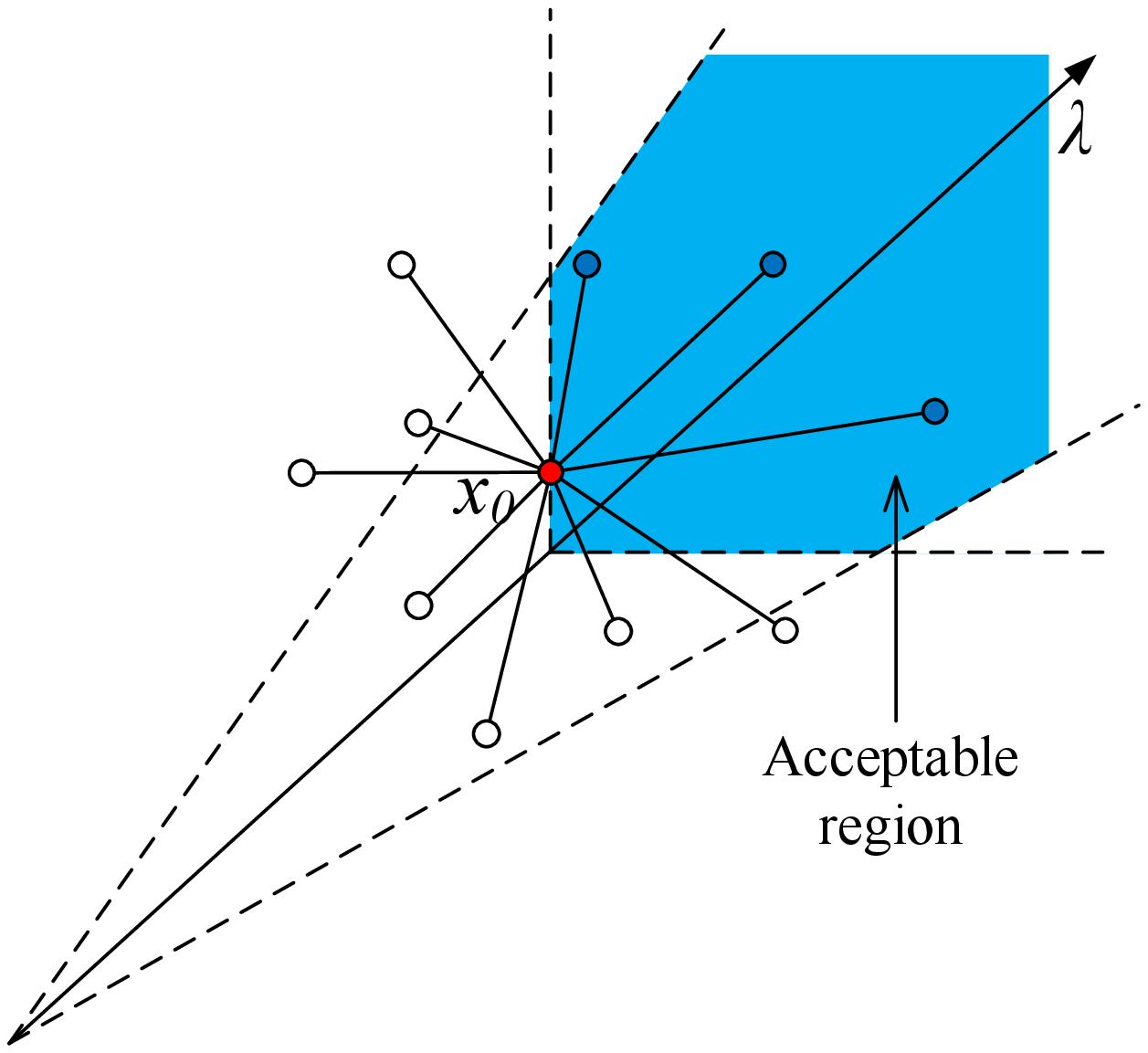}}
  \subfigure[in a narrow subregion]{
    \label{fig:narrow_b} %% label for first subfigure
    \includegraphics[width=0.48\linewidth]{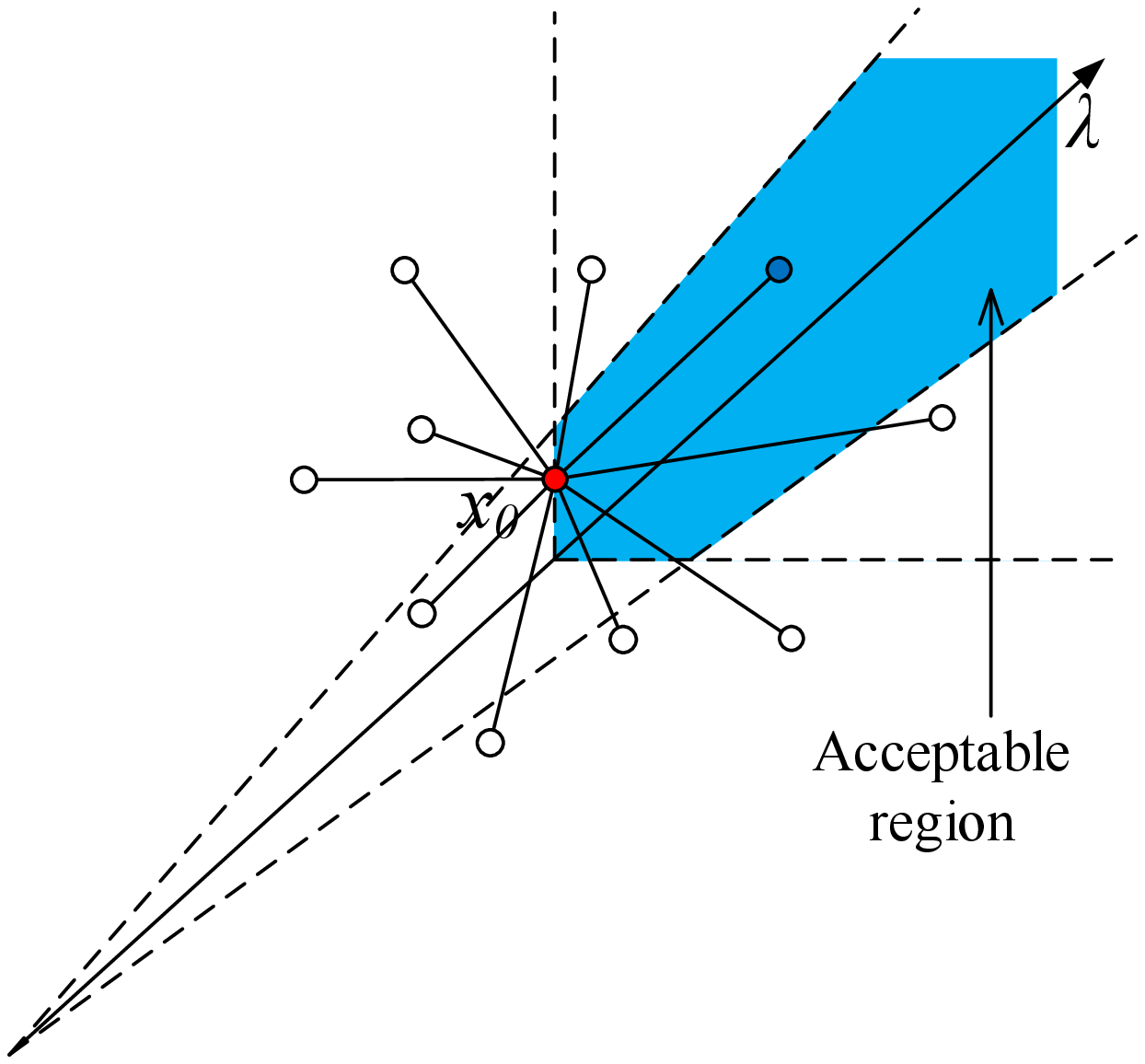}}\\
  %\vspace{-0.1in}
  \caption{Influence of the width of the subregion in the 2-D space. The current solution $x_0$ has ten neighboring solutions. Under the acceptance criterion 1, the acceptable region and the acceptable solutions are marked by blue. (a) in a wide subregion, three neighboring solutions are acceptable, (b) in a narrow subregion, only one neighboring solution is acceptable.} \label{fig:narrow}
  %%\vspace{-0.2in}
\end{figure}

In Fig.~\ref{fig:errbar} there is an interesting phenomenon that the PPLS/D algorithm with a lower $H$ value has a higher hypervolume value at the early stage. This phenomenon is very obvious on the mUBQP and mTSP instances with more than two objectives. For example, on mubqp\_4\_200 (Fig.~\ref{fig:errbar_mubqp4}) PPLS/D(H=6) gets the highest hypervolume among all the three PPLS/D algorithms at the 0.1s. Considering that a low $H$ value means a small process number $L$, this phenomenon seems to be contrary to the common knowledge that more parallel processes can bring a higher hypervolume value. A possible explanation is that a higher parallel process number means a narrower search region for each process. As illustrated in Fig.~\ref{fig:narrow}, solution $x_0$ is the current solution of a PPLS/D process running on a bi-objective problem. We assume that $x_0$ totally has 10 neighboring solutions which are uniformly distributed around $x_0$ in the objective space. In Fig.~\ref{fig:narrow_a}, the search region is relatively wide and three of $x_0$'s neighboring solutions are acceptable. Meanwhile in Fig.~\ref{fig:narrow_b}, the search region is relatively narrow and only one neighboring solution is acceptable. Under the first improvement rule, the PPLS/D process in Fig.~\ref{fig:narrow_b} needs more solution evaluation times to find an improving solution compared to the PPLS/D process in Fig.~\ref{fig:narrow_a}. Hence the search progress is relatively slow when the search region is narrow. On the other hand, as the PPLS/D process approaches the PF, the search region becomes wider. Hence the PPLS/D algorithm with more processes can get a higher hypervolume value after the early stage, as shown in Fig.~\ref{fig:errbar}.
\begin{figure}
  %\vspace{-0.1in}
  \centering
  \subfigure[mubqp\_3\_200]{
    \label{fig:accept_r_m3} %% label for first subfigure
    \includegraphics[width=0.48\linewidth]{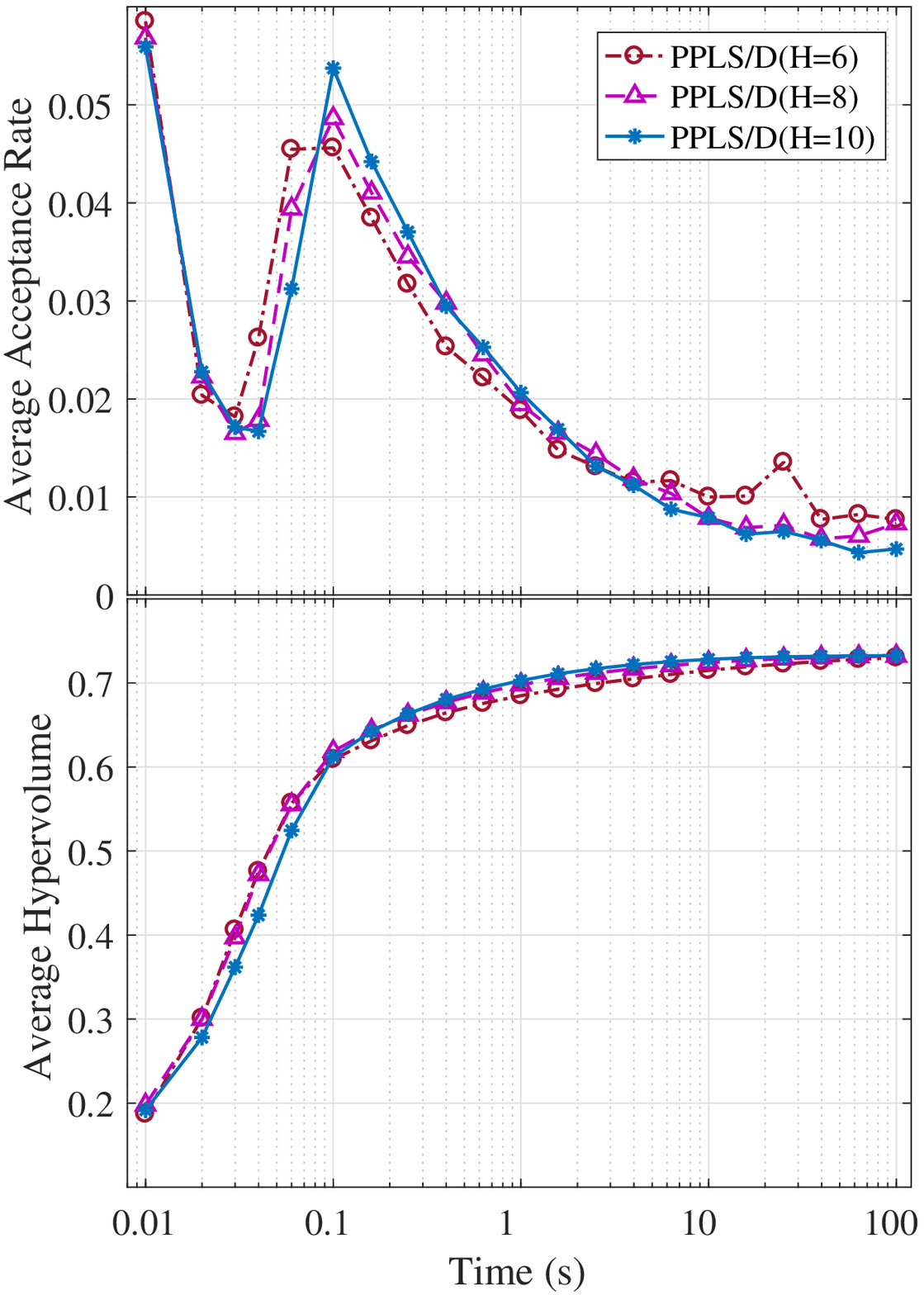}}
  \subfigure[mubqp\_4\_200]{
    \label{fig:accept_r_m4} %% label for first subfigure
    \includegraphics[width=0.48\linewidth]{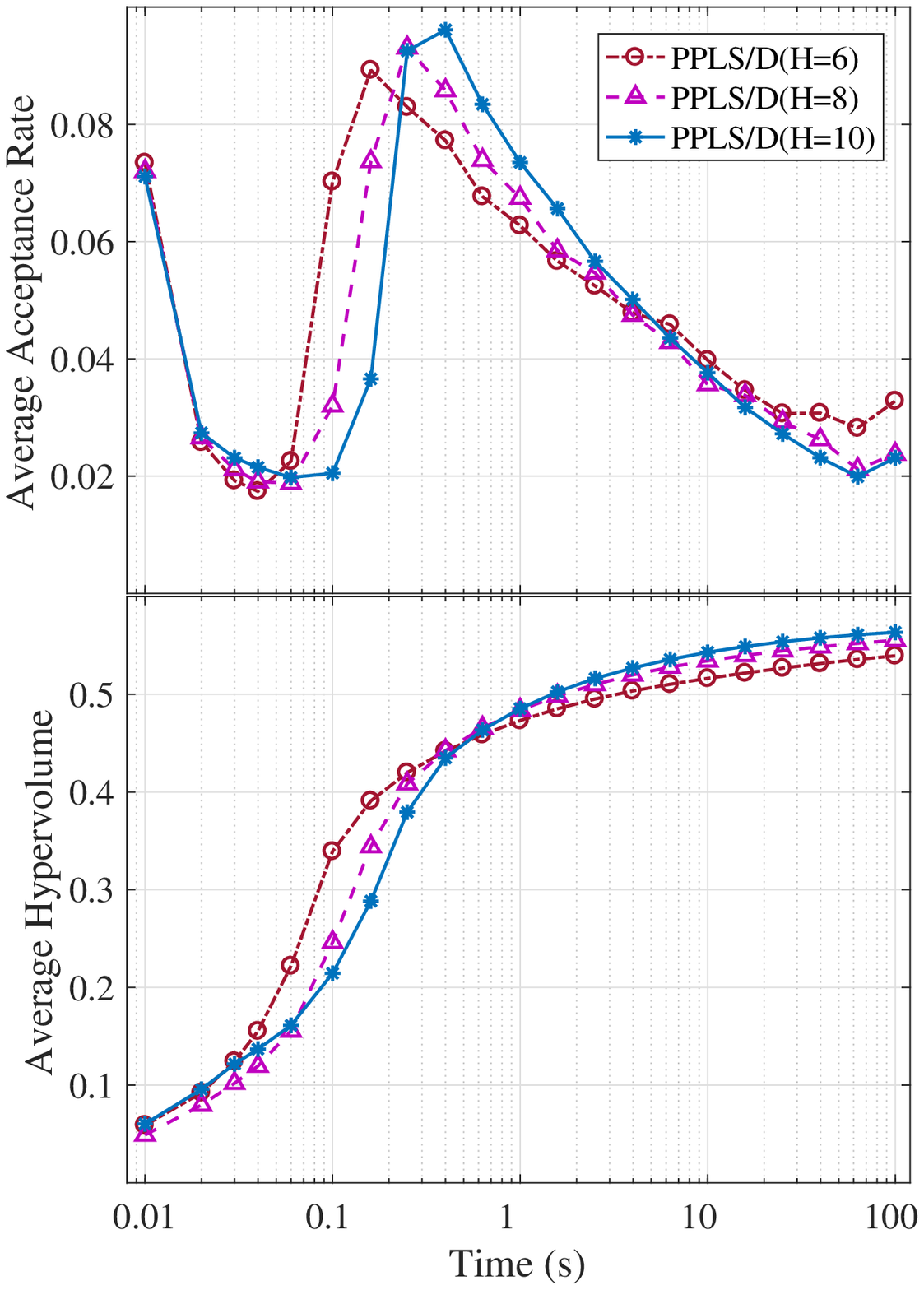}}\\
  %\vspace{-0.1in}
  \caption{The average acceptance rate and the average hypervolume attained by the three PPLS/D algorithms over time. (a) the results on the three-objective instance mubqp\_3\_200, (b) the results on the four-objective instance mubqp\_4\_200.}\label{fig:accept_r} %epm_201707311047
  %\vspace{-0.2in}
\end{figure}

To verify the above explanation, we define a metric called \emph{acceptance rate}, which is the number of newly accepted solutions divided by the total solution evaluation times. Obviously a high acceptance rate means a high search efficiency. We count the average acceptance rate per 0.01s of PPLS/D(H=6), PPLS/D(H=8) and PPLS/D(H=10) on mubqp\_3\_200 and mubqp\_4\_200. Fig.~\ref{fig:accept_r} shows the recorded average acceptance rate and the average hypervolume over time. On the instance mubqp\_3\_200 (Fig.~\ref{fig:accept_r_m3}), we can see that PPLS/D(H=6) gets the highest acceptance rate at most of the time in the first 0.1s and it also gets the highest hypervolume in the first 0.1s. After the 0.1s, PPLS/D(H=10) gets a higher acceptance rate than PPLS/D(H=6) and its hypervolume also becomes higher than that of PPLS/D(H=6). The same trend can be observed on mubqp\_4\_200 (Fig.~\ref{fig:accept_r_m4}) too. It is clearly that the hypervolume of a PPLS/D algorithms is related to the acceptance rate and the PPLS/D algorithm with fewer parallel processes can get a higher acceptance rate at the early stage because its subregion is relatively wide.

From Fig.~\ref{fig:accept_r} we can see that the acceptance rate follows a ``decrease $\to$ increase $\to$ decrease'' curve. A possible reason is that at the early stage the PPLS/D conducts scalar objective search. It can easily find a neighboring solution that improves the current best $f^{te}$ value because the current best $f^{te}$ value is relatively low at the beginning. As the current best $f^{te}$ value increases, it becomes hard to find an acceptable solution so that the acceptance rate decreases. At the middle stage, PPLS/D switches to accept non-dominated solutions, so the acceptance rate increases. At the final stage, since PPLS/D is close to the PF, finding the non-dominated solutions also becomes hard, so the acceptance rate decreases again.

\section{Conclusion}\label{sec:conclusion}
In this paper, a speed-up PLS variant called PPLS/D is proposed. Compared to the existing PLS variants, PPLS/D uses the techniques of parallel computation and problem decomposition. In PPLS/D, the search space is divided into several subregions. In each subregion, a unique PPLS/D process is executed. During each PPLS/D process, the algorithm is guided by a scalar objective function in a fine-grained manner to fast approach the PF. In the experimental study, the test suites are the instances of mUBQP and mTSP with two, three and four objectives and the initial solutions are randomly generated solutions and high quality solutions. The experimental results show that PPLS/D significantly outperforms the basic PLS and a recently proposed PLS variant on all test instances and both initial conditions. In the future, a promising alternative could be to introduce communication between the parallel processes. A well-designed parallel cooperation strategy may further improve the overall performance of PPLS/D. In addition, it has been shown in some literatures that limiting the archive size of PLS can bring performance improvement. Hence the other research direction is to develop a better archiving strategy for PPLS/D.

\medskip

{\small \noindent
\textbf{Acknowledgments.}
The authors gratefully acknowledge Bilel Derbel, Arnaud Liefooghe and S\'ebastien Verel for fruitful discussions. The work described in this paper was supported by a grant from ANR/RGC Joint Research Scheme sponsored by the Research Grants Council of the Hong Kong Special Administrative Region, China and France National Research Agency (Project No. A-CityU101/16).}



\bibliographystyle{IEEEtran}
% \bibliography{Ref_Shi}


%\end{IEEEbiography}


%\newpage



%\vfill


\begin{appendices}
\section*{Appendix}
\begin{figure*}%[H]
  %\vspace{-0.1in}
  \centering
  \subfigure[\tiny{PLS on mubqp\_2\_200}]{
    \label{fig:tree_m2_PLS_HQI} %% label for first subfigure
    \includegraphics[width=0.16\linewidth]{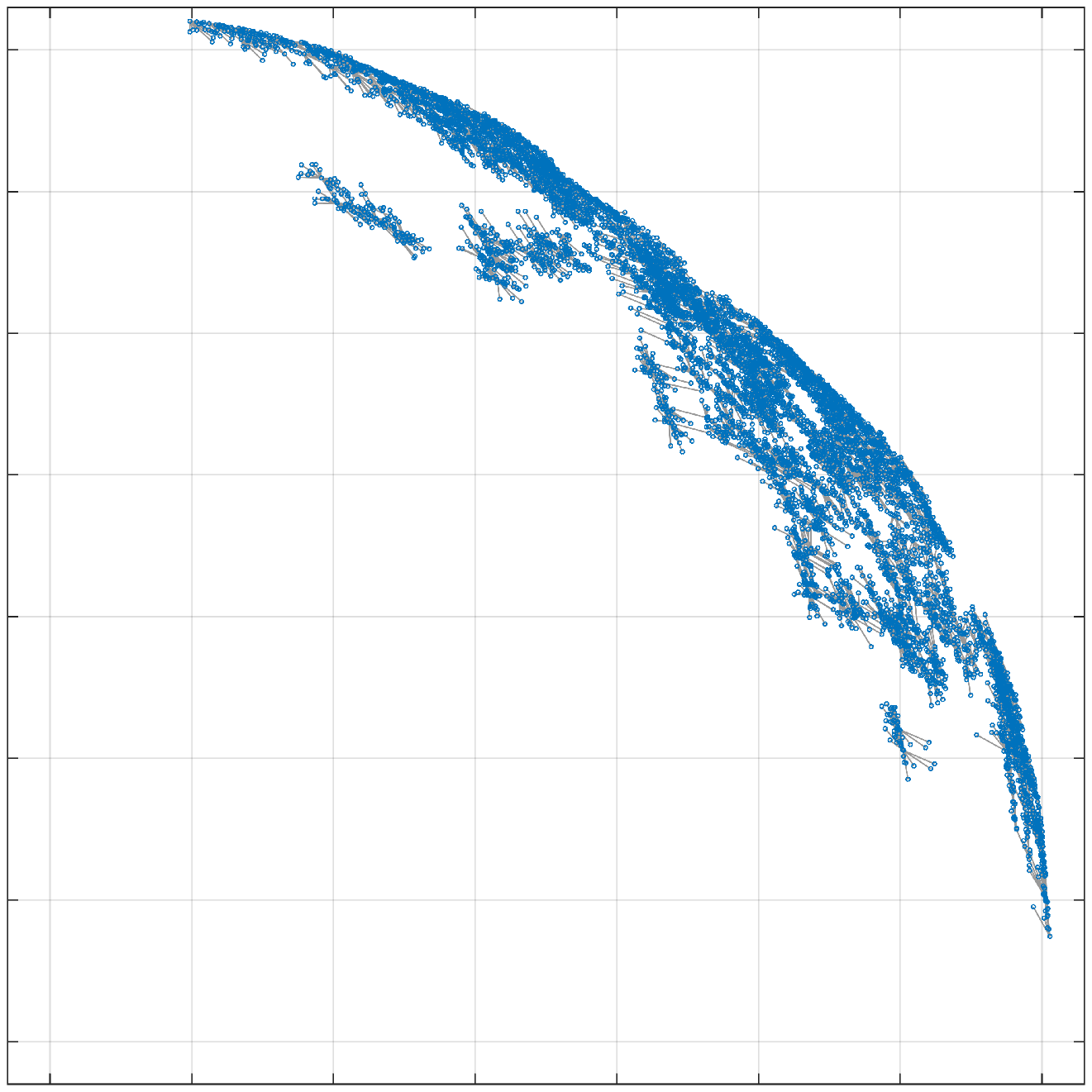}}%epm_201802051149
    %\hspace{-0.08in}
  \subfigure[\tiny{PLS-ABI on mubqp\_2\_200}]{
    \label{fig:tree_m2_PLSABI_HQI} %% label for first subfigure
    \includegraphics[width=0.16\linewidth]{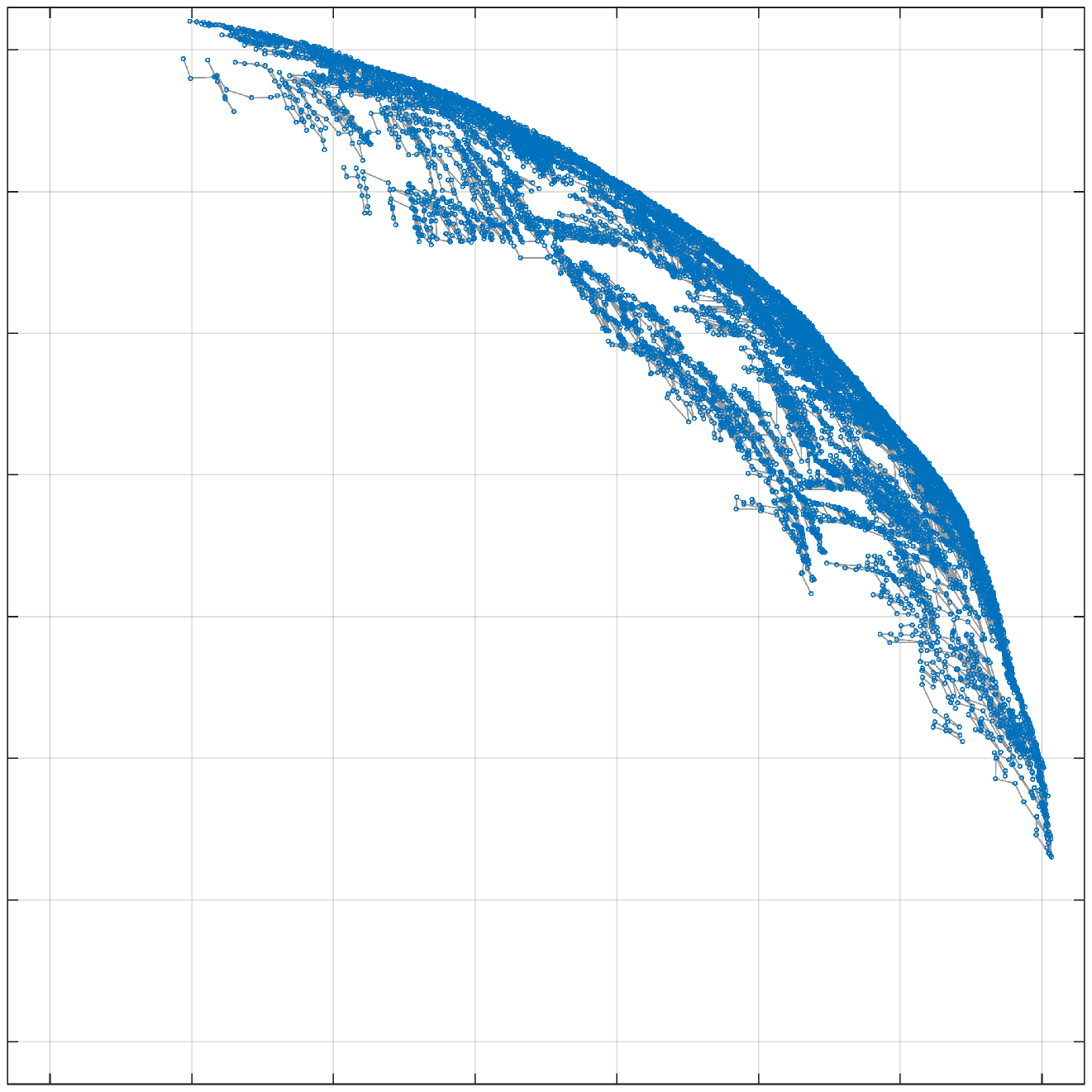}}%epm_201802051149
    %\hspace{-0.08in}
  \subfigure[\tiny{PPLS/D(H=6) on mubqp\_2\_200}]{
    \label{fig:tree_m2_PPLSD_H6_HQI} %% label for first subfigure
    \includegraphics[width=0.16\linewidth]{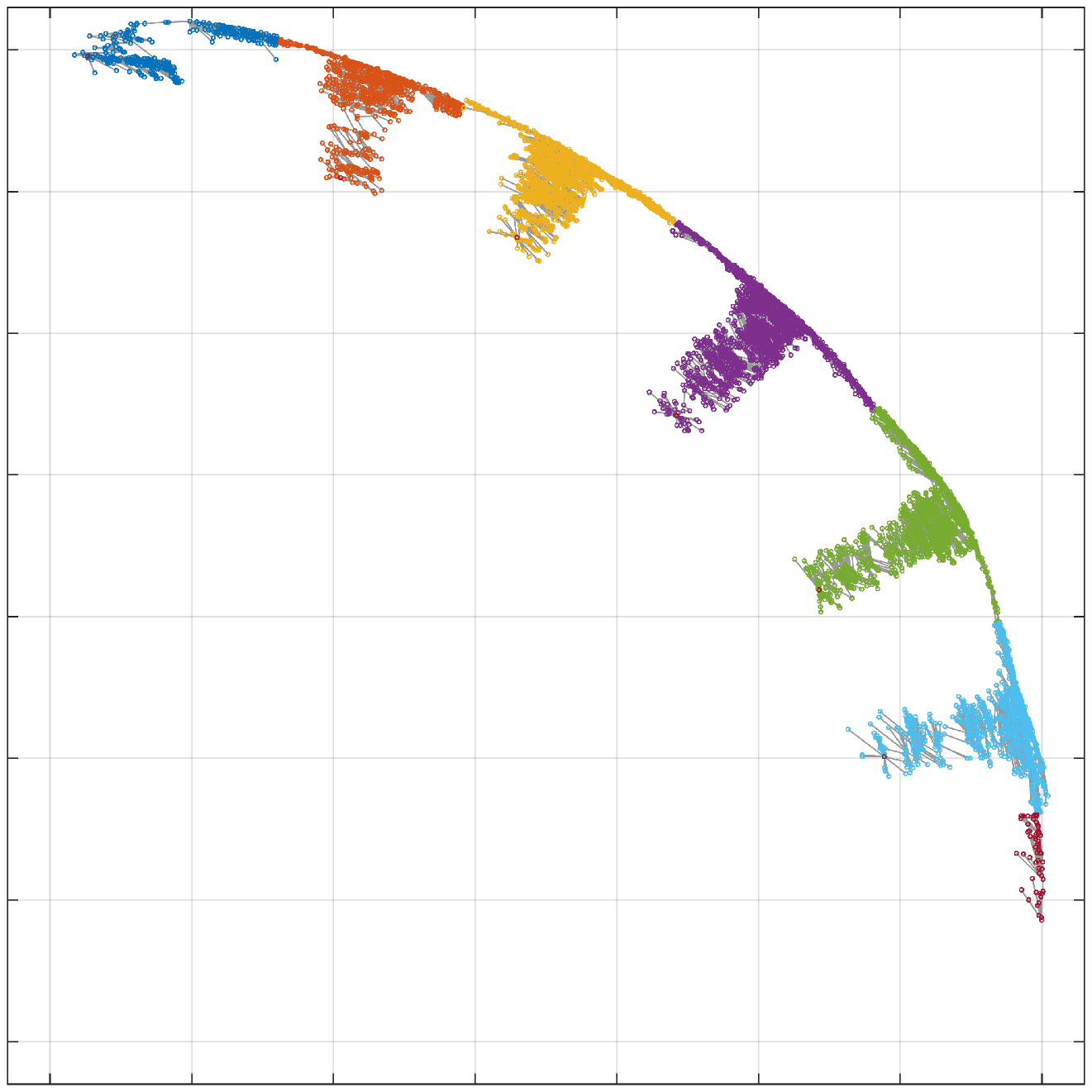}}%epm_201802051149
  \subfigure[\tiny{PLS on kroAB100}]{
    \label{fig:tree_kroAB100_PLS_HQI} %% label for first subfigure
    \includegraphics[width=0.16\linewidth]{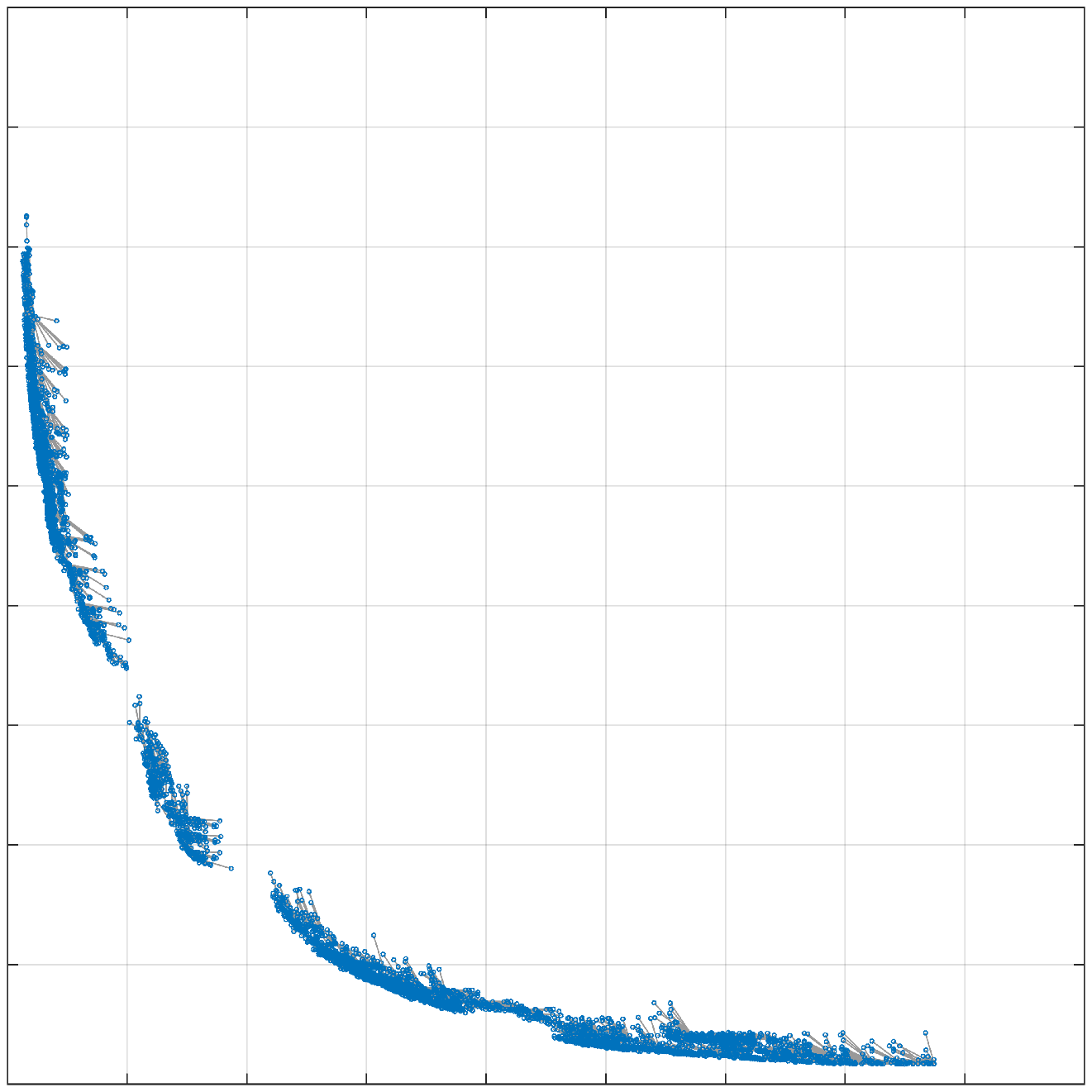}}%epm_201802261945
    %\hspace{-0.08in}
  \subfigure[\tiny{PLS-ABI on kroAB100}]{
    \label{fig:tree_kroAB100_PLSABI_HQI} %% label for first subfigure
    \includegraphics[width=0.16\linewidth]{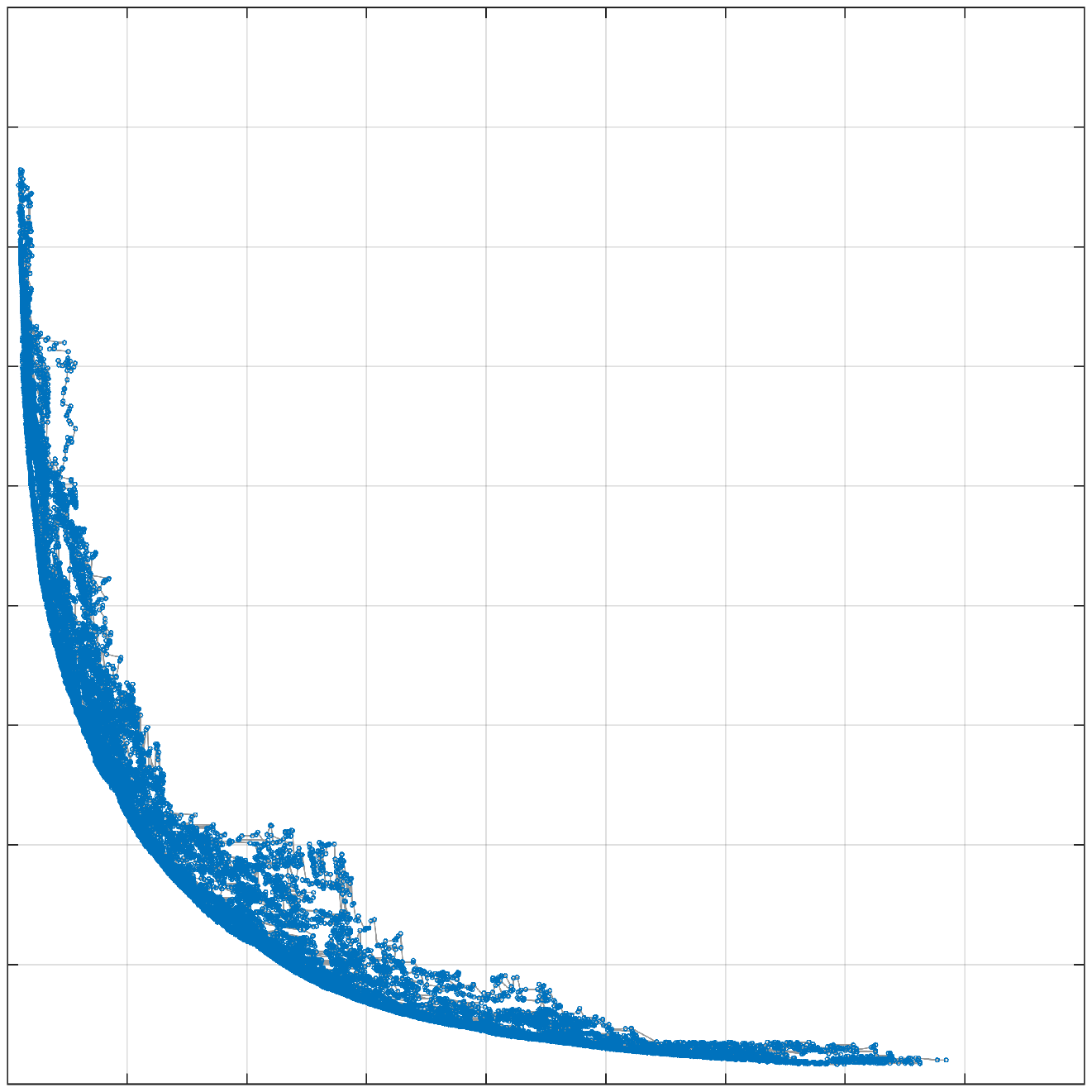}}%epm_201802261945
    %\hspace{-0.08in}
  \subfigure[\tiny{PPLS/D(H=6) on kroAB100}]{
    \label{fig:tree_kroAB100_PPLSD_H6_HQI} %% label for first subfigure
    \includegraphics[width=0.16\linewidth]{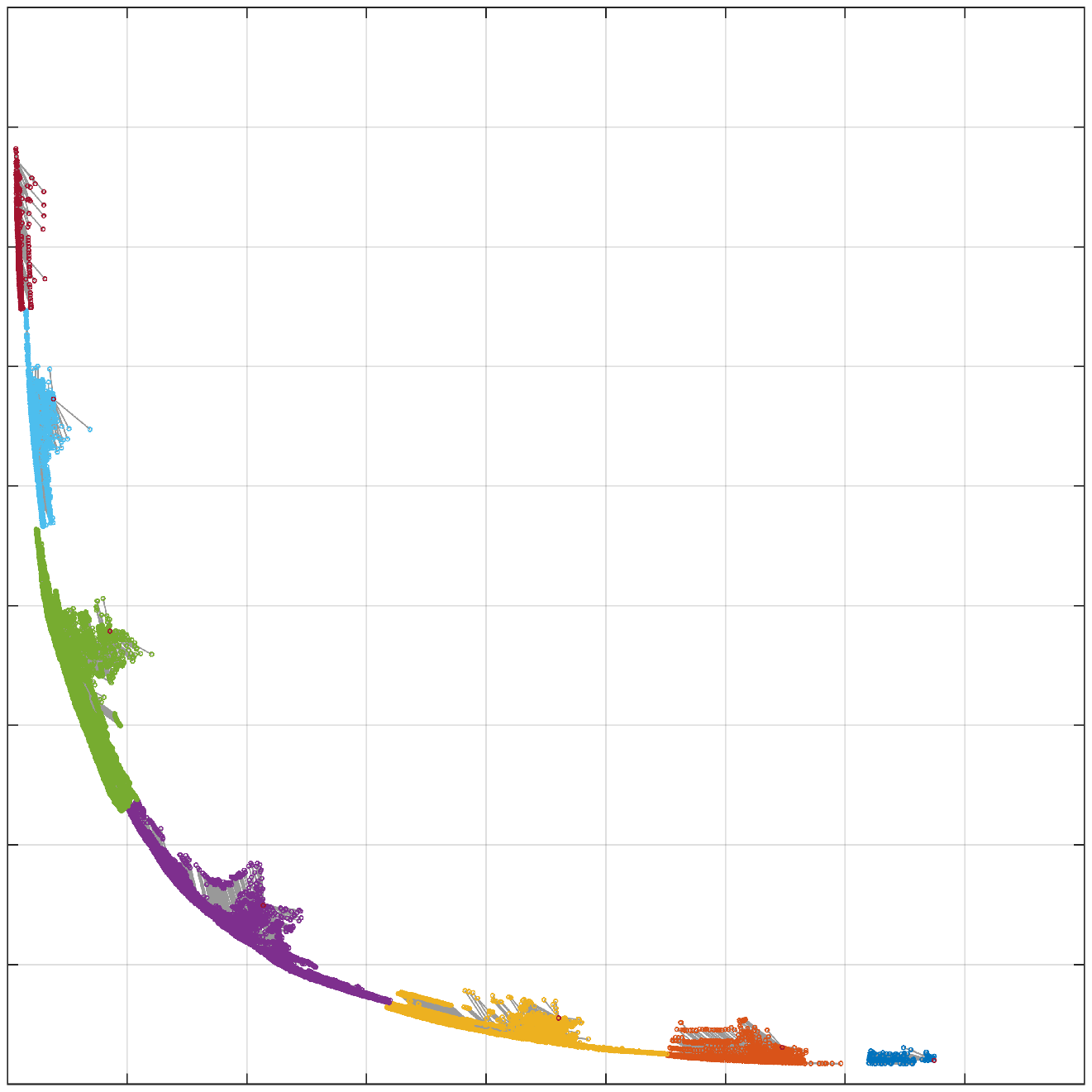}}\\%epm_201802261945
  %\vspace{-0.1in}
  \caption{The trajectory trees of a run of PLS, a run of PLS-ABI and a run of PPLS/D on the bi-objective mUBQP instances mubqp\_2\_200 and the bi-objective mTSP instance kroAB100. The algorithms are started from high quality solutions. }\label{fig:trees_m2_HQI}
  %\vspace{-0.2in}
\end{figure*}

Figure~\ref{fig:trees_m2} shows the example trajectory trees of PLS, PLS-ABI and PPLS/D(H=6) from randomly generated solutions on the bi-objective mUBQP instance mubqp\_2\_200 and the bi-objective mTSP instance kroAB100. Here Fig.~\ref{fig:trees_m2_HQI} shows the example trajectory trees when the algorithms are started from high quality solutions on the two instances. A very interesting phenomenon in Fig.~\ref{fig:trees_m2_HQI} is that on the mTSP instance kroAB100 PLS's tree has fewer branches than PLS-ABI's tree (see Fig~\ref{fig:tree_kroAB100_PLS_HQI} and Fig~\ref{fig:tree_kroAB100_PLSABI_HQI}). This does not mean that PLS performs better than PLS-ABI. It is because that when starting from high quality solutions, in each round PLS needs more time to find an acceptable solution than PLSABI. Hence, when the runtime is fixed, the tree of PLS has fewer branches than that of PLS-ABI and of course the final archive of PLS is worse than that of PLS-ABI.

\begin{figure*}%[H]
  %\vspace{-0.1in}
  %\centering
  \subfigure[\tiny{PLS on mubqp\_3\_200}]{
    \label{fig:tree_m3_PLS_1000} %% label for first subfigure
    \includegraphics[width=0.16\linewidth]{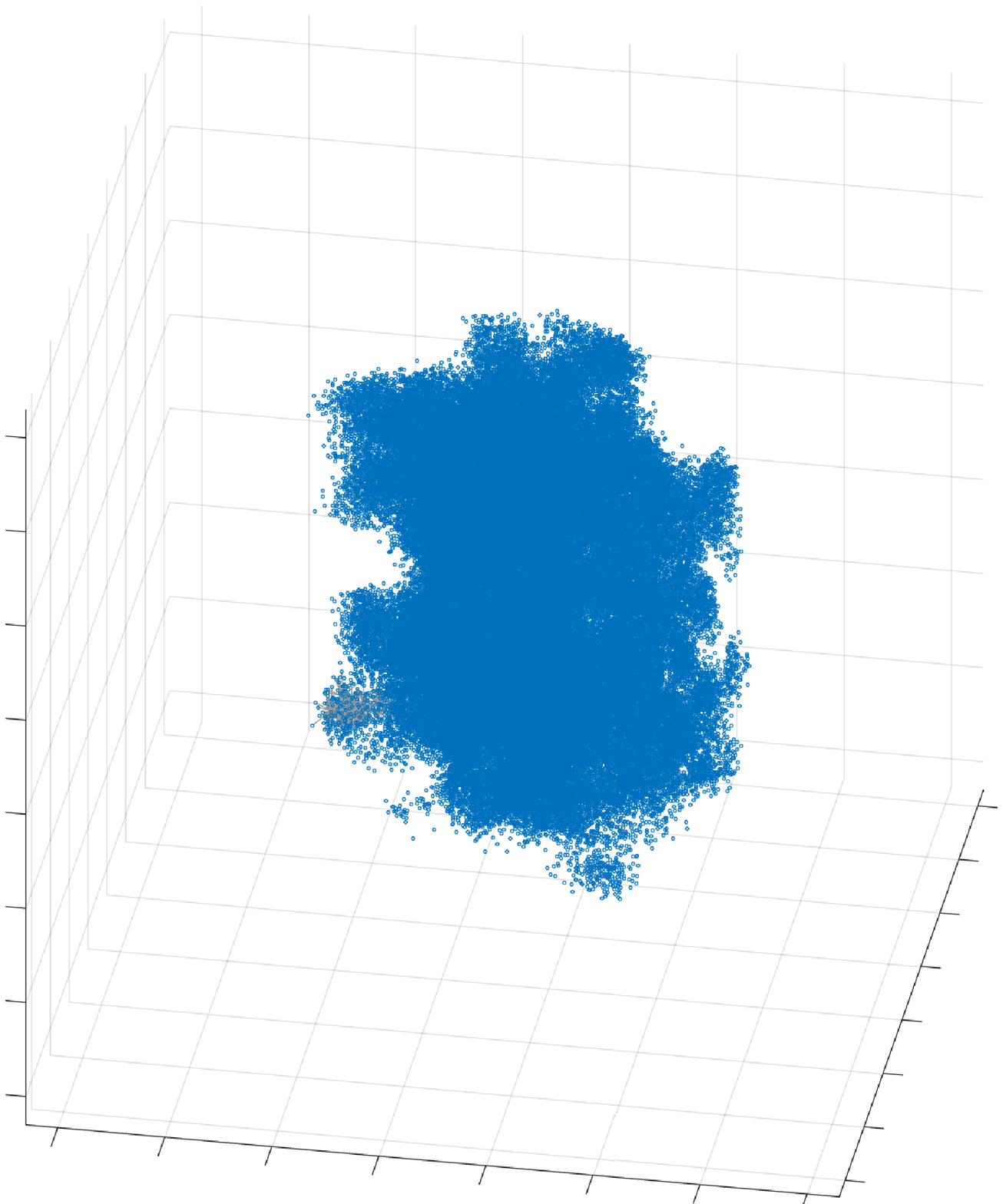}}%epm_201803052052
    %\hspace{-0.08in}
  \subfigure[\tiny{PLS-ABI on mubqp\_3\_200}]{
    \label{fig:tree_m3_PLSABI_1000} %% label for first subfigure
    \includegraphics[width=0.16\linewidth]{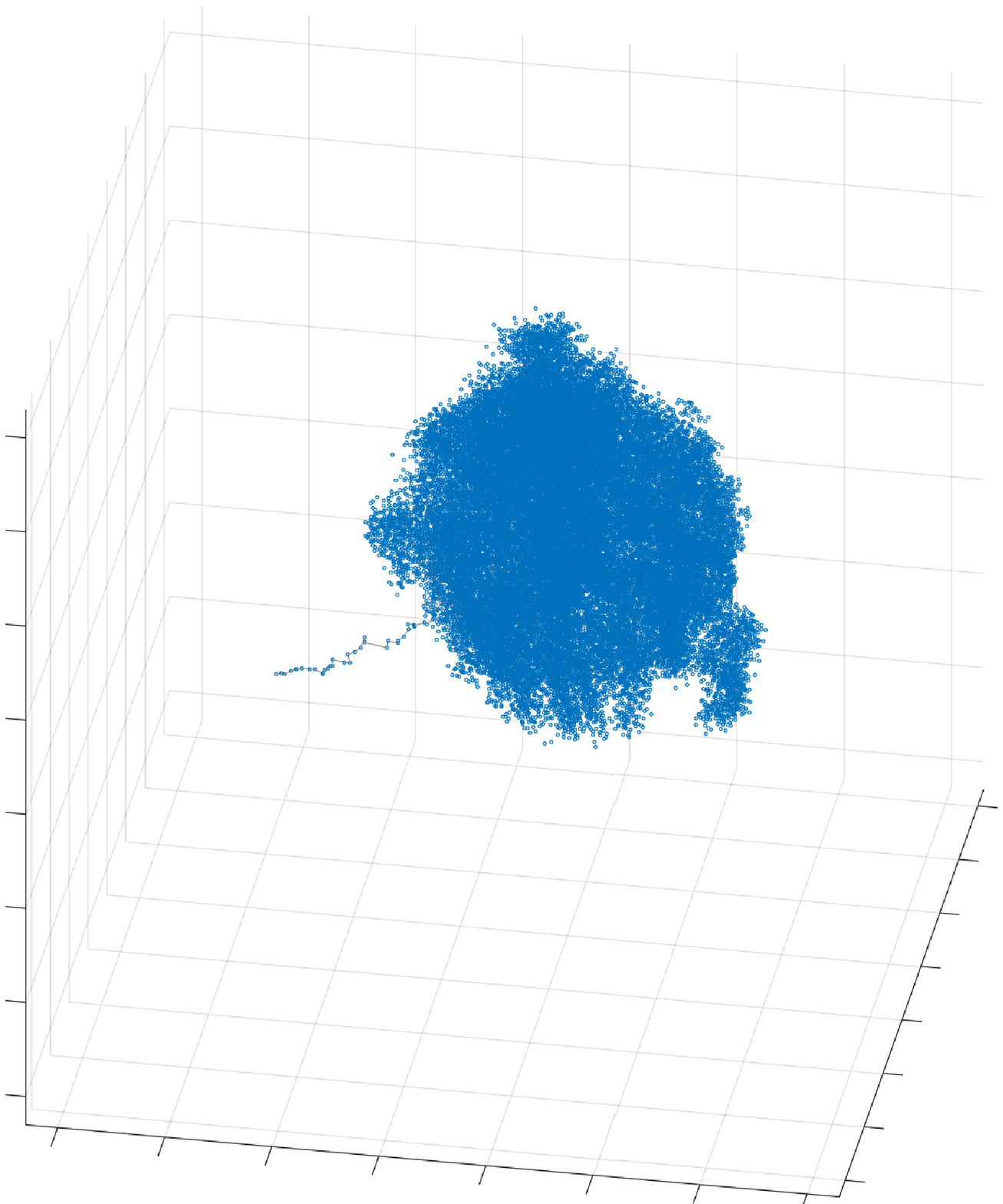}}%epm_201803052052
    %\hspace{-0.08in}
  \subfigure[\tiny{PPLS/D(H=10) on mubqp\_3\_200}]{
    \label{fig:tree_m3_PPLSD_1000} %% label for first subfigure
    \includegraphics[width=0.16\linewidth]{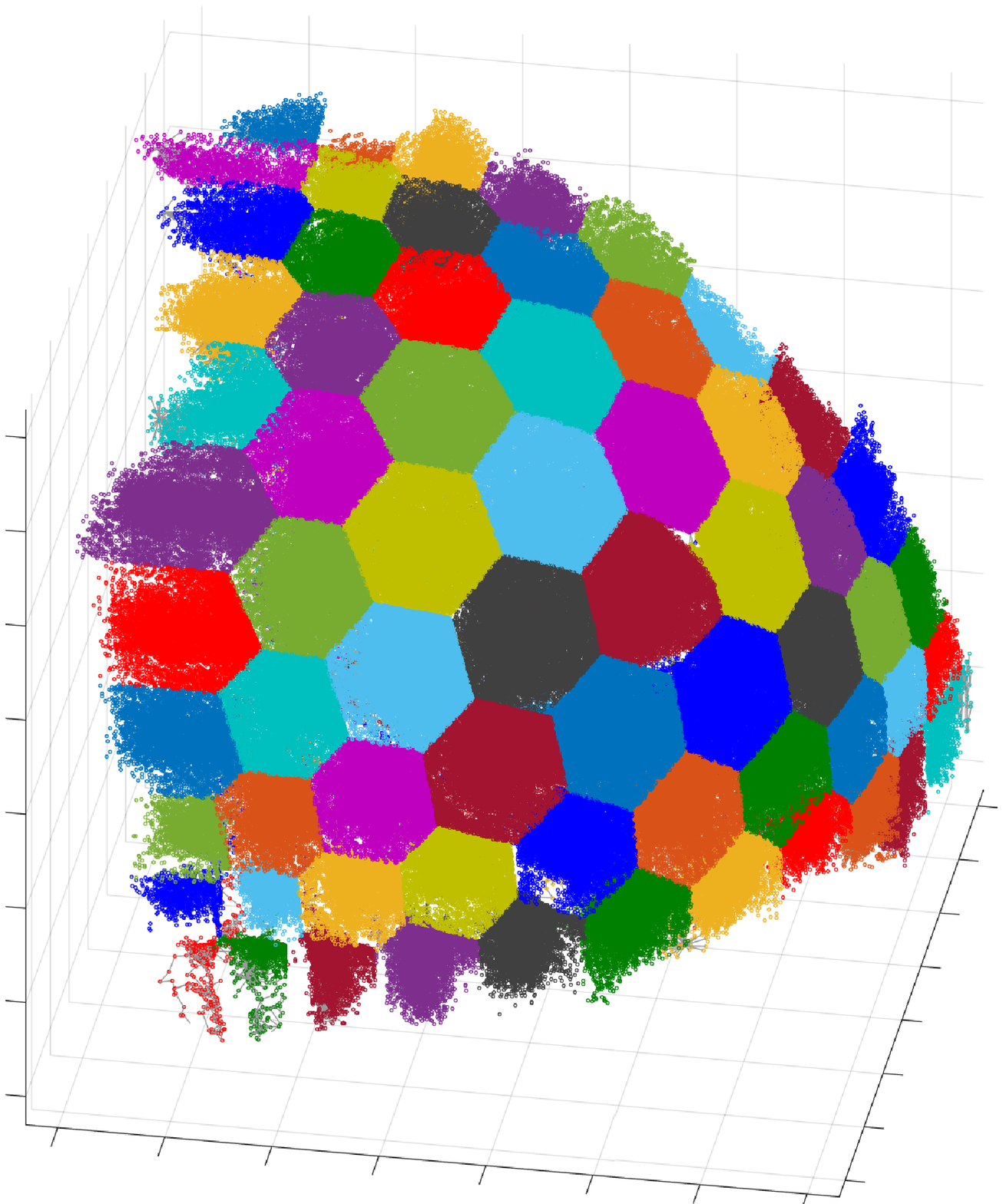}}%epm_201803052052
    %\hspace{-0.08in}
  \subfigure[\tiny{PLS on kroABC100}]{
    \label{fig:tree_kroABC100_PLS_100} %% label for first subfigure
    \includegraphics[width=0.16\linewidth]{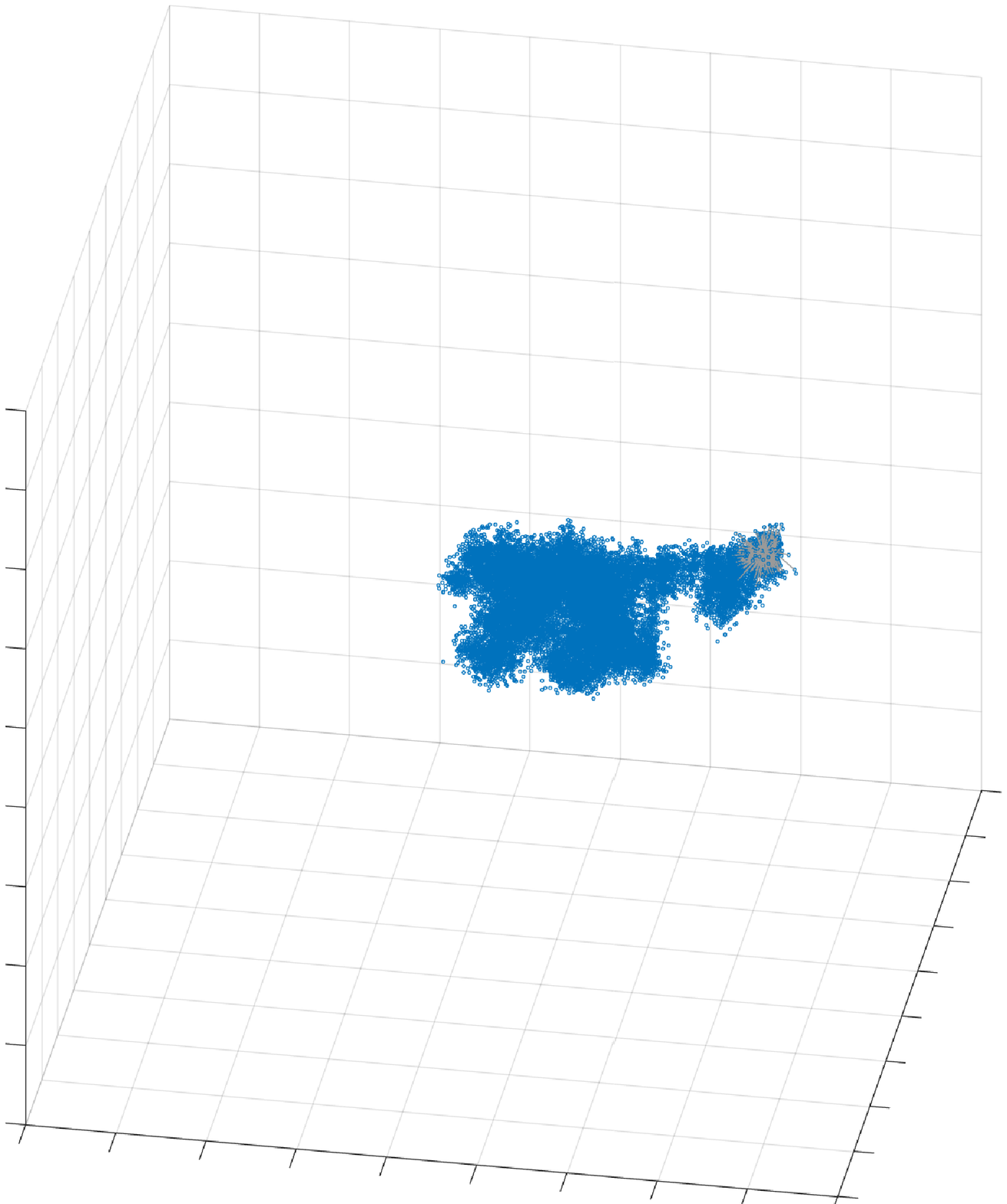}}%epm_201801291655
    %\hspace{-0.08in}
  \subfigure[\tiny{PLS-ABI on kroABC100}]{
    \label{fig:tree_kroABC100_PLSABI_100} %% label for first subfigure
    \includegraphics[width=0.16\linewidth]{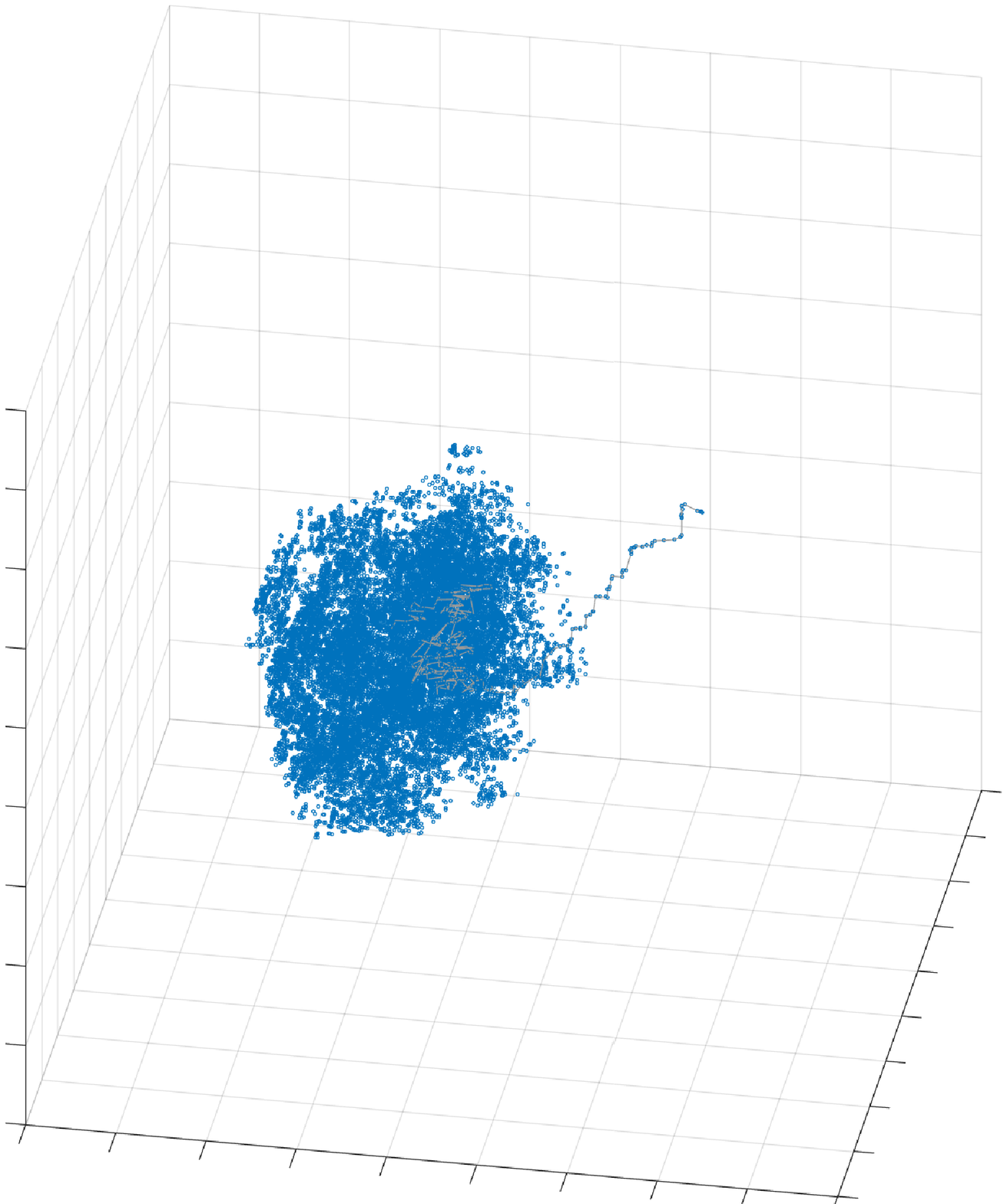}}%epm_201801291655
    %\hspace{-0.08in}
  \subfigure[\tiny{PPLS/D(H=10) on kroABC100}]{
    \label{fig:tree_kroABC100_PPLSD_100} %% label for first subfigure
    \includegraphics[width=0.16\linewidth]{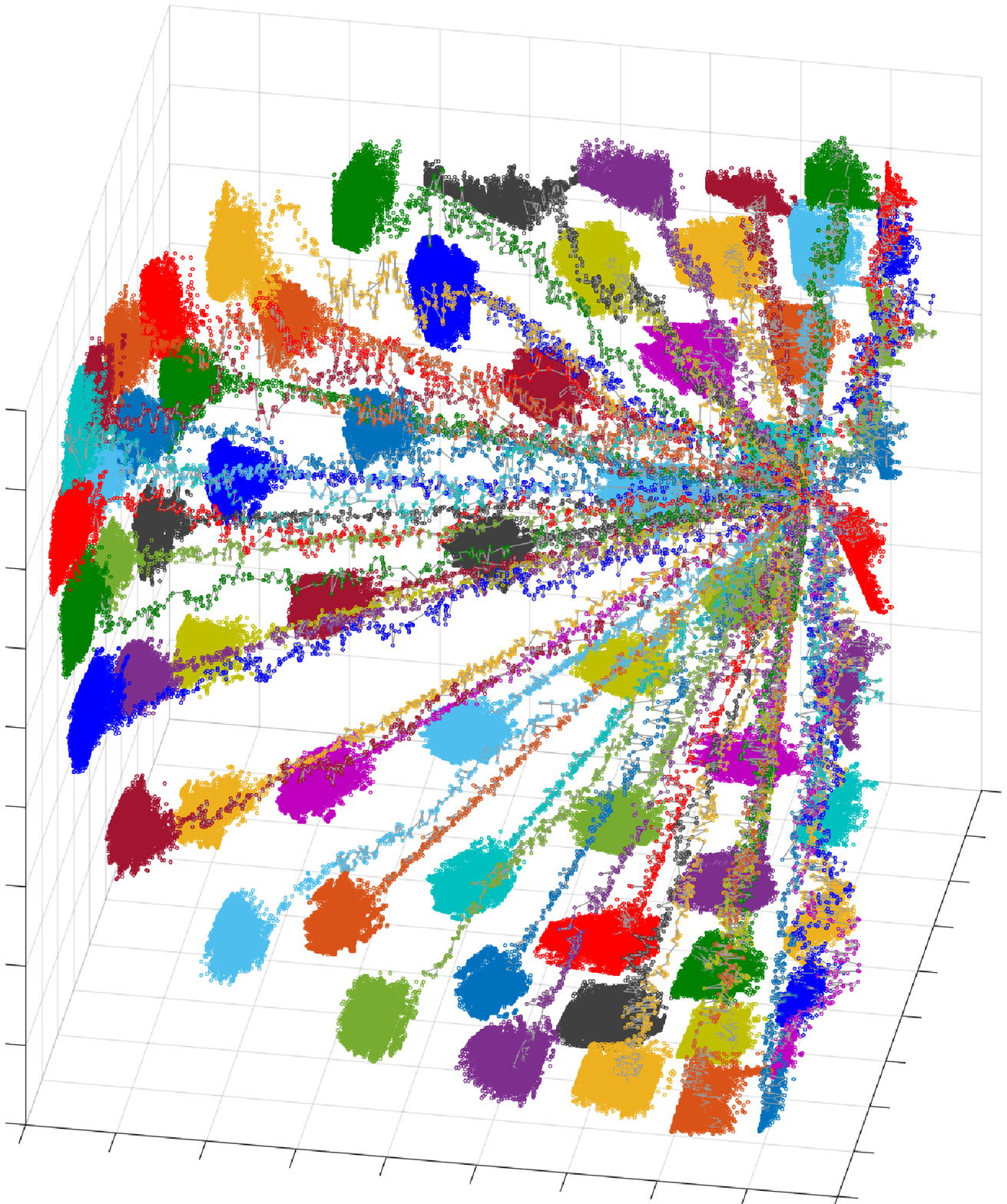}}\\%epm_201801291655
  %\vspace{-0.1in}
  \caption{The trajectory trees of a run of PLS, a run of PLS-ABI and a run of PPLS/D(H=10) on the three-objective mUBQP instances mubqp\_3\_200 and the three-objective mTSP instances kroABC100. The algorithms are started from randomly generated solutions. }\label{fig:trees_m3}
  %\vspace{-0.2in}
\end{figure*}

\begin{figure*}%[H]
  %\vspace{-0.1in}
  %\centering
  \subfigure[\tiny{PLS on mubqp\_3\_200}]{
    \label{fig:tree_m3_PLS_100_HQI} %% label for first subfigure
    \includegraphics[width=0.16\linewidth]{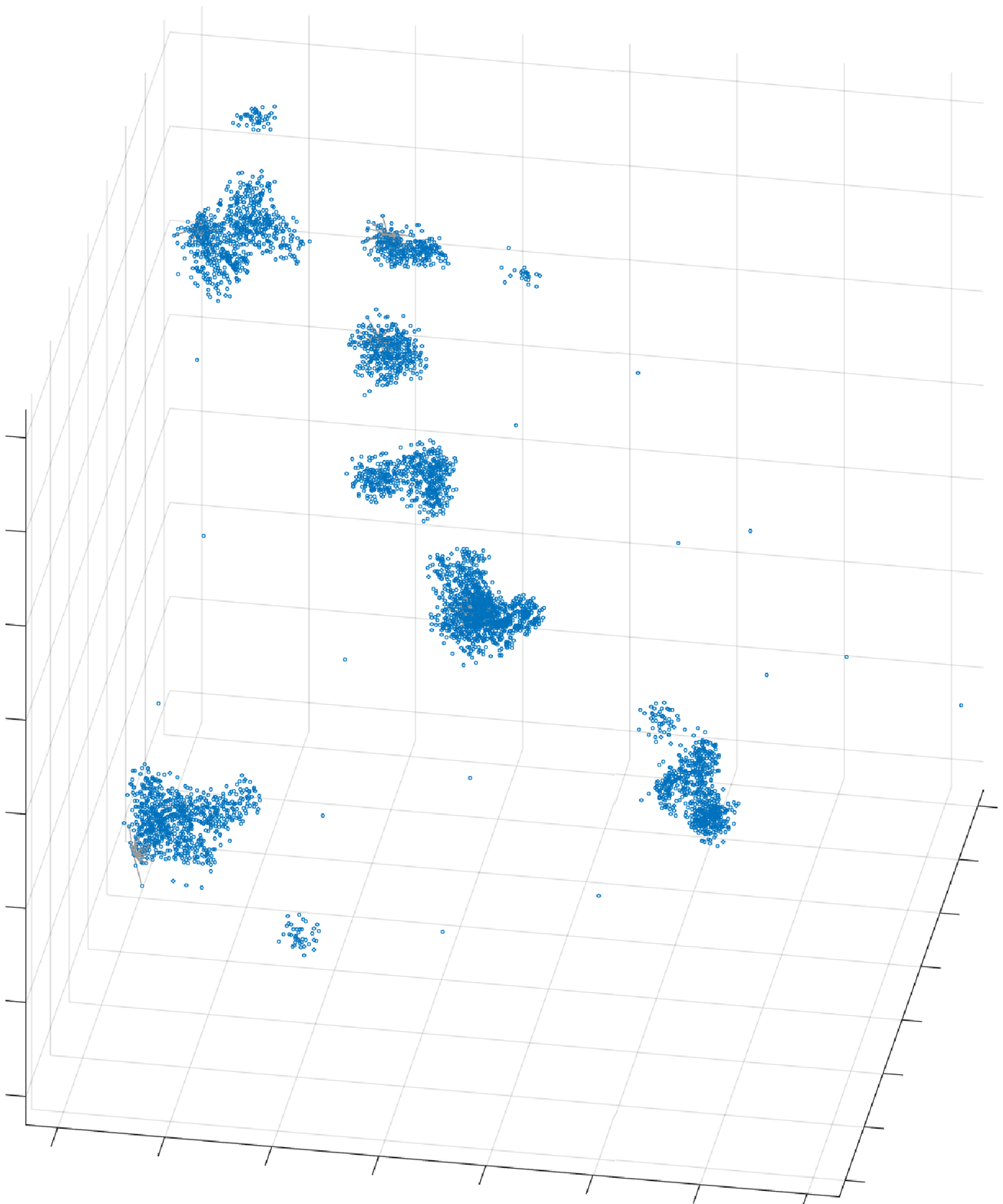}}%epm_201802051149
    %\hspace{-0.08in}
  \subfigure[\tiny{PLS-ABI on mubqp\_3\_200}]{
    \label{fig:tree_m3_PLSABI_100_HQI} %% label for first subfigure
    \includegraphics[width=0.16\linewidth]{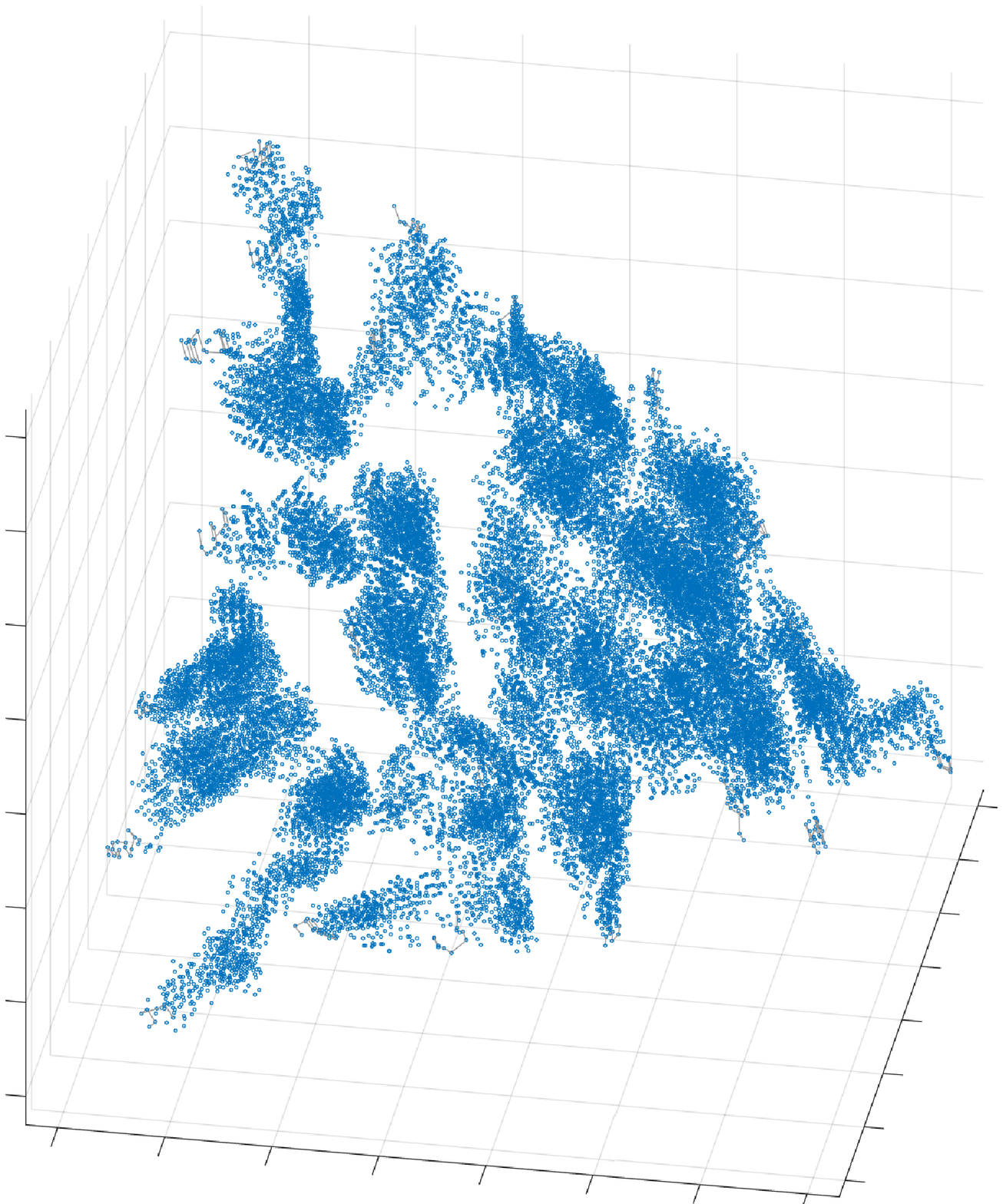}}%epm_201802051149
    %\hspace{-0.08in}
  \subfigure[\tiny{PPLS/D(H=6) on mubqp\_3\_200}]{
    \label{fig:tree_m3_PPLSD_100_HQI} %% label for first subfigure
    \includegraphics[width=0.16\linewidth]{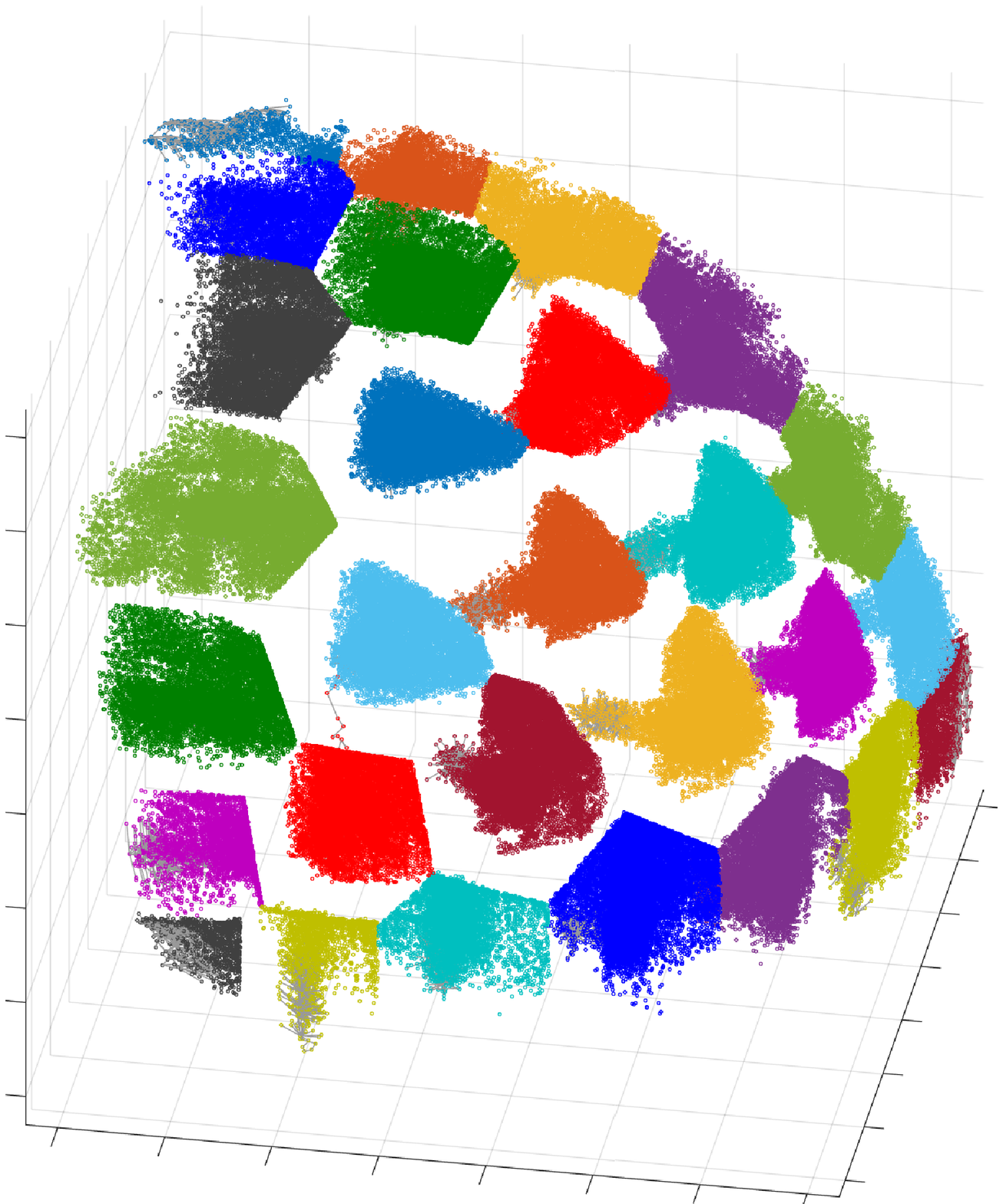}}%epm_201802051149
    %\hspace{-0.08in}
  \subfigure[\tiny{PLS on kroABC100}]{
    \label{fig:tree_kroABC100_PLS_100_HQI} %% label for first subfigure
    \includegraphics[width=0.16\linewidth]{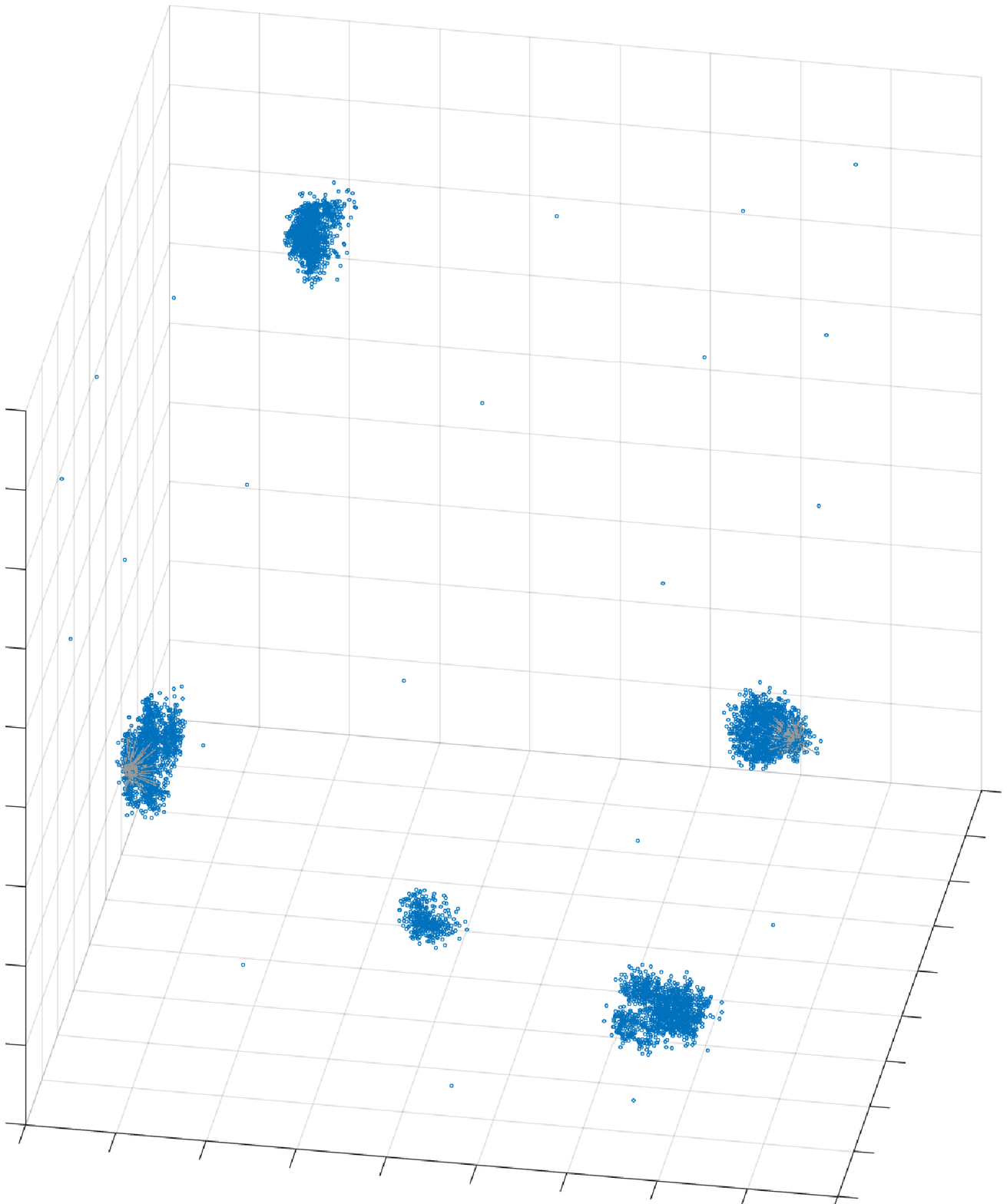}}%epm_201802121153
    %\hspace{-0.08in}
  \subfigure[\tiny{PLS-ABI on kroABC100}]{
    \label{fig:tree_kroABC100_PLSABI_100_HQI} %% label for first subfigure
    \includegraphics[width=0.16\linewidth]{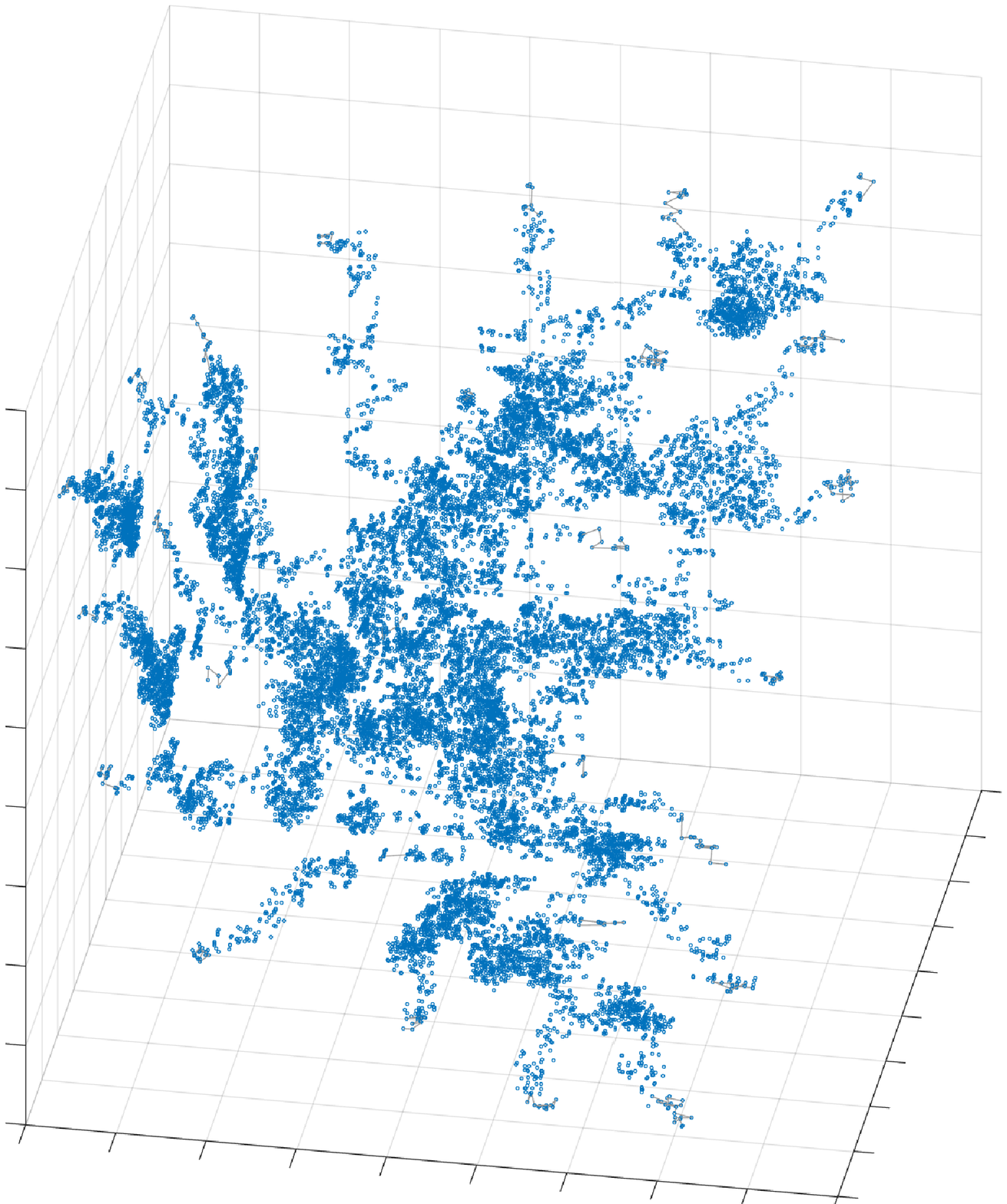}}%epm_201802121153
    %\hspace{-0.08in}
  \subfigure[\tiny{PPLS/D(H=6) on kroABC100}]{
    \label{fig:tree_kroABC100_PPLSD_100_HQI} %% label for first subfigure
    \includegraphics[width=0.16\linewidth]{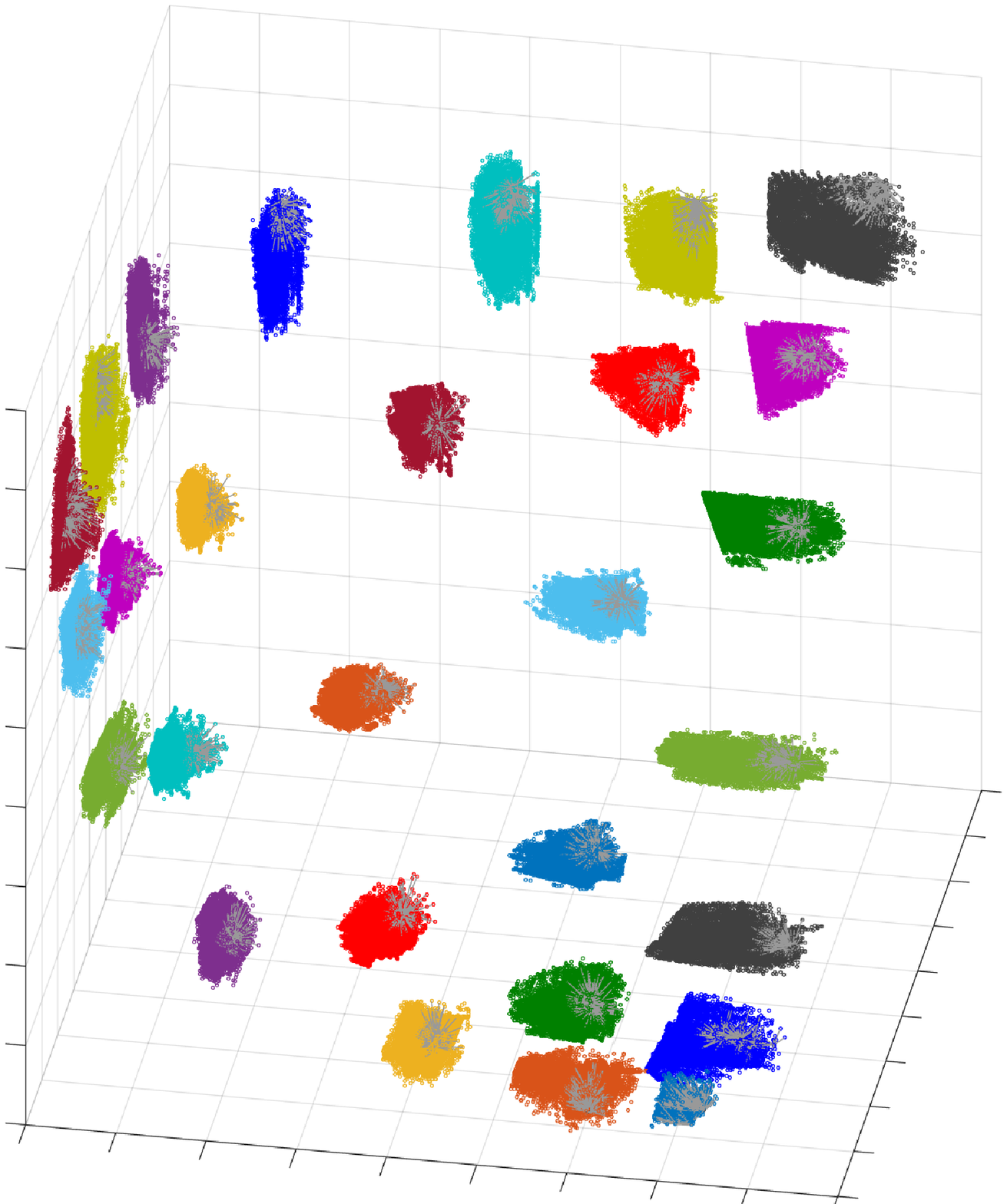}}\\%epm_201802121153
  %\vspace{-0.1in}
  \caption{The trajectory trees of a run of PLS, a run of PLS-ABI and a run of PPLS/D(H=6) on the three-objective mUBQP instances mubqp\_3\_200 and the three-objective mTSP instances kroABC100.  The algorithms are started from high quality solutions. }\label{fig:trees_m3_HQI}
  %\vspace{-0.2in}
\end{figure*}

The trajectory tree also can be plotted on the three-objective MCOPs. Fig.~\ref{fig:trees_m3} shows the trajectory trees of PLS, PLS-ABI and PPLS/D(H=10) on the three-objective mUBQP instances mubqp\_3\_200 and the three-objective mTSP instance kroABC100. In Fig.~\ref{fig:trees_m3} the algorithms are started from randomly generated solutions. On mubqp\_3\_200 the runtime budget is 1000s and on kroABC100 the runtime budget is 100s. Fig.~\ref{fig:trees_m3_HQI} shows the trajectory trees of PLS, PLS-ABI, PPLS/D(H=6) when they are started from high quality solutions on the mUBQP instance mubqp\_3\_200 and the mTSP instance kroABC100. On both mubqp\_3\_200 and kroABC100 the runtime budget is 100s.

\end{appendices}

\end{document}